\documentclass[twocolumn,twocolappendix]{aastex631}

\usepackage{amsmath}

\hypersetup{linkcolor=red,citecolor=green,filecolor=cyan,urlcolor=magenta}

\newcommand{\p}[1]{ ^{ \left ( #1 \right ) } } 
\newcommand{\avg}[1]{\left \langle #1 \right \rangle} 
\newcommand{\eq}[1]{\begin{equation} #1 \end{equation}} 

\newcommand{\abs}[1]{\left | #1 \right |} 

\newcommand{\parens}[1]{\left ( #1 \right )} 
\newcommand{\brackets}[1]{\left [ #1 \right ]} 

\renewcommand{\eqref}[1]{Eq.~\ref{#1}}
\newcommand{\secref}[1]{Sec.~\ref{#1}}
\newcommand{\figref}[1]{Fig.~\ref{#1}}
\newcommand{\tabref}[1]{Tab.~\ref{#1}}

\newcommand{\unitVec}[1]{\mathbf{\hat{#1}}} 

\newcommand{\nhat}{\unitVec{n}}
\newcommand{\ellvec}{\boldsymbol{\ell}}

\newcommand{\HI}{H{\textsc i}}

\newcommand{\rev}[1]{#1}

\shorttitle{Magnetic misalignment}
\shortauthors{Cukierman, Clark \& Halal}
\graphicspath{{./}{}}

\begin{document}

\title{Magnetic Misalignment of Interstellar Dust Filaments}

\correspondingauthor{Ari J. Cukierman}
\email{ajcukier@caltech.edu}

\author[0000-0002-7471-719X]{Ari J. Cukierman}
\affiliation{Department of Physics, Stanford University, Stanford, California 94305, USA}
\affiliation{Kavli Institute for Particle Astrophysics and Cosmology,
SLAC National Accelerator Laboratory, 2575 Sand Hill Rd, Menlo Park, California 94025, USA}
\affiliation{Department of Physics, California Institute of Technology, Pasadena, CA 91125, USA}

\author[0000-0002-7633-3376]{S.~E. Clark}
\affiliation{Department of Physics, Stanford University, Stanford, California 94305, USA}
\affiliation{Kavli Institute for Particle Astrophysics and Cosmology,
SLAC National Accelerator Laboratory, 2575 Sand Hill Rd, Menlo Park, California 94025, USA}

\author[0000-0003-2221-3018]{George Halal}
\affiliation{Department of Physics, Stanford University, Stanford, California 94305, USA}

\begin{abstract}

We present evidence for scale-independent misalignment of interstellar dust filaments and magnetic fields.  We estimate the misalignment by comparing millimeter-wave dust-polarization measurements from \emph{Planck} with filamentary structures identified in neutral-hydrogen~(\HI) measurements from \HI4PI. We find that the misalignment angle displays a scale independence (harmonic coherence) for features larger than the \HI4PI beam width~($16.2'$). We additionally find a spatial coherence on angular scales of~$\mathcal{O}(1^\circ)$. We present several misalignment estimators formed from the auto- and cross-spectra of dust-polarization and \HI-based maps, and we also introduce a map-space estimator. Applied to large regions of the high-Galactic-latitude sky, we find a global misalignment angle of~$\sim 2^\circ$, which is robust to a variety of masking choices. By dividing the sky into small regions, we show that the misalignment angle correlates with the parity-violating~$TB$ cross-spectrum measured in the \emph{Planck} dust maps.  The misalignment paradigm also predicts a dust $EB$~signal, which is of relevance in the search for cosmic birefringence but as yet undetected; the measurements of~$EB$ are noisier than of~$TB$,  and our correlations of~$EB$ with misalignment angle are found to be weaker and less robust to masking choices.  We also introduce an \HI-based dust-polarization template constructed from the Hessian matrix of the \HI~intensity, which is found to correlate more strongly than previous templates with \emph{Planck} dust $B$~modes.

\end{abstract}



\section{Motivation}

We continue the investigation of \emph{magnetic misalignment} from \citet{Clark2021}, which sought an explanation for the parity-violating $TB$~correlation measured in Galactic dust polarization by the \emph{Planck} satellite at millimeter wavelengths~\citep{Planck2018XI}.  A polarization field can be generically decomposed into parity-even $E$~modes and parity-odd $B$~modes~~\citep{Seljak1997,Kamionkowski1997}. The $TB$~cross-spectrum is a measure of the correlation between the total intensity~$T$ and the $B$-mode polarization and indicates a net chirality in the polarization field.  The $TE$~cross-spectrum is a correlation with the $E$-mode polarization and is non-chiral. Using the dust-dominated frequency channel centered at~$353~\mathrm{GHz}$, \citet{Planck2018XI} reported $TB/TE \sim 0.1$ in the multipole range $40 < \ell < 600$\rev{, which roughly corresponds to angular scales of $1$-$10^\circ$}. 

Parity-probing cross-spectra such as~$TB$ and~$TE$ are of interest both in studies of the interstellar medium~(ISM),  for which the observed cross spectra may constrain magnetohydrodynamic~(MHD) models~\citep{Caldwell2017,Kandel2017,Kritsuk2018,Kim2019}, and in measurements of the cosmic microwave background~(CMB), for which asymmetries in the Galactic foregrounds can bias polarization calibration~\citep{Abitbol2016} and confound searches for cosmic birefringence~\citep{Minami2020}.  A magnetic helicity in the local ISM~\citep{Brandenburg2005,Blackman2015} could produce a nonzero $TB$~correlation, and \citet{Bracco2019} produced toy models with positive~$TB$ and~$TE$ on large scales~(multipoles $\ell \lesssim 20$).  

The polarization of interstellar dust emission is a probe of Galactic magnetic fields, the observed polarization orientation being perpendicular to the plane-of-sky (POS) magnetic field~\citep{Stein1966,Hildebrand1988,Martin2007}.  At the same time, the dust in the diffuse ISM is organized partially in filamentary structures that are preferentially aligned to the magnetic field~\citep{PlanckXXXII2016,Planck2016XXXVIII}.  Filamentary structures can also be identified in neutral hydrogen~(\HI),  which is well mixed with dust~\citep{Lenz2017} and has the advantage of three-dimensional information from spectroscopic separation into velocity bins.  The alignment between \HI\ filaments and magnetic-field orientations has been confirmed by comparison to millimeter-wave polarization and optical starlight polarization~\citep{McClure-Griffiths2006,Clark2014,Clark2015,Martin2015,Kalberla2016}.

Whereas previous work, e.g., \citet{ClarkHensley2019}, has assumed a perfect alignment between interstellar dust filaments and magnetic-field lines,  it was suggested in \citet{Huffenberger2020} that a small misalignment could act as a mechanism for parity violation\rev{, i.e., a tendency toward features of one chirality (or handedness) over the other}.  In \citet{Clark2021}, this idea was extended to allow dust filaments and magnetic-field orientations to display a \emph{scale-dependent} misalignment, which could potentially account for the observed~$TB$.  

In this work, we directly compute the misalignment angle in many regions of the sky and in many multipole bins for $\ell > 100$. We find evidence for scale \emph{independence} of the misalignment angle and also for a correlation with the observed dust~$TB$.

\subsection{Observed~$TB$ \label{sec:IntroTB}}

The dust~$TB$ was reported in \citet{PlanckXXX2016}, where it was noted that a positive signal in the multipole range $60 < \ell < 130$ became more significant as the sky area was increased.  The investigation was continued in \citet{Planck2018XI} with the observation that $TB/TE \sim 0.1$ for $40 < \ell < 600$.  The $EB$~signal was reported to be consistent with null. 

In \citet{Weiland2020}, the $TB$~signal was further confirmed by using WMAP $K$-band polarization~\citep{Page2007} in place of the \emph{Planck} $B$~modes and also by using the magnetic-field template from~\citet{Page2007} that is based on optical starlight-polarization catalogs~\citep{Heiles2000,Berdyugin2001,Berdyugin2002,Berdyugin2004}. The $K$-band measurement is dominated by synchrotron rather than dust emission but is also a probe of Galactic magnetic fields. The starlight measurements largely probe the same magnetic dust-grain alignment that produces polarized millimeter-wave emission. Both choices are independent of the \emph{Planck} polarization calibration, and both show positive~$TB$.

\subsection{Magnetic misalignment \label{sec:IntroMagMisalign}}

Magnetic misalignment is a discrepancy between the orientation of filamentary density structures and the polarization-inferred magnetic-field lines. In the case of perfect magnetic alignment, we expect $TE>0$ and $TB=0$~\citep{Zaldarriaga2001}. A misalignment of~$45^\circ$ would produce $TE=0$ and $TB \not = 0$, where the sign depends on the chirality of the misalignment. The robustly positive~$TE$ measured by \emph{Planck} can be interpreted as supportive evidence for magnetic alignment of dust filaments~\citep{Clark2015,Planck2016XXXVIII,Kalberla2016}.  In \citet{Planck2018XI}, the $TE$~correlation over sky regions and multipoles is reported as $r_\ell^{TE} = D_\ell^{TE} / \sqrt{D_\ell^{TT} D_\ell^{EE}} = 0.357 \pm 0.003$,\rev{ where $D_\ell^{XY}$~denotes the cross-spectrum of~$X$ and~$Y$; }the $TB$~correlation is reported as $r_\ell^{TB} = D_\ell^{TB} / \sqrt{D_\ell^{TT} D_\ell^{BB}} \approx 0.05$.  Since the $TB$~correlation is much smaller than the $TE$~correlation, the magnetically \emph{aligned} model need only be perturbed a small but coherent amount in order to produce the observed~$TB$, and this perturbation would also produce a positive~$EB$~\citep{Huffenberger2020}\rev{, though this~$EB$ would be obscured by \emph{Planck} noise~\citep{Clark2021}.}

In \citet{Clark2021}, it was suggested that an \HI-based filamentary polarization template could be used as a comparison point in the search for magnetic misalignment.  The template of \citet{ClarkHensley2019} is constructed by 1)~quantifying the orientation of linear \HI\ structures with the Rolling Hough Transform~\citep[RHT,][]{Clark2014} in velocity-channel maps from \HI4PI~\citep{HI4PI2016},  2)~assuming perfect alignment between the RHT-measured \HI\ orientation and the POS magnetic-field orientation and thereby obtaining a prediction for the dust polarization angle,  3)~applying weights based on the \HI~intensity, and 4)~integrating the channel maps to form a template that can be compared to the measured millimeter-wave dust polarization. A strong correlation with the \emph{Planck} 353-GHz maps is detected in both $E$~and $B$~modes up to the \HI4PI beam scale of~$16.2'$ ($\ell \lesssim 1000$).  The \HI4PI-based template was used in \citet{Clark2021}, but \citet{ClarkHensley2019} also constructed polarization templates with observations from the Galactic Arecibo L-Band Feed Array \HI\ Survey~\citep[GALFA-\HI,][]{Peek2018}, which has higher angular resolution~($\mathrm{FWHM} = 4.1'$) but smaller sky coverage~(32\%~of the celestial sphere).

Given the alignment between \HI\ and dust filaments, a difference between the \emph{Planck}-measured dust polarization angles and the \HI-inferred angles is a potential indication of magnetic misalignment and could be used as a tracer of dust~$TB$ and~$EB$~\citep{Clark2021}. In extending the work of \citet{Clark2021}, we measure the aggregate misalignment angle in different sky regions by using the \HI~template as a reference. We study the observed properties of this misalignment angle and its correlation with the measured dust~$TB$ and~$EB$.

\citet{Clark2021}~also introduced a scale-dependent effective misalignment angle~$\psi_\ell$, which is a function of multipole~$\ell$. This effective misalignment angle is given explicitly by~\citep[\rev{cf. }Eq.~11 of][] {Clark2021}
\eq{
	\psi_\ell \equiv \frac{1}{2} \operatorname{arctan}\frac{D_\ell^{TB}}{D_\ell^{TE}} ,  \label{eq:psiellDef}
}
where we see that the ratio~$TB/TE$ is the controlling quantity.\footnote{\rev{We use the notation~$D_\ell^{XY}$ to denote the cross-spectrum of~$X$ and~$Y$, but we will often refer to this quantity in the text with the shorthand~$XY$.}} As noted in \citet{Planck2018XI}, the ratio~$TB/TE$ is approximately constant across a broad range of multipoles at high Galactic latitudes, and this is related to the observation of \citet{Clark2021} that $\psi_\ell \sim 5^\circ$ in the range $100 \lesssim \ell \lesssim 500$ on a similar sky area.  

Equation~\ref{eq:psiellDef} provides an estimator for the effective misalignment angle. Because it is formed from the $TB$~and $TE$~cross-spectra, we will call this type of estimator ``spectrum-based''.  In this work, we present several additional spectrum-based estimators by considering the auto- and cross-spectra of the \emph{Planck} dust maps and the \HI~templates. We also present a \emph{map-based} estimator that is similar to the projected Rayleigh statistic~(PRS) of~\citet{Jow2017}.  We test for consistency among these estimators.

Although \citet{Clark2021}~allowed for scale dependence in the misalignment angle, we find in this work that $\psi_\ell$~tends to display a scale \emph{independence} even when measured on small regions of sky. Equivalently, we find that $\psi_\ell$~is roughly constant with~$\ell$, which we will occasionally refer to as \emph{harmonic coherence}. 

It is important to note that the dust is likely \rev{organized} \emph{only partially} \rev{in} filaments, which are in turn \emph{only partially} captured by the \HI~template. We expect, therefore, that there are contributions to the dust polarization that are unrelated to the \HI~template and, more generally, unrelated even to the true underlying filamentary structure.  An estimator like that of \eqref{eq:psiellDef} may be influenced by these non-filamentary contributions, since it depends only on the~$TB$ and~$TE$ cross-spectra of the \emph{full} dust maps. Some of the estimators we will introduce in later sections will be defined by reference to the \HI~template, which will partially but imperfectly restrict the analysis to modes which are related to filaments.  

In contrast to the previous paragraph,  the {\tt DUSTFILAMENTS} code of \citet{HerviasCaimapo2022} constructs a phenomenological dust model, which is composed \emph{entirely} of filaments and which reproduces the main features of the angular power spectra measured by \emph{Planck}. Using this model,  it was recently shown in \citet{Huang2022} that the measured~$TB$ is unlikely to be a statistical fluctuation of an underlying parity-even distribution, if the assumptions of the {\tt DUSTFILAMENTS} code represent the true sky.

\subsection{Cosmic birefringence \label{sec:IntroCB}}

Cosmic birefringence is an observable consequence of certain types of parity-violating physics beyond the Standard Model and manifests as a rotation of the plane of linear polarization of photons~\citep{Carroll1990,Harari1992,Carroll1998}. A popular source of cosmic birefringence is an electromagnetically-coupled axion-like field, which can behave as both dark matter and dark energy~\citep{Marsh2016}.   In the CMB, the polarization rotation can be detected as an $EB$~correlation~\citep{Lue1999,Feng2005,Feng2006,Liu2006}. A $TB$~correlation should also be produced, but it is typically a less sensitive observable on account of the large cosmic variance in~$T$. 

There are several species of cosmic birefringence that have been investigated in the literature. An isotropic, static cosmic birefringence manifests as an overall polarization rotation by the same angle along every line of sight. This observable is, unfortunately, degenerate with a miscalibration of the instrumental polarization orientation~\citep{Yadav2010}.  The degeneracy is sometimes exploited as a means of ``self-calibration'' by \emph{assuming} a standard cosmology in which the true~$EB$ vanishes~\citep{Keating2013}. Although this type of calibration removes sensitivity to an isotropic, static cosmic birefringence, it is still possible to search for cosmic birefringence which is \emph{anisotropic}~\citep{PB2015,BKIX,Namikawa2020,Bianchini2020} or \emph{time-variable}~\citep{BKXII,BKXIV,Ferguson2022}.
Through a campaign of modeling and calibration, it is possible to account for instrumental systematics and measure the isotropic, static cosmic-birefringence angle. Recent measurements of this kind are consistent with a standard cosmology~\citep{Kaufman2014,Gruppuso2016,PlanckIntXLIX2016,Choi2020}. 

A new technique was proposed in \citet{Minami2019}, which exploits the fact that the Galactic foregrounds are subject only to polarization miscalibration and \emph{not} to cosmic birefringence. The observed CMB is rotated by \emph{both} miscalibration and a possible cosmic birefringence. With measurements at multiple frequencies,  the calibration angles and the cosmic-birefringence angle can be extracted simultaneously. Applied to \emph{Planck} 2018 polarization data~\citep{Planck2018III}, a cosmic-birefringence angle~\rev{$\beta = 0.35 \pm 0.14^\circ$}, a discrepancy with the null hypothesis with a significance of~$2.4\sigma$, was reported in \citet{Minami2020} under the assumption of a vanishing dust~$EB$.  With the newer \emph{Planck} maps produced by the {\tt NPIPE} pipeline~\citep{PlanckIntLVII2020}, the same prescription produced~\rev{$\beta = 0.30 \pm 0.11^\circ$} as reported in \citet{DiegoPalazuelos2022}.  Recently, a similar analysis that includes WMAP polarization data~\citep{Bennett2013} produced the consistent but stronger result $\beta = {0.342^\circ}^{+0.094^\circ}_{-0.091^\circ}$~\citep{Eskilt2022}.
In these two recent cosmic-birefringence analyses, the impact of a possible foreground $EB$~correlation was incorporated by two different approaches, one of which was based on the filamentary misalignment paradigm of \citet{Huffenberger2020} and \citet{Clark2021}.  When accounting for a possible foreground~$EB$, the birefringence angle varies as a function of sky fraction but remains positive.  \citet{DiegoPalazuelos2022} refrain from an estimate of statistical significance due to the currently limited understanding of foreground polarization, while \citet{Eskilt2022} quote a significance of~$3.6\sigma$ but acknowledge that the foreground polarization must be better understood to be confident that the measured~$EB$ is cosmological rather than Galactic. The study of magnetic misalignment is, therefore, of central importance in the search for cosmic birefringence.

\subsection{Outline}

In \secref{sec:data}, we describe the data products used throughout the analysis. In \secref{sec:HessianMethod}, we introduce a new filamentary polarization template that relies on the Hessian matrix of \HI~intensity maps. In \secref{sec:misalignmentAnsatz}, we present our misalignment \emph{ansatz}, i.e., our assumptions of how misalignment perturbs the dust polarization in both map space and harmonic space.  We derive misalignment estimators in terms of the auto- and cross-spectra of the \emph{Planck} dust maps and the \HI-based polarization templates, and we test some immediate consequences of these relations.  In \secref{sec:simulations}, we describe a set of \rev{mock skies} that we have used to check our estimators.  These \rev{mock skies} are constructed to match the 2-point statistics of the \emph{Planck} dust maps including cross-spectra with the \HI~template.  In \secref{sec:misalignmentEstimator}, we introduce a map-based misalignment estimator and present tentative evidence for a global misalignment angle of~$\sim 2^\circ$.  In \secref{sec:misalignment}, we divide the sky into small patches and present evidence for scale independence~(harmonic coherence) of magnetic misalignment as well as evidence of spatial coherence.  In \secref{sec:parityViolatingCrossSpectra}, we present evidence for a scale-independent relation between magnetic misalignment and parity-violating cross-spectra such as~$TB$ and~$EB$.  We close in \secref{sec:conclusion} with suggestions for improvements in our analysis and new directions to further the investigation of parity violation in Galactic dust polarization.

\section{Data \label{sec:data}}

We use the \emph{Planck} {\tt Commander} dust maps~\citep{PlanckIV2018} as our fiducial measurements of the on-sky thermal dust emission in Stokes~$T$,  $Q$ and~$U$.  The maps are constructed by component separation applied to the nine \emph{Planck} frequency maps, whose passband centers span $30$-$857~\mathrm{GHz}$, though polarization is available only for the seven bands spanning $30$-$353~\mathrm{GHz}$. \rev{\emph{Half-mission} maps are available and are constructed using data exclusively from either the first or the second half of the \emph{Planck} observation period. When forming cross-spectra, we will often use these half-mission splits in order to avoid positive-definite noise biases.}

In this work, our \HI~template is derived from the \HI4PI survey~\citep{HI4PI2016}, a set of full-sky maps of the 21-cm hyperfine transition with an angular resolution of~$16.2'$, a sensitivity~$\sigma_\mathrm{rms} = 43~\mathrm{mK}$ and a velocity (spectral) resolution of~$\delta v = 1.49~\mathrm{km}/\mathrm{s}$. The \HI4PI survey is a combination of the Parkes Galactic All-Sky Survey~\citep[GASS,][]{McClureGriffiths2009} and the Effelsberg-Bonn \HI\ Survey~\citep[EBHIS,][]{Winkel2016}. The GASS observations cover the southern sky in the velocity range $-470 \leq v_\mathrm{lsr} \leq 470~\mathrm{km}/\mathrm{s}$, and the EBHIS observations cover the northern sky in the velocity range $-600 \leq v_\mathrm{lsr} \leq 600~\mathrm{km}/\mathrm{s}$.  At high Galactic latitudes, nearby dust is generally expected to be associated with lower-velocity \HI~emission, i.e., with small~$|v_\mathrm{lsr}|$, so our dust-polarization template is drawn from the range $-15 \leq v_\mathrm{lsr} \leq 4~\mathrm{km}/\mathrm{s}$, a choice that is motivated in more detail in \secref{sec:HessianPrescription}.

We compute purified power spectra with {\tt NaMaster}~\citep{Alonso2019} using a $C^2$~apodization window~\citep{Grain2009} with a scale of~$1^\circ$. Before computing power spectra, we smooth the {\tt Commander} maps to $16.2'$,  the \HI4PI beam width. We use a {\tt HEALPix} pixelization scheme~\citep{Gorski2005} and downgrade all maps to $N_\mathrm{side} = 256$ for faster power-spectrum estimation.  We spot-checked some of our results at higher~$N_\mathrm{side}$ and find that they are consistent. 

\subsection{Galaxy masks \label{sec:galaxyMasks}}

\begin{figure}
\includegraphics[width=\columnwidth]{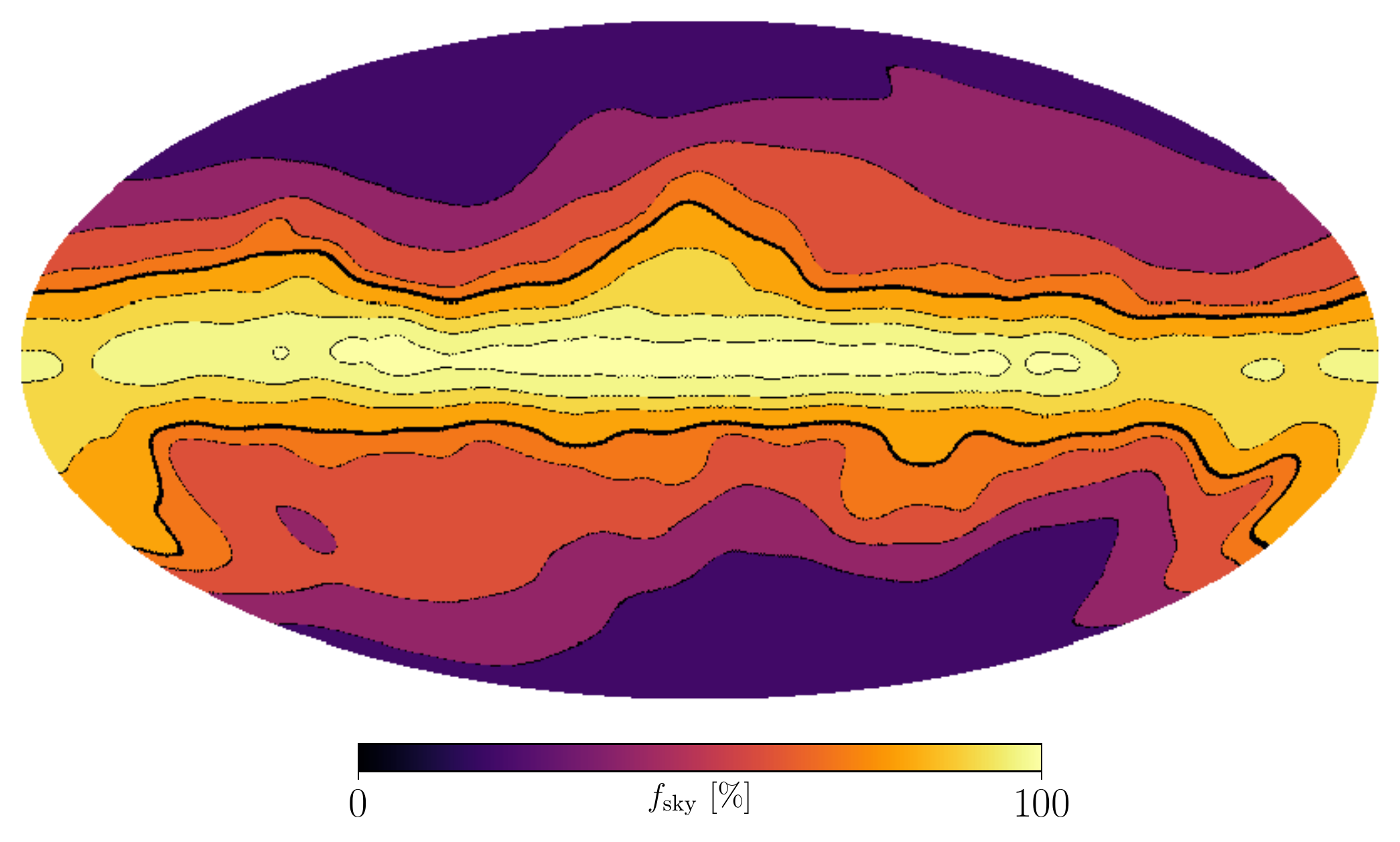}
\caption{The Galaxy masks described in \secref{sec:galaxyMasks}.  Each color indicates the sky fraction~$f_\mathrm{sky}$, where the darker colors are meant to be subsumed by the masks associated with the lighter colors.  \rev{The 70\%~mask, which is our fiducial choice in much of the analysis, is indicated by the thicker black boundaries. } \label{fig:maskMap}}
\end{figure}

We use the Galaxy masks provided by the \emph{Planck} Legacy Archive.\footnote{\url{pla.esac.esa.int}} These masks are constructed to limit Galactic emission to varying levels. The masks with smaller sky fraction~$f_\mathrm{sky}$ restrict the analysis to relatively high latitudes. The masks with larger~$f_\mathrm{sky}$ allow more contributions from nearer the Galactic plane. The set of Galaxy masks is shown in \figref{fig:maskMap}.
 Our fiducial mask in much of the analysis is defined by $f_\mathrm{sky} = 70\%$, and we will refer to it as the ``70\% Galaxy mask''. 

\subsection{Notation}

We use the subscript~``$\mathrm{\HI}$'' to denote quantities derived from the \HI-based polarization template.  For example, the \HI-based prediction for dust $E$~modes is denoted by~$E_\mathrm{\HI}$. It is important to note that these quantities are describing \HI-based predictions for the polarization of \emph{dust} rather than polarization properties of the \HI~itself. The \HI~is measured in total (unpolarized) intensity, and prescriptions like the Hessian method of \secref{sec:HessianMethod} convert those intensity maps into dust-polarization \emph{templates}.

We use the subscript~``$\mathrm{d}$'' to denote quantities related to Galactic dust.  Usually, this will refer specifically to the \emph{Planck} {\tt Commander} maps described above.

\subsection{Bandpass filtering \label{sec:bandpassFiltering}}

Much of our analysis is restricted to $\ell > 100$, and we often form maps which are bandpass filtered.  We filter by applying an $\ell$-dependent Tukey window to the spherical-harmonic representation of the maps. We use a taper of length~$t_\ell = 50$, which produces a flat-topped passband when the window width is larger than~$2 t_\ell$. We tested these filters on full-sky \emph{Planck} dust maps and find the out-of-band response to be suppressed by a factor of more than~$10^4$. In particular, the out-of-band leakage is below the level of the high-latitude dust power (computed on the \emph{Planck} 70\%~Galaxy mask), even when the filtered power spectra are computed on the full sky, i.e., including the Galactic Plane.

\section{Hessian method \label{sec:HessianMethod}}

We introduce a new Hessian-based filament-finding algorithm~\citep[similar to those of, e.g.,][]{Planck2016XXXVIII,Kalberla2021}.  Whereas previous work on misalignment \citep{Clark2021} used a filamentary model based on the Rolling Hough Transform~\citep[RHT,][]{Clark2014,ClarkHensley2019},   we find that our new Hessian-based polarization template correlates more strongly with \emph{Planck} measurements of $B$-mode dust polarization for \rev{$\ell \gtrsim 100$~(\figref{fig:RHTvsHess}).  Furthermore, whereas the RHT loses its correlation with \emph{Planck} $B$~modes for $\ell \gtrsim 400$, the Hessian maintains a correlation of $\sim 10\%$ up to our highest multipoles~($\ell_\mathrm{max} = 767$).} In $E$~modes, the two methods correlate with \emph{Planck} at roughly equivalent levels.  

\rev{We additionally prefer the Hessian method for its relative computational efficiency.  The Hessian method requires only two operations in spherical-harmonic space, while the RHT requires a suite of convolutions to sample polarization angles. Direct comparisons will be presented in future work~\citep{HalalPrep}.}

The Hessian matrix contains information about the local second derivatives. By searching for regions of \emph{negative curvature} in an intensity map, we find candidate filaments. Negative curvature implies that at least one of the Hessian eigenvalues is negative. The orientation of the filament is determined by the local eigenbasis. As in, e.g., \citet{Clark2015}, we assume that the plane-of-sky (POS) filament is aligned with the POS magnetic field. The dust polarization is taken to be orthogonal to the filament.  With these assumptions, we can convert an intensity map into a polarization \emph{template}.

Hessian-based filament identifications have been performed, e.g., on 353-GHz maps in \citet{PlanckXXXII2016,Planck2016XXXVIII}, on \HI4PI \HI~and \emph{Planck} 857-GHz maps in \citet{Kalberla2021}, on \emph{Herschel} images of molecular clouds~\citep{Polychroni2013} and on simulations of the cosmic web~\citep{Colombi2000,ForeroRomero2009}. In addition to the RHT, some non-Hessian filament-finding algorithms that have been applied to studies of the ISM include {\tt DisPerSE}~\citep{Sousbie2011,Arzoumanian2011} and {\tt getfilaments}~\citep{Menshchikov2013}. See Sec.~3.10 of \citet{Hacar2022} for a more comprehensive review.

\subsection{Prescription \label{sec:HessianPrescription}}

We use the Hessian matrix to identify filament orientations. To construct \HI-based templates for dust polarization,  we form weights from the Hessian \emph{eigenvalues}.  

We analyze the \HI~maps in individual velocity bins. Our final polarization template is produced by summing over velocities. The \HI~intensity in velocity channel~$v$ is denoted~$I_v$. We work in spherical coordinates with polar angle~$\theta$ and azimuthal angle~$\phi$. The local Hessian matrix is given by 
\eq{ H \equiv \parens{ \begin{array}{cc} H_{xx} & H_{xy} \\
							    H_{yx} & H_{yy} \end{array} } , }
where		    
\begin{eqnarray}
		 H_{xx}  & = & \frac{\partial^2 I_v}{\partial \theta^2} ,  \\
		 H_{yy} & = & \frac{1}{\sin^2 \theta} \frac{\partial^2 I_v}{\partial \phi^2},  \\ 
		 H_{xy} = H_{yx} & = & - \frac{1}{\sin\theta} \frac{\partial^2 I_v}{\partial \phi \partial \theta} . 
\end{eqnarray}
The eigenvalues are
\eq{ \lambda_{\pm} = \frac{1}{2} \parens{ H_{xx} + H_{yy} \pm \alpha } , }
where
\eq{ \alpha \equiv \sqrt{ \parens{ H_{xx} - H_{yy} }^2 + 4 H_{xy}^2 } . }
The candidate polarization angle is then
\eq{ \theta_v = \arctan \parens{ \frac{H_{xx} - H_{yy} + \alpha}{2 H_{xy}} } , \label{eq:thetav} }
but we will enforce conditions below to ensure this identification is sensible. 

First, for the local curvature to be negative along at least one axis, we need $\lambda_{-} < 0$. Second, we want this negative curvature to be the \emph{dominant} local morphology, so we require $\lambda_{-}$~to be the larger of the two eigenvalues in magnitude. Define
\eq{ \Delta \lambda \equiv \abs{ \lambda_{-}} - \abs{\lambda_{+} } . }
Then we define the velocity-dependent weight
\eq{ w_v \equiv \abs{ \lambda_{-} } \brackets{ \Delta \lambda > 0 } \brackets{ \lambda_{-} < 0 } , }
where the \emph{Iverson brackets} on the right-hand side enforce conditions on~$\Delta\lambda$ and~$\lambda_{-}$.\footnote{For a statement~$p$, the Iverson bracket~$\brackets{p}$ is~$1$ when $p$~is true and~$0$ when $p$~is false~\citep{Iverson1962,Knuth1992}.} 

Our velocity-dependent Stokes templates are given by
\begin{eqnarray}
T_v(\nhat) & \equiv & w_v(\nhat) ,  \label{eq:Hessian Tv} \\
Q_v(\nhat) & \equiv & w_v(\nhat) \cos \brackets{ 2 \theta_v (\nhat) } ,  \\ 
U_v(\nhat) & \equiv & w_v(\nhat) \sin \brackets{ 2 \theta_v (\nhat) } .   
\end{eqnarray}
The Hessian method is susceptible to small-scale noise and scan artifacts, so we restrict our analysis to the \HI4PI velocity bins with greatest sensitivity. We start with the binning of \citet{ClarkHensley2019}. As a proxy for noisiness, we search for pixels with intensities that are reported to be \emph{negative}. We remove any velocity slice that contains negative-intensity pixels on the \emph{Planck} 70\% Galaxy mask. This leaves a continuous range between~$-15$ and~$4~\mathrm{km}/\mathrm{s}$.  The velocity selection is intended to avoid numerical pathologies and should be revisited in a future iteration of the Hessian algorithm.  Most of the \HI~emission is at low velocities \citep[see, e.g., Fig.~1 of][]{ClarkHensley2019}, so we are retaining the dominant contributions even with the current velocity cuts. \rev{The velocity cut affects our analysis mainly in terms of sensitivity, since there is potentially useful information about dust filaments in the velocity slices that are discarded.  In general, sensitivity is greater when the \HI~template correlates more strongly with \emph{Planck} dust polarization.  We defer to future work an investigation of the potential improvements from differently chosen velocity cuts~\citep{HalalPrep}.}   We repeated a majority of the following calculations with the RHT-based template~\rev{(\secref{sec:HessianSupp})} that uses the much broader velocity range of~$-90$ to~$90~\mathrm{km}/\mathrm{s}$ as in \citet{ClarkHensley2019}, and we find consistent results.
The full templates are given by
\eq{ X_\mathrm{\HI}(\nhat) \equiv \sum_v X_v(\nhat) \label{eq:XsumDef} }
for $X \in \{T,Q,U\}$.

\begin{figure*}
\includegraphics[width=\textwidth]{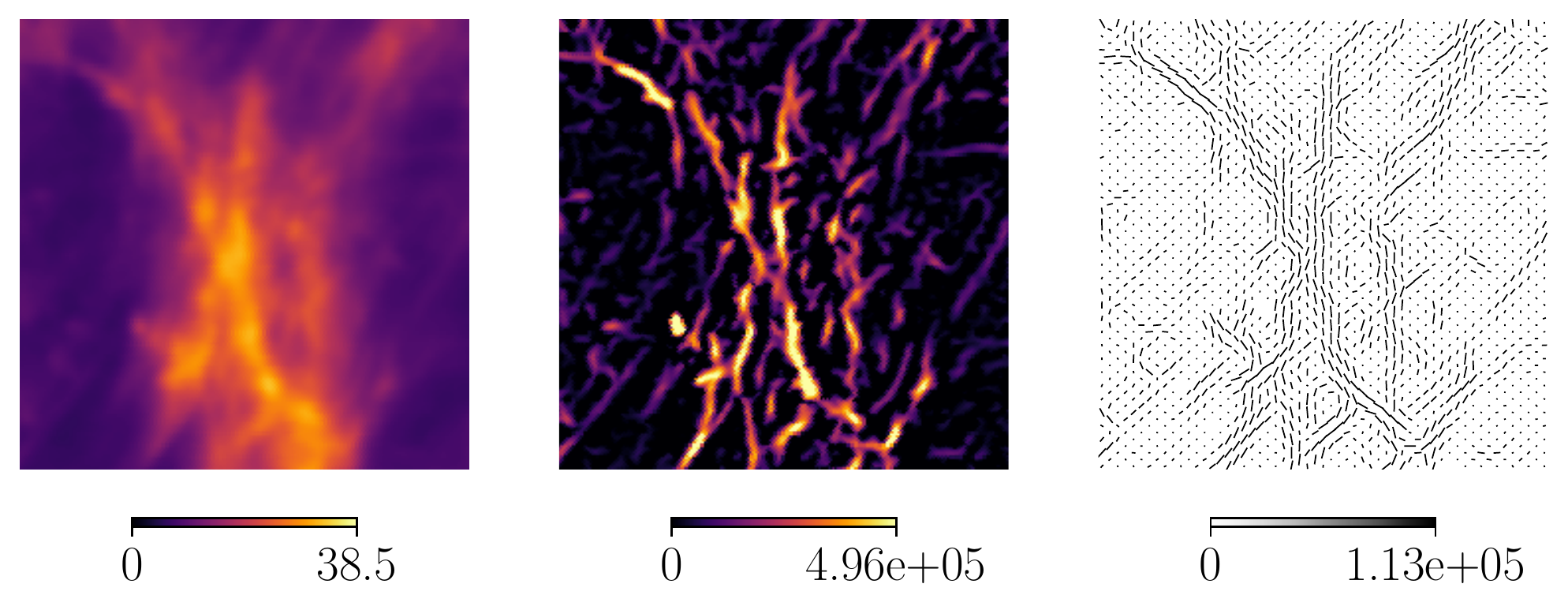}
\caption{Illustration of the Hessian-based filament-finding algorithm (\secref{sec:HessianMethod}) on a sky region of area $10^\circ \times 10^\circ$ centered on $(l,b) = (12^\circ,45^\circ)$ in a velocity bin centered on $-4.4~\mathrm{km}/\mathrm{s}$ with a width of~$1.3~\mathrm{km}/\mathrm{s}$. The units for all figures are~$\mathrm{K}~\mathrm{km}/\mathrm{s}$, but note that the right two figures are produced by taking second derivatives of intensity with respect to angular coordinates in radians. (\emph{Left}) Raw \HI4PI intensity. (\emph{Middle}) Hessian intensity~$T_v$  (\eqref{eq:Hessian Tv}), which upweights regions of negative curvature and produces structure that is visually filamentary. (\emph{Right}) Hessian-derived magnetic-field orientations (orthogonal to the polarization angle~$\theta_v$ from \eqref{eq:thetav}) tracing the orientation of the filaments. \label{fig:HessianDemo}}
\end{figure*}

An illustration of our Hessian method is provided in \figref{fig:HessianDemo}, where we analyze a region with area $10^\circ \times 10^\circ$ centered on $(l,b) = (12^\circ,45^\circ)$.
The velocity bin is centered on $-4.4~\mathrm{km}/\mathrm{s}$ with a width of~$1.3~\mathrm{km}/\mathrm{s}$.  The panels of \figref{fig:HessianDemo} show how the raw \HI4PI intensity map is transformed into a filamentary intensity~$w_v$ and how the \rev{filament orientations determine the inferred magnetic-field orientations.}

Additional material related to our Hessian method is provided in \secref{sec:HessianSupp}.

\section{Misalignment ansatz \label{sec:misalignmentAnsatz}}

As an ansatz for the observable signature of magnetic misalignment, we assume a multipole-dependent rotation angle~$\psi_\ell$ as in \citet{Clark2021}.  We denote the \emph{observed} $E$~and $B$~modes by~$E(\ellvec)$ and~$B(\ellvec)$ and the unphysical modes that would be observed in the \emph{absence} of misalignment by~$\tilde{E}(\ellvec)$ and~$\tilde{B}(\ellvec)$, where $\ellvec$~identifies a particular spherical harmonic with multipole moment~$\ell$. Our ansatz takes the form
\eq{ \parens{ \begin{array}{c} E(\ellvec) \\ B(\ellvec) \end{array} } = 
			\parens{ \begin{array}{cc} \cos\parens{2 \psi_\ell} & -\sin\parens{2 \psi_\ell} \\
							\sin\parens{2 \psi_\ell} & \cos\parens{2 \psi_\ell} \end{array} }
			\parens{ \begin{array}{c} \tilde{E}(\ellvec) \\ \tilde{B}(\ellvec) \end{array} } . \label{eq:EBmixing} }
For the purposes of the ansatz, we are imagining that \emph{all} of the dust polarization participates in the misalignment. As mentioned in \secref{sec:IntroMagMisalign}, this assumption is likely inaccurate, since some of the dust \rev{morphology} is non-filamentary.  Later, we will form estimators by comparing the observed dust polarization with the predictions of the \HI~template, and this will restrict the analysis to the filamentary modes that we do expect to be described by the ansatz of \eqref{eq:EBmixing} (in the misalignment paradigm). 

\begin{figure}
\includegraphics[width=\columnwidth]{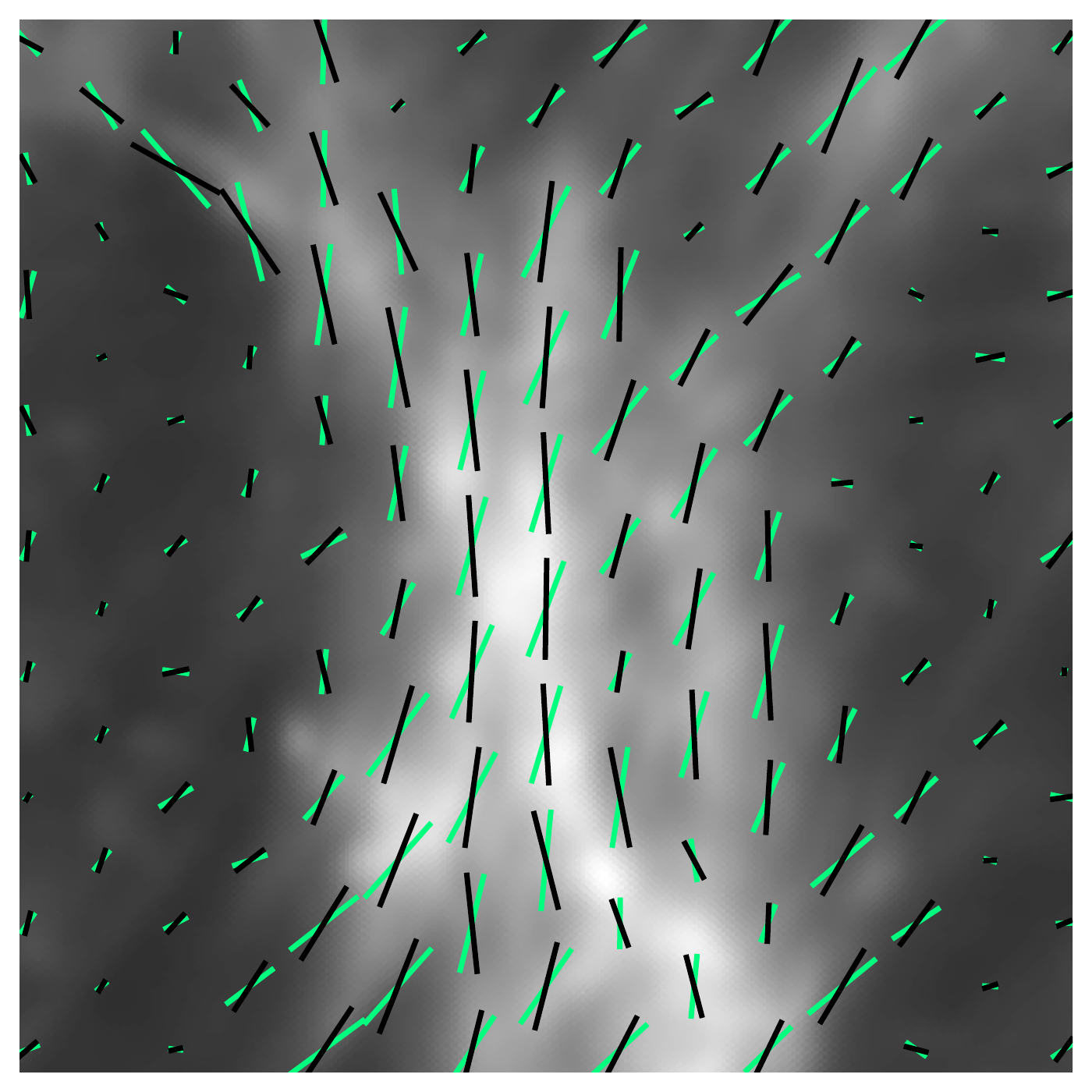}
\caption{Illustration of magnetic misalignment. (\emph{Black})~Magnetic-field orientations derived from our \HI~template for the same sky region and velocity bin as in \figref{fig:HessianDemo} but downgraded to $N_\mathrm{side} = 64$. The pseudovector lengths are proportional to the template-implied polarization intensity.  (\emph{Green})~The same after applying a global misalignment angle $\psi = 20^\circ$, which is an unrealistically large amplitude for better visualization.  (\emph{Greyscale colormap})~The raw \HI~intensity map~(identical to the leftmost panel of \figref{fig:HessianDemo}).  \label{fig:mapmisalign}}
\end{figure}

To make magnetic misalignment less abstract, we provide an illustration in \figref{fig:mapmisalign}.  
We consider perfect alignment~(black) and scale-independent misalignment~(green).  Perfect alignment is assumed by the \HI-based filamentary model of dust polarization~(\secref{sec:IntroMagMisalign}).  For scale-independent misalignment, we take $\psi_\ell = 20^\circ$ for all~$\ell$. This is a much larger misalignment than we expect to measure on the true sky, but the exaggeration is useful for visualization.  In this case, the magnetic field shows a consistent rotation by the same amount and with the same sense relative to the \HI~template. 

\subsection{Assumptions}

The \HI-based filamentary model of dust polarization assumes \emph{perfect alignment}.   We observe in \secref{sec:intrinsic parity} that the \HI~model displays no intrinsic parity violation.  We, therefore, assume $E_\mathrm{\HI}$~correlates with~$\tilde{E}_\mathrm{d}$ but \emph{not} with~$\tilde{B}_\mathrm{d}$, and we make a symmetric assumption for~$B_\mathrm{\HI}$. Our assumptions are summarized by
\eq{ D_\ell^{E_\mathrm{\HI} \tilde{B}_\mathrm{d}} = D_\ell^{B_\mathrm{\HI} \tilde{E}_\mathrm{d}} = D_\ell^{T_x \tilde{B}_\mathrm{d}} = 0 \label{eq:EBTB assumptions} }
for $x \in \{\mathrm{\HI},\mathrm{d}\}$.

\subsection{Implications for cross-spectra \label{sec:implicationsCrossSpectra} }

With Eqs.~\ref{eq:EBmixing} and~\ref{eq:EBTB assumptions}, we can derive the following relations between observable cross-spectra in terms of the misalignment angle~$\psi_\ell$:
\begin{eqnarray}
D_\ell^{E_\mathrm{\HI} B_\mathrm{d}} & = & \tan\parens{2 \psi_\ell} D_\ell^{E_\mathrm{\HI} E_\mathrm{d}} ,  \label{eq:EHBpsi relation} \\
D_\ell^{B_\mathrm{\HI} E_\mathrm{d}} & = & -\tan\parens{2 \psi_\ell}  D_\ell^{B_\mathrm{\HI} B_\mathrm{d}} ,  \label{eq:BHEpsi relation} \\
D_\ell^{T_x B_\mathrm{d}} & = & \tan\parens{2 \psi_\ell} D_\ell^{T_x E_\mathrm{d}} \label{eq:TBpsi relation}
\end{eqnarray}
and
\eq{ D_\ell^{E_\mathrm{d} B_\mathrm{d}} = \frac{1}{2} \tan\parens{4 \psi_\ell} \parens{ D_\ell^{E_\mathrm{d} E_\mathrm{d}} - D_\ell^{B_\mathrm{d} B_\mathrm{d}} } .  \label{eq:EBrelation} }
From the known positive~$T_\mathrm{d} B_\mathrm{d}$ and~$T_\mathrm{d} E_\mathrm{d}$ measured by \emph{Planck}~(\secref{sec:IntroTB}), we expect~$\psi_\ell$ to be mostly positive in the range $100 \lesssim \ell \lesssim 500$ on large sky areas away from the Galactic plane, e.g., with a 70\%~Galaxy mask~\citep{Clark2021}. We also know that $E_\mathrm{\HI} E_\mathrm{d} > 0$, $B_\mathrm{\HI} B_\mathrm{d} > 0$~\citep{ClarkHensley2019} and $E_\mathrm{d} E_\mathrm{d} > B_\mathrm{d} B_\mathrm{d}$~\citep{PlanckXXX2016} across the same multipole range and on the same sky area. Our qualitative expectations, then, are to find $E_\mathrm{\HI} B_\mathrm{d}$, $T_\mathrm{\HI} B_\mathrm{d}$ and~$E_\mathrm{d} B_\mathrm{d}$ to be \emph{positive} but to find $B_\mathrm{\HI} E_\mathrm{d}$ to be \emph{negative}.

We can make simple estimates of~$\psi_\ell$ with Eqs.~\ref{eq:EHBpsi relation}, \ref{eq:BHEpsi relation} and~\ref{eq:TBpsi relation}, though each is potentially biased by noise in the denominator:
\eq{ \tan(2 \psi_\ell) = \frac{D_\ell^{E_\mathrm{\HI} B_\mathrm{d}}}{D_\ell^{E_\mathrm{\HI} E_\mathrm{d}}} = - \frac{D_\ell^{B_\mathrm{\HI} E_\mathrm{d}}}{D_\ell^{B_\mathrm{\HI} B_\mathrm{d}}} = \frac{D_\ell^{T_\mathrm{\HI} B_\mathrm{d}}}{D_\ell^{T_\mathrm{\HI} E_\mathrm{d}}} = \frac{D_\ell^{T_\mathrm{d} B_\mathrm{d}}}{D_\ell^{T_\mathrm{d} E_\mathrm{d}}} . \label{eq:tanpsi estimates} }
We could form a similar estimate from \eqref{eq:EBrelation}, but the $E_\mathrm{d} B_\mathrm{d}$~measurement from \emph{Planck} is especially noisy, so we ignore it for the remainder of this section. 
The four cross-spectrum ratios in \eqref{eq:tanpsi estimates} allow for tests of the misalignment ansatz without explicit calculation of~$\psi_\ell$.

While positive~$T_\mathrm{\HI} B_\mathrm{d}$ and~$E_\mathrm{\HI} B_\mathrm{d}$ might be anticipated on account of the known positive~$T_\mathrm{d} B_\mathrm{d}$,  $T_\mathrm{d}T_\mathrm{\HI}$, $T_\mathrm{d} E_\mathrm{d}$ and~$E_\mathrm{\HI} E_\mathrm{d}$,  it is, in principle, possible for the $T_\mathrm{d} B_\mathrm{d}$~signal to be entirely decoupled from the \HI-correlated components of the dust maps. In \secref{sec:simulations}, we describe how to construct \rev{mock skies} with exactly this property. These \rev{mock skies} show positive~$T_\mathrm{d} B_\mathrm{d}$ but zero~$T_\mathrm{\HI} B_\mathrm{d}$ and zero~$E_\mathrm{\HI} B_\mathrm{d}$.  While we consider $T_\mathrm{\HI}B_\mathrm{d},E_\mathrm{HI}B_\mathrm{d} > 0$ to be the most plausible expectation, it is formally nontrivial.

\begin{figure}
\includegraphics[width=\columnwidth]{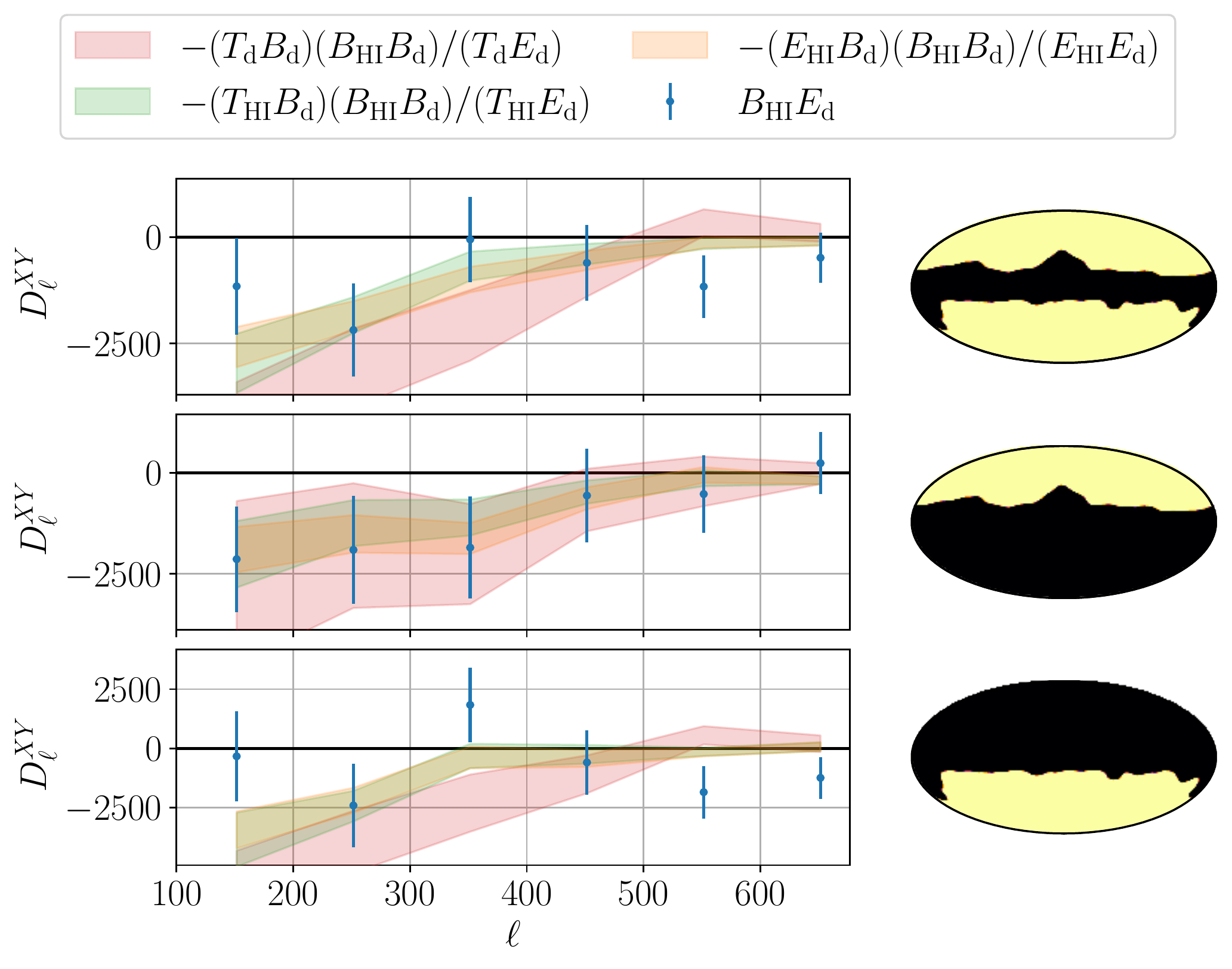}
\caption{The $B_\mathrm{\HI} E_\mathrm{d}$~cross-spectrum measured on the \emph{Planck} 70\%~Galaxy mask.  The spectra are shown for the full mask (top), the northern hemisphere of that mask (middle) and the southern hemisphere (bottom).  The shaded bands show three different expectations for~$B_\mathrm{\HI} E_\mathrm{d}$ based on other measured spectra~(\eqref{eq:tanpsi estimates}).  The units are $\mu\mathrm{K}_\mathrm{RJ}~\mathrm{K}~\mathrm{km}/\mathrm{s}$, and the error bars and bands are derived from Gaussian variances.  The expectations and the measurements show a broad consistency, in particular, the tendency for $B_\mathrm{\HI} E_\mathrm{d} < 0$, which persists in each hemisphere independently. \label{fig:EBfsky3}}
\end{figure}

The negativity of~$B_\mathrm{\HI} E_\mathrm{d}$ is a new prediction of the misalignment ansatz. We can make quantitative predictions for this signal~(\eqref{eq:tanpsi estimates}), and comparisons are shown in \figref{fig:EBfsky3} for our fiducial 70\%~Galaxy mask~(\secref{sec:galaxyMasks}).
We mentioned above that we expect $\psi_\ell$~to be smooth over the multipole range $100 \lesssim \ell \lesssim 500$, so we expect $B_\mathrm{\HI} E_\mathrm{d}$~to be smooth over similar multipoles.  We can, therefore, gain in per-bandpower sensitivity by using the relatively large bin width of~$\Delta\ell = 100$.

In \figref{fig:EBfsky3}, we find that $B_\mathrm{\HI} E_\mathrm{d}$~tends negative and is broadly consistent with the expectations of \eqref{eq:tanpsi estimates} over the full mask and in the northern and southern hemispheres independently.  Due to the unavailability of suitable dust and \HI~simulations, we do not attempt a statistical evaluation of the consistency. The plotted error bars are derived from Gaussian variances. As the dust field displays both non-Gaussianity and statistical anisotropy, these variances are meant only as a rough indication of the fidelity of the measurements.

We expect $B_\mathrm{\HI} E_\mathrm{d}$~to be noisier than~$E_\mathrm{\HI} B_\mathrm{d}$ and~$T_\mathrm{\HI} B_\mathrm{d}$, because $r_\ell^{B_\mathrm{\HI} B_\mathrm{d}}$~is smaller (by roughly a factor of~$2$-$3$ for $\ell > 100$) than $r_\ell^{E_\mathrm{\HI} E_\mathrm{d}}$ and $r_\ell^{T_\mathrm{\HI} T_\mathrm{d}}$, i.e., $B_\mathrm{\HI}$~is a less accurate representation of~$B_\mathrm{d}$ than $E_\mathrm{\HI}$~or $T_\mathrm{\HI}$~is of~$E_\mathrm{d}$ or~$T_\mathrm{d}$, respectively.  The \HI-based polarization template is, therefore, more sensitive to $\tilde{E}_\mathrm{d}$~modes mixed into the observed~$B_\mathrm{d}$ than to $\tilde{B}_\mathrm{d}$~modes mixed into~$E_\mathrm{d}$ (\eqref{eq:EBmixing}).

We consider the results of \figref{fig:EBfsky3} to be a first step in confirming that the misalignment ansatz of \eqref{eq:EBmixing} is at least a partial description of the true sky.  These results avoided an explicit calculation of the misalignment angle~$\psi_\ell$. In later sections, we will compute~$\psi_\ell$ directly.

\section{\rev{Mock skies} \label{sec:simulations} }

We construct a set of mock sky realizations in order to check for biases and spurious signals in the estimators that we will introduce in subsequent sections. We maintain the 2-point statistics of the true sky including correlations with the \HI-based polarization templates.  These \rev{mock skies} are phenomenological in the sense that they produce realistic observables without explicit appeal to the underlying ISM physics; in particular, these are \emph{not} numerical ISM simulations.

Our \rev{mock skies} include Gaussian noise, an \HI-based filamentary component and Gaussian dust.  We arrange for all of the 2-point statistics to be the same as for the true sky, i.e., the \rev{mock skies} replicate the measured~$X_a Y_b$ for $X,Y \in \{T,E,B\}$ and $a,b \in \{\mathrm{d},\mathrm{\HI}\}$. The \HI-based component is the same for all realizations and is derived from the true-sky Hessian template~(\secref{sec:HessianMethod}).

In harmonic space, we express the \rev{mock-sky}~($S$) map as a linear combination of an $\ell$-filtered \HI~template, a Gaussian dust component~($G$) and a Gaussian noise component~($N$):
\eq{ X_S(\ellvec) = k_\ell^{(X)} X_\mathrm{\HI}(\ellvec) + X_G(\ellvec) + X_N(\ellvec) \label{eq:sim linear combo} }
for $X \in \{T,E,B\}$, where $X_\mathrm{\HI}(\ellvec)$~is the harmonic-space representation of the Hessian template~(\secref{sec:HessianMethod}). The $\ell$-dependent coefficient in the \HI~term is necessary, because $D_\ell^{X_\mathrm{\HI} X_\mathrm{\HI}} \not \propto D_\ell^{X_\mathrm{\HI} X_\mathrm{d}}$.  To maintain $D_\ell^{X_\mathrm{\HI} X_S} = D_\ell^{X_\mathrm{\HI} X_\mathrm{d}}$,  we modify with the transfer function~$k_\ell^{(X)}$~(\secref{sec:TransferFunction}), which ensures consistency with the true \HI-\emph{Planck} cross-spectra.  While the \HI\ term is constant across realizations and based on the true sky, the Gaussian dust and noise are stochastic.

The power-spectra of the Gaussian dust and noise components are estimated from the measured dust and \HI\ power-spectra.  We calculate these spectra after applying the 70\%~Galaxy mask.  We compute~$X_\mathrm{\HI}(\ellvec)$ from a masked map as well.  As a result, the \rev{mock skies} are well-defined \emph{only} on the unmasked 70\%~of the \rev{celestial sphere}.

Unlike the CMB,  Galactic dust emission is statistically anisotropic, i.e., the statistics of the dust are different in different regions of sky. We approximate the non-stationarity by beginning with Gaussian noise and Gaussian dust that are \emph{isotropic} and then \emph{modulating} based on the statistical anisotropies in the {\tt Commander} maps.  The modulation is performed on scales much larger than those used in our analysis. The modulation field is smoothed to~$14.7^\circ$, twice the side length of a pixel with $N_\mathrm{side} = 8$, while most of our analysis is concerned with multipoles $\ell > 100$, i.e., degree scales and smaller. We, therefore, expect negligible mode mixing. 

\begin{figure*}
\includegraphics[width=\textwidth]{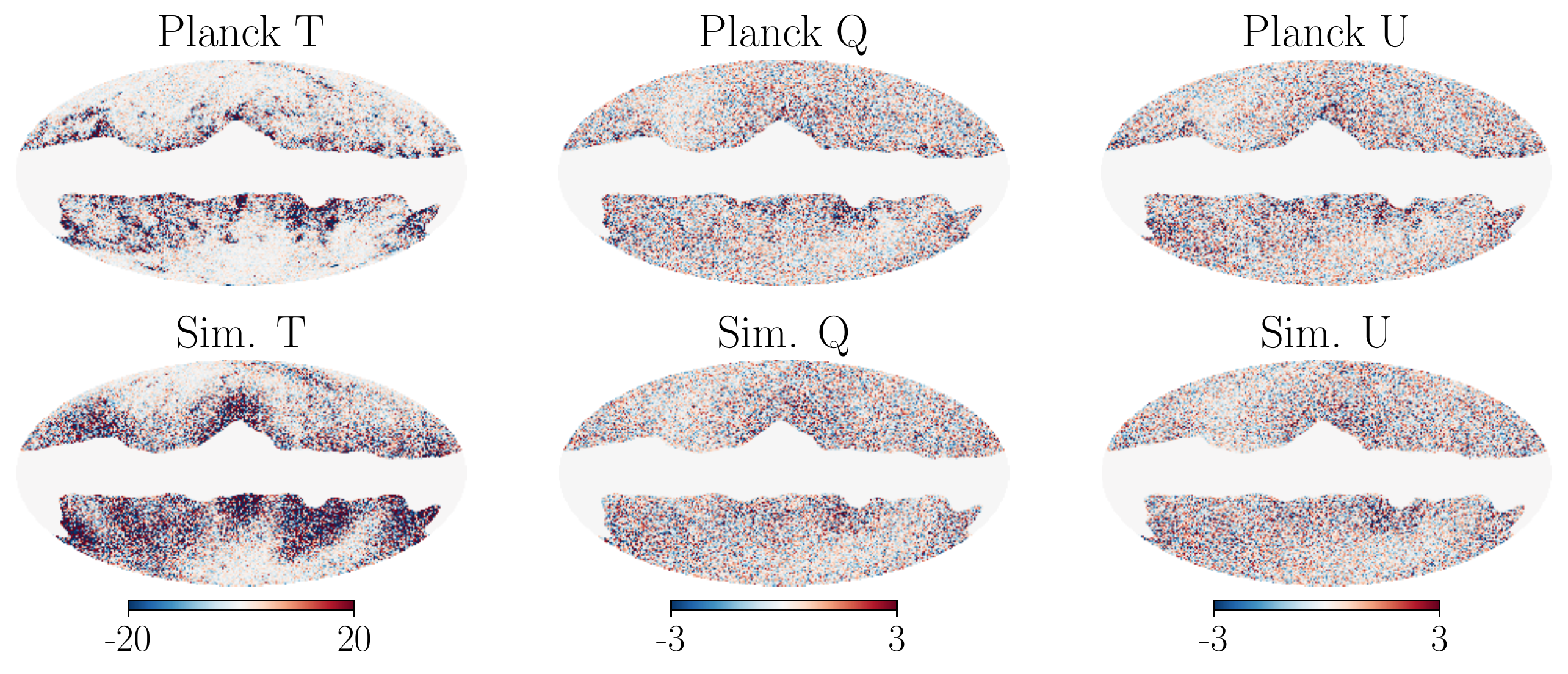}
\caption{A realization of the modulated Gaussian \rev{mock skies}~(\secref{sec:simulations}) after highpass filtering to $\ell > 100$.  The units of the color scales are~$\mu\mathrm{K}_\mathrm{RJ}$.  The modulated Gaussianity adds an additional level of realism to the \rev{mock skies}, but the strongly non-Gaussian features, especially in the \emph{Planck} $T_\mathrm{d}$~map, are not captured.  These \rev{mock skies} are used only for basic estimator tests, \emph{not} for statistical inference. \label{fig:mod sim map} }
\end{figure*}

A realization of the modulated dust \rev{mock sky} is shown in \figref{fig:mod sim map} after highpass filtering to $\ell > 100$, the multipole range targeted by most of our analysis. 
Before filtering, the \rev{mock skies} are dominated by large-scale modes, which are bright and relatively poorly estimated, but those large-scale modes are irrelevant for most of our analysis.  Visually, we find greater non-Gaussianity in the real $T_\mathrm{d}$~map than in the \rev{mock sky}. The polarization maps bear a greater resemblance to each other.  A higher level of realism is unnecessary, since we use these \rev{mock skies} only to check for biases in our estimators.

\begin{figure}
\includegraphics[width=\columnwidth]{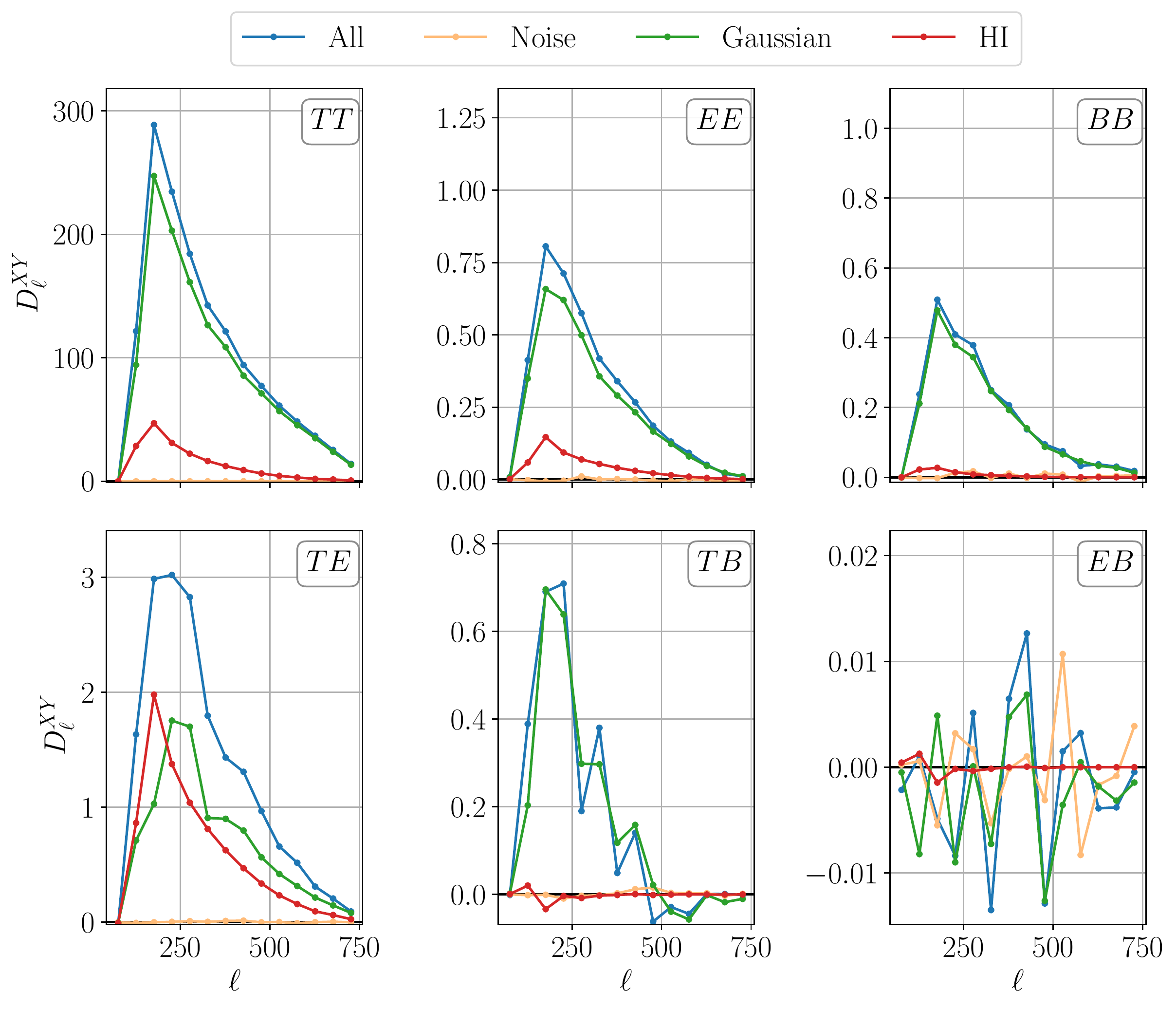}
\caption{Breakdown of \rev{mock-sky} components in power-spectrum space for the modulated-Gaussian realization shown in \figref{fig:mod sim map}~(with the same highpass filtering to $\ell > 100$).  The vertical-axis units are~$\mu\mathrm{K}_\mathrm{RJ}^2$. The annotations indicate the spectrum type, e.g., $TT$~for the upper-left plot.  Blue shows the half-mission cross-spectrum~$\tilde{D}_\ell^{XY}$ (\eqref{eq:unbiased Dell}). Orange shows the half-mission noise cross-spectrum. The Gaussian-dust spectrum is shown in green, and the \HI~spectrum is shown in red. We note that the \HI-correlated component represents a minority contribution to all of the spectra except~$TE$, which is a characteristic signature of a filamentary magnetically-aligned polarization model. \label{fig:sim components}}
\end{figure}

A breakdown of the \rev{mock-sky} components is shown in \figref{fig:sim components}, where we see that most of the dust power is in the Gaussian rather than the \HI~component.
We note that the unbiased power-spectrum estimator
\eq{\tilde{D}_\ell^{XY} \equiv \frac{1}{2} \parens{ D_\ell^{X^{(1)} Y^{(2)}} + D_\ell^{X^{(2)} Y^{(1)}} } \label{eq:unbiased Dell}}
 is crucial for avoiding a large noise bias in polarization, especially for $\ell \gtrsim 200$.  The \HI~component shows negligible~$TB$ and~$EB$ but strong~$TE$, which is arguably a defining characteristic of the filamentary polarization model.  Although the Gaussian component dominates in~$TT$, $EE$ and~$BB$, the \HI~component accounts for roughly \emph{half} of the $TE$~signal. 

We verified that the 2-point statistics of the \rev{mock skies} are approximately equivalent to those of the true dust maps. In particular, we check that $X_S Y_S$~(where ``$S$''~denotes a \emph{\rev{mock sky}}) matches~$X_\mathrm{d} Y_\mathrm{d}$ and that $X_\mathrm{\HI} Y_S$~matches~$X_\mathrm{\HI} Y_\mathrm{d}$.
The agreement is sufficient to test the estimators that we will introduce below.

We emphasize that our \rev{mock-sky} framework is \emph{not} intended to represent a null hypothesis for the purposes of statistical inference.  In particular, the \rev{mock skies} are missing much of the non-Gaussian structure in the true sky, even beyond the \HI-correlated component. Instead, because no aggregate misalignment has been input, these \rev{mock skies} are useful for testing our estimators for spurious signals.

\section{Misalignment estimator \label{sec:misalignmentEstimator}}

We present an estimator for the misalignment angle of a region of sky containing multiple pixels.  In \citet{Clark2021}, the angle difference between the dust and the \HI~template was computed by
\eq{ \Delta\theta   \equiv  \frac{1}{2} \operatorname{atan2} \brackets{ c_\mathrm{\HI} s_\mathrm{d} - s_\mathrm{\HI} c_\mathrm{d} , c_\mathrm{\HI} c_\mathrm{d} + s_\mathrm{HI} s_\mathrm{d} } , 
\label{eq:deltatheta} }
where $c_x \equiv \cos\parens{2\theta_x}$ and $s_x \equiv \sin\parens{2\theta_x}$.\footnote{\rev{We use the 2-argument arctangent~($\operatorname{atan2}$) to avoid quadrant ambiguities in the angle determination.}}
While \eqref{eq:deltatheta} measures the misalignment angle of a single pixel,  care must be taken in computing the mean over multiple pixels, because $\Delta\theta$~is a circular statistic. The values of~$\Delta\theta$ are restricted to~$[-90^\circ,90^\circ]$, but the endpoints of this range are physically identical. Naively averaging random values from this range will produce mean values that cluster near~$0$ instead of being uniformly distributed. As a result, noise fluctuations produce a multiplicative bias that suppresses the magnitude of~$\avg{\Delta\theta}$.

To account for the circularity of~$\Delta\theta$, we use a modified version of the projected Rayleigh statistic~\citep{Jow2017}, which is itself a form of alignment estimator~\citep[cf.~Sec.~6.2 of][]{ClarkHensley2019}. The essence of the method is to consider terms of the form $\cos\brackets{2\parens{\theta_\mathrm{d}(\nhat) - \theta_\mathrm{\HI}(\nhat)} - \psi}$, where $\theta_\mathrm{d}(\nhat)$~is the polarization angle measured by \emph{Planck}, $\theta_\mathrm{\HI}(\nhat)$~is the angle predicted by the \HI~template and $\psi$~is a free parameter independent of~$\nhat$ and representing the \emph{misalignment angle}. We sum such terms over the selected map pixels~$\nhat$ and \emph{maximize} with respect to~$\psi$. Denote the maximizing value by~$\hat{\psi}$.  When $\theta_\mathrm{d}(\nhat) - \theta_\mathrm{\HI}(\nhat)$ is random, $\hat{\psi}$~is also \emph{random}. This is a plausibility argument that $\hat{\psi}$~is unbiased, but we will describe an explicit test below.

Rather than simply summing the cosine terms described above, we upweight pixels with higher signal-to-noise ratio in polarization. 
The weights are proportional to the product of the signal-to-noise ratios for \emph{Planck} and \HI4PI. Denote the per-pixel weight by~$w(\nhat)$.

We form the alignment metric 
\eq{ \xi(\psi) \equiv \frac{1}{W} \sum_\nhat w(\nhat) \cos \brackets{ 2 \parens{ \theta_\mathrm{d}(\nhat) - \theta_\mathrm{\HI}(\nhat) - \psi } } , }
where $\psi$~is a free parameter and $W = \sum_{\nhat} w(\nhat)$.  The non-uniform weighting of the contributing pixels distinguishes our alignment metric from that of Sec.~6.2 of \citet{ClarkHensley2019}, but it is an estimator for the same quantity.  We maximize~$\xi(\psi)$ with respect to~$\psi$ and denote the maximizing value by~$\hat{\psi}$. 

We can calculate~$\hat{\psi}$ analytically with the following prescription.  Form 
\begin{eqnarray} A & \equiv & \frac{1}{W} \sum_\nhat w(\nhat) \parens{ c_\mathrm{\HI} c_\mathrm{d} + s_\mathrm{HI} s_\mathrm{d} } , \\
B & \equiv & \frac{1}{W} \sum_\nhat w(\nhat) \parens{ c_\mathrm{\HI} s_\mathrm{d} - s_\mathrm{\HI} c_\mathrm{d} } ,  \end{eqnarray}
where $c_x  \equiv Q_x/P_x$ and $s_x \equiv U_x/P_x$.
Then we can express the alignment metric as
\eq{ \xi(\psi) = A \cos(2 \psi) + B \sin(2 \psi) ,}
from which the maximizing value can be found to be
\eq{ \hat{\psi} = \frac{1}{2} \operatorname{atan2}(B,A) ,  \label{eq:psihat from B,A} }
where, because $A(\theta_\mathrm{d} = \theta_\mathrm{\HI}) > 0$ and $B(\theta_\mathrm{d} = \theta_\mathrm{\HI}) = 0$, this choice of arctangent ensures $\hat{\psi} = 0$ in the case of perfect alignment.  In the limit of a single pixel~$\nhat$, the estimator~$\hat{\psi}$ is equivalent to \eqref{eq:deltatheta}, the~$\Delta\theta$ used in \citet{Clark2021}. The added benefit of~$\hat{\psi}$ is in aggregating pixels into patches without biasing the estimates low (as described at the beginning of this section).

\subsection{Misalignment maps \label{sec:misalignmentMaps}}

\begin{figure*}
\gridline{\fig{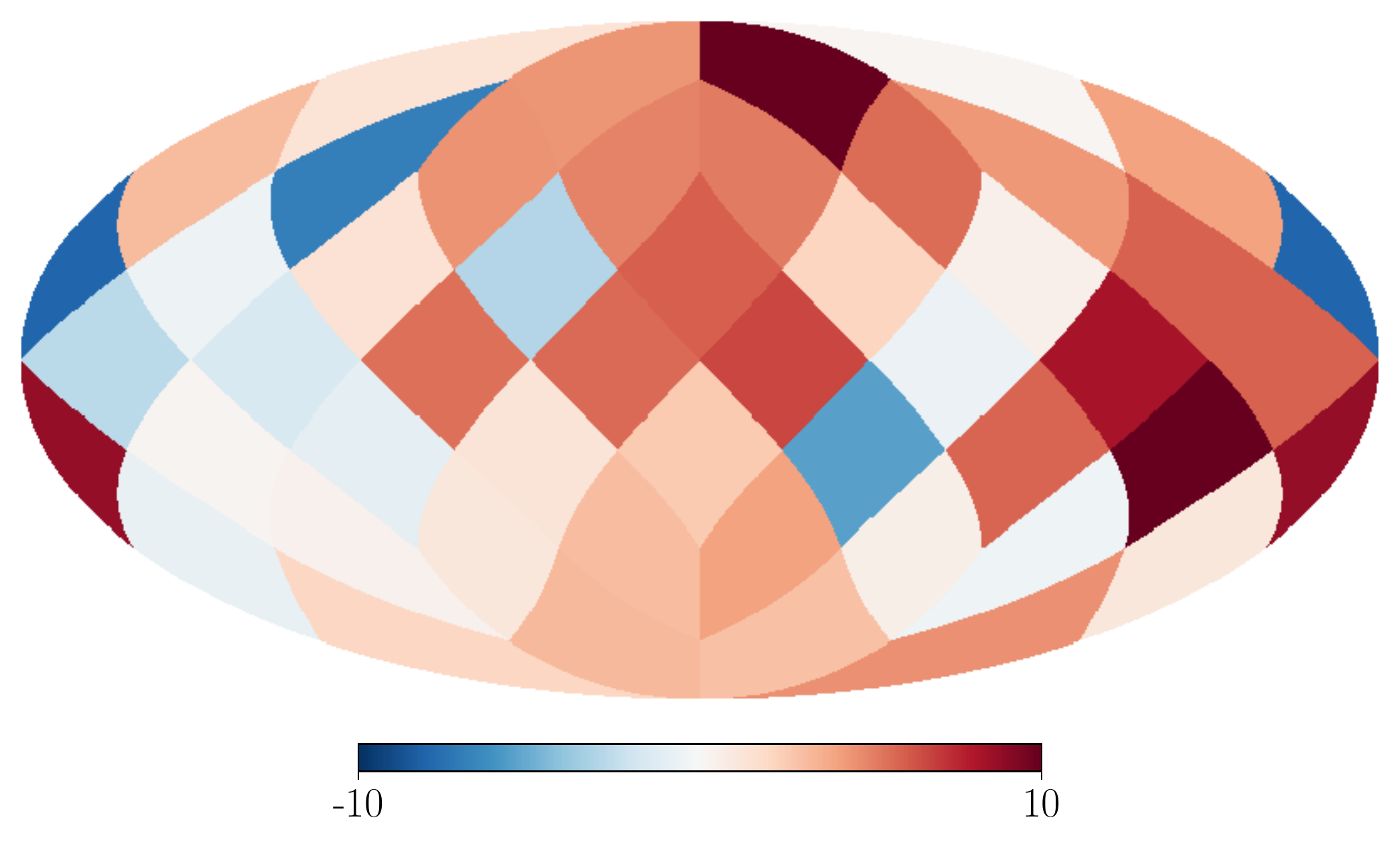}{\columnwidth}{$\ell < 702$, $N_\mathrm{side}=2$} \fig{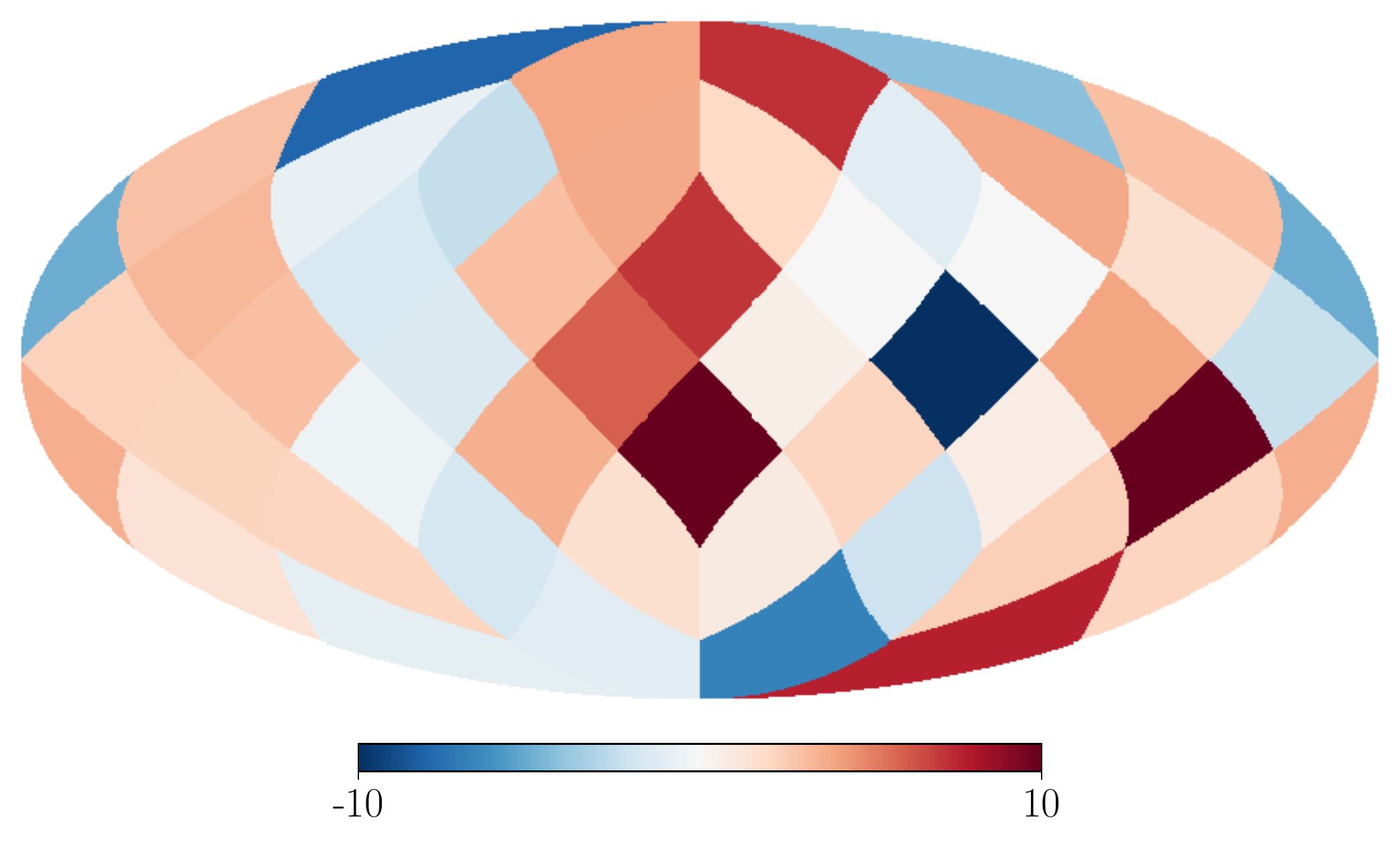}{\columnwidth}{$101 < \ell < 702$, $N_\mathrm{side}=2$}}
\gridline{\fig{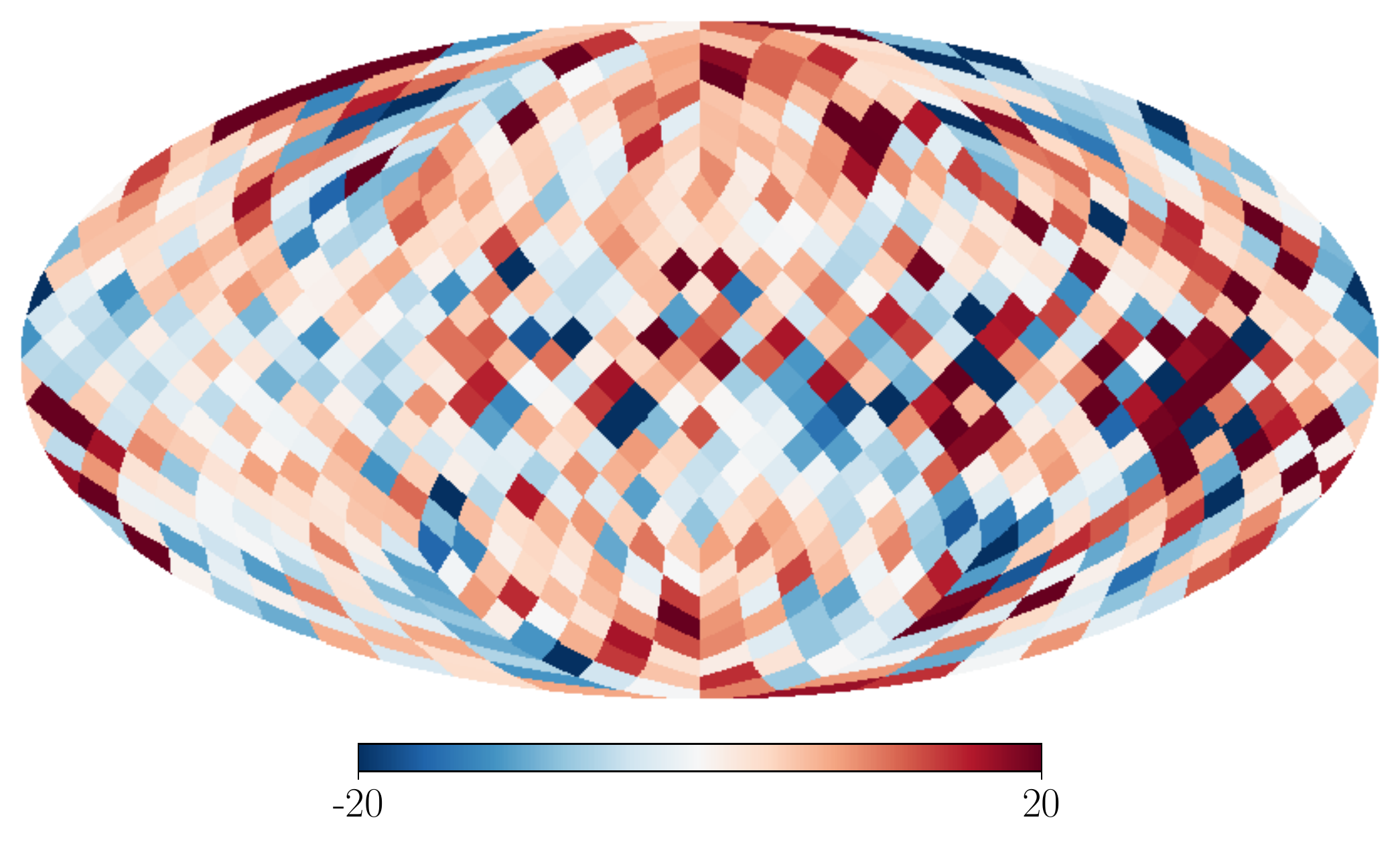}{\columnwidth}{$\ell < 702$, $N_\mathrm{side}=8$} \fig{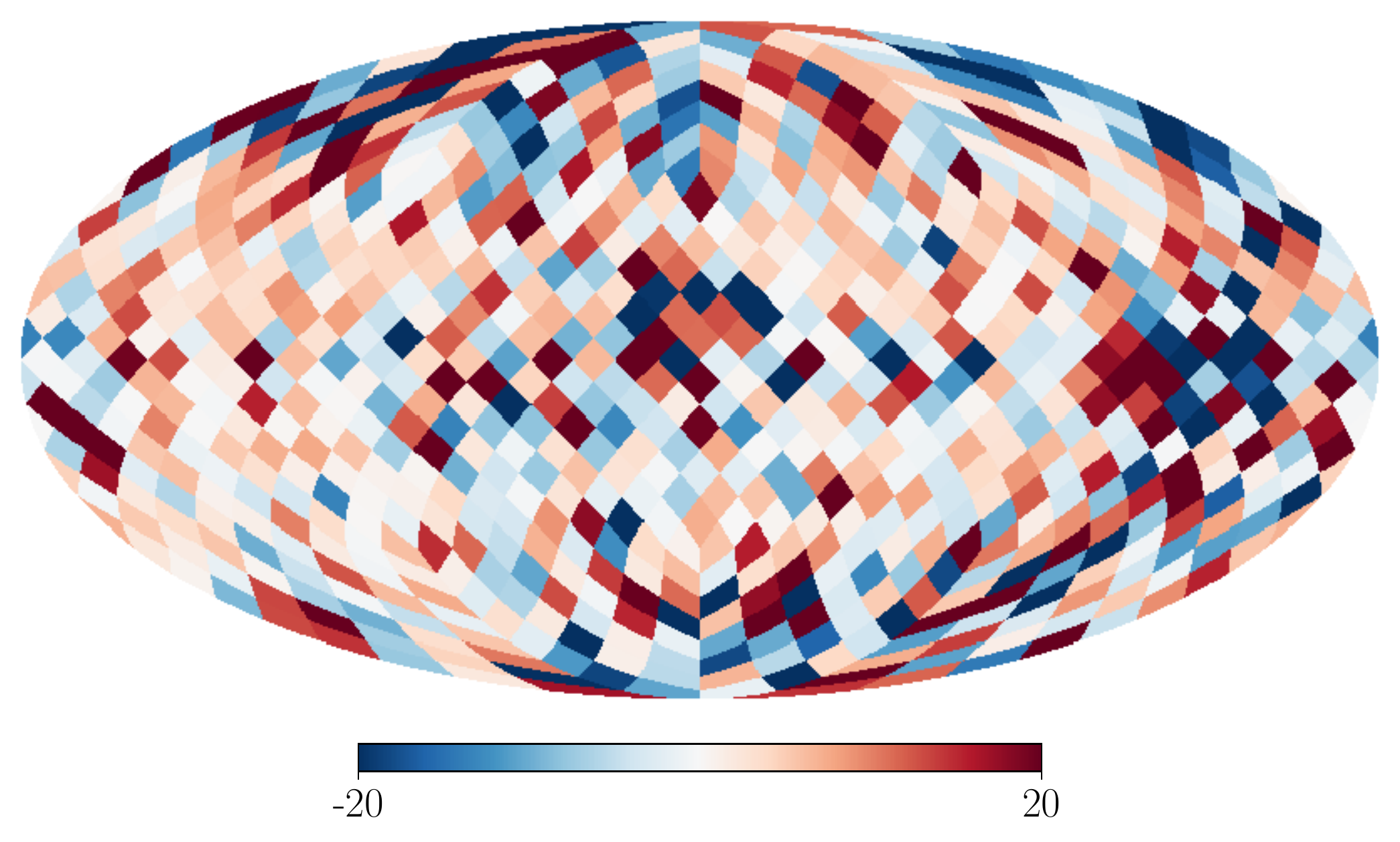}{\columnwidth}{$101 < \ell < 702$, $N_\mathrm{side}=8$}}
\gridline{\fig{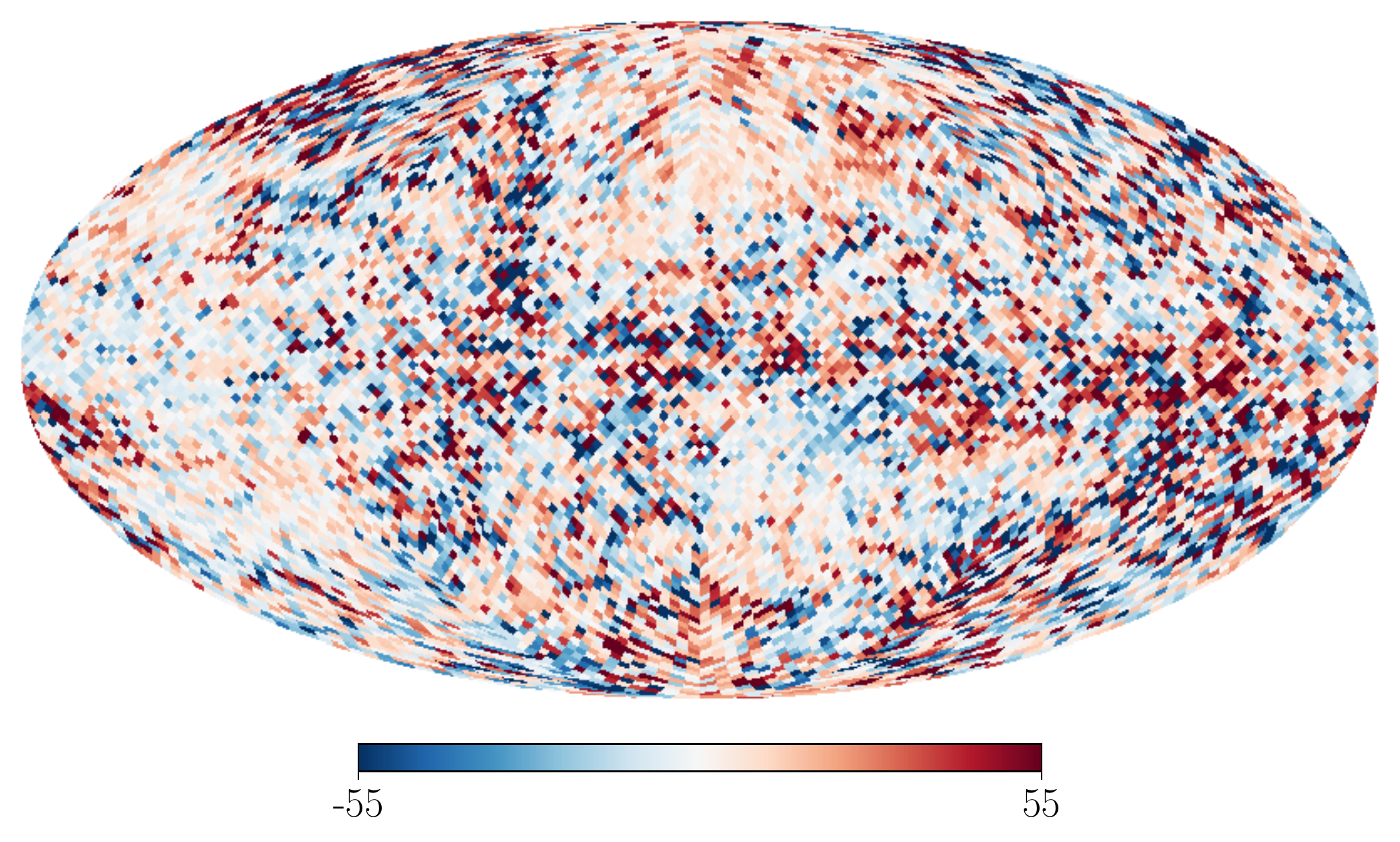}{\columnwidth}{$\ell < 702$, $N_\mathrm{side}=32$} \fig{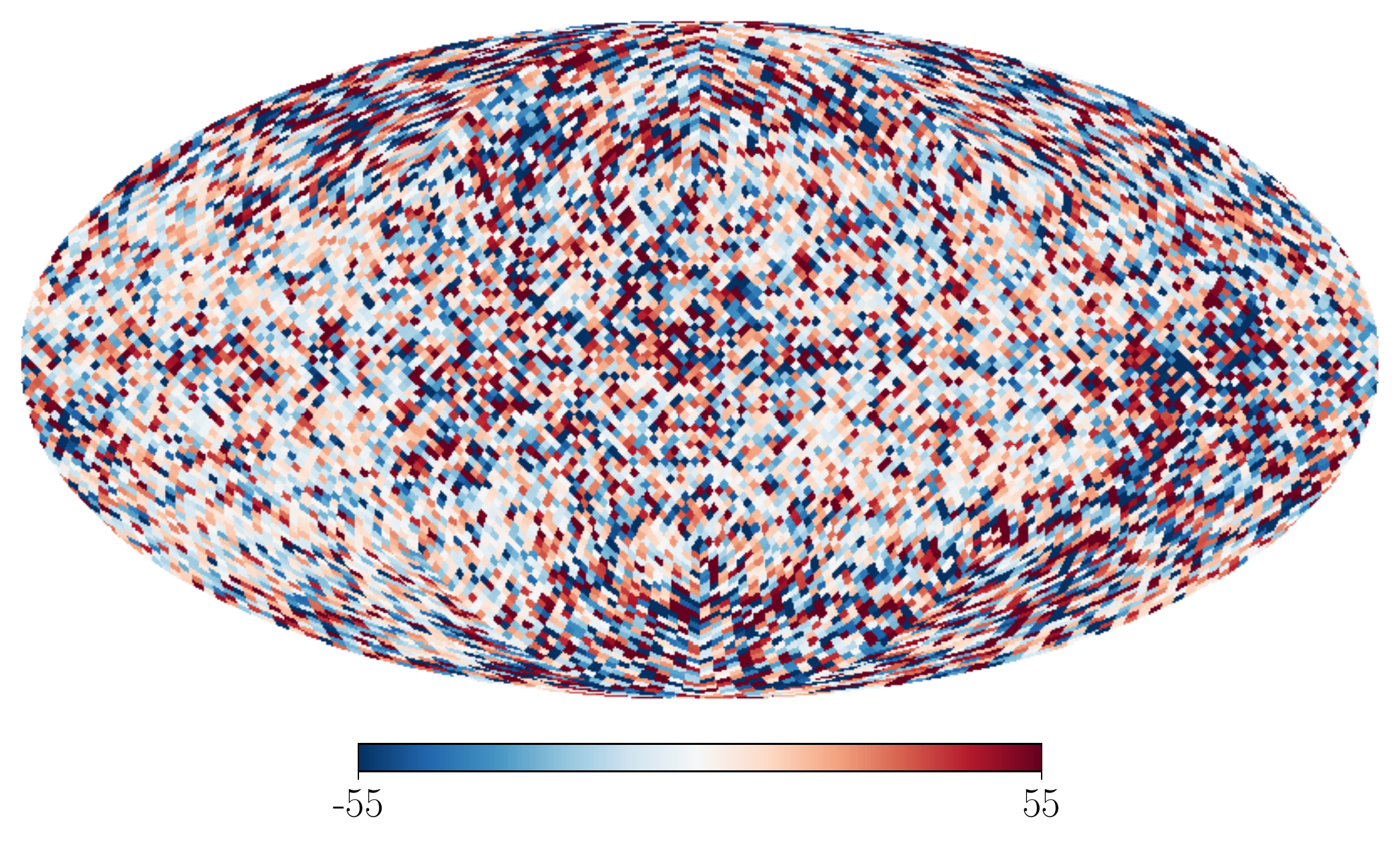}{\columnwidth}{$101 < \ell < 702$, $N_\mathrm{side}=32$}}
\caption{Maps of misalignment angle~$\hat{\psi}$ (\eqref{eq:psihat from B,A}) on masks defined by pixels with various values for~$N_\mathrm{side}$ and with different filtering options.  In the left column, the \emph{Planck} and \HI~maps have been lowpass filtered to $\ell < 702$, which is approximately the \HI4PI beam scale.  In the right column, the maps have been bandpass filtered to $101 < \ell < 702$, which is the multipole range used for much of the following analysis. The color scales are in degrees.  We see greater variance in the northeast~(top left) and southwest~(bottom right), which are the regions with the lowest dust intensity, and we see a tendency for the misalignment angles to be \emph{positive}.  \label{fig:psimap}}
\end{figure*}

We present maps of~$\hat{\psi}$ in \figref{fig:psimap}. 
We compute~$\hat{\psi}$ on masks defined by {\tt HEALPix} pixels with various values of~$N_\mathrm{side}$. For the lowpass-filtered~($\ell < 702$) results in the \emph{left} column of \figref{fig:psimap}, the misalignment angles are partially correlated between patches due to the presence of large-scale polarization modes. Part of the motivation for the bandpass filtering~($101 < \ell < 702$) implemented for the \emph{right} column of \figref{fig:psimap} is to remove these correlations and acquire approximately independent estimates in each patch. For $N_\mathrm{side} = 32$, the smallest patch size we consider, the side length of each mask is~$1.8^\circ$, which means that the above-mentioned bandpass filtering suppresses modes with wavelengths larger than a single patch. Most of the patch-to-patch correlations are removed by the bandpass filtering. (We will show in \secref{sec:psiSpatialCoherence}, however, that there is evidence for nontrivial spatial coherence of~$\hat{\psi}$ that cannot be simply attributed to large-scale modes.)

As the patch size decreases, regions of higher and lower variance emerge at all latitudes in a pattern that is similar to that of~$\Delta\theta$ in Fig.~3 of \citet{Clark2021}.
The above observations are broadly consistent between the lowpass- ($\ell < 702$) and bandpass-filtered ($101 < \ell < 702$) maps, but the former are visually smoother.

\subsection{Test for estimator bias}

To check for biases in our misalignment estimator, we measure~$\hat{\psi}$ on masks defined by {\tt HEALPix} pixels as described above,  artificially rotate the angles of the \emph{Planck} polarization map by a known amount~$\psi_0 \in [-90^\circ,90^\circ]$ and then recompute~$\hat{\psi}$.  We track the median of the distribution and find that it follows~$\psi_0$.  We conclude that $\hat{\psi}$~is an \emph{unbiased} estimator of misalignment angle.

\subsection{Positive misalignment tendency}

We observe a tendency toward \emph{positive} misalignment angles in \figref{fig:psimap}.  
To estimate the statistical significance, it is tempting to appeal to the central limit theorem. Unfortunately, the values of~$\hat{\psi}(\nhat)$, where $\nhat$~here represents a particular patch, are neither completely independent nor identically distributed. By bandpass filtering to $101 < \ell < 702$ as in the right column of \figref{fig:psimap}, we can achieve approximate independence of the estimates for different~$\nhat$. We cannot, however,  guarantee that the estimates are identically distributed.

Nevertheless, because the calculation is simple, we estimate a mean and standard error by appealing to the central limit theorem. For $N_\mathrm{side} = 32$,  we find $\hat{\psi}_{\ell < 702} = 1.9 \pm 0.3^\circ$, but we caution that the patches are nontrivially correlated with each other by the bright, large-scale polarization modes and, therefore, refrain from claiming any statistical significance.  Restricting to the sky area allowed by our fiducial 70\%~Galaxy mask (cf.~\figref{fig:patches}), we find $\hat{\psi}_{\ell < 702} = 1.7 \pm 0.3^\circ$. After bandpass filtering, the patches are more (but not completely) independent, and we find $\hat{\psi}_{101 < \ell < 702} = 0.9 \pm 0.3^\circ$, which implies a statistical significance of~$3\sigma$.  Restricting to our fiducial 70\%~Galaxy mask,  we find $\hat{\psi}_{101 < \ell < 702} = 0.8 \pm 0.4^\circ$, which implies a significance of~$2\sigma$. We have deliberately limited ourselves to a single significant figure, because we consider these calculations to be crude estimates.  

\subsection{Relationship to dust properties}

\begin{figure*}
\begin{center}
\includegraphics[width=\textwidth]{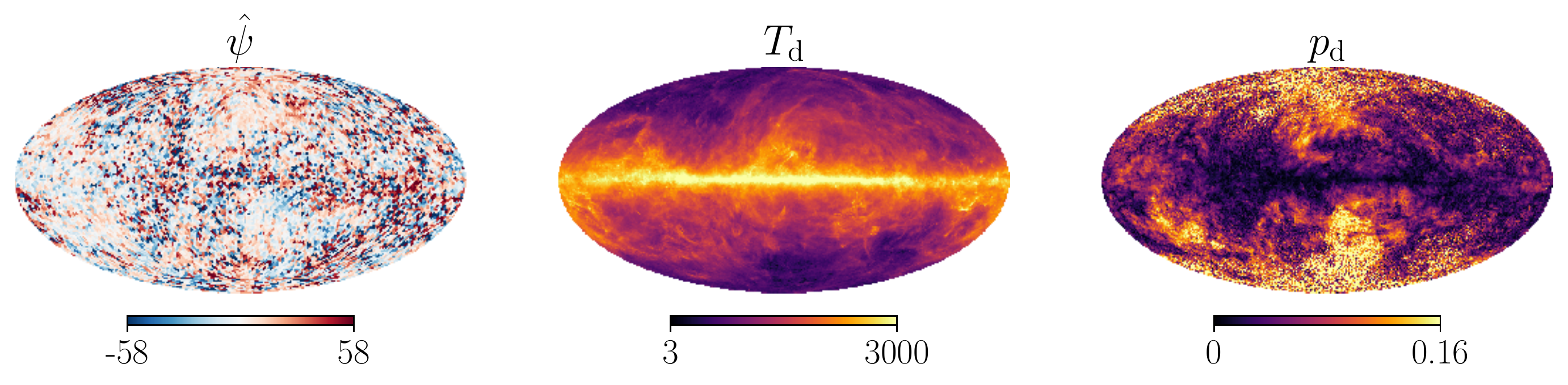}
\caption{Maps of misalignment angle~$\hat{\psi}$ (\emph{left}), dust intensity~$T_\mathrm{d}$ (\emph{middle}) and dust polarization fraction~$p_\mathrm{d}$ (\emph{right}).  The $\hat{\psi}$~map has been created with $N_\mathrm{side} = 32$ and is identical to the lower left map of \figref{fig:psimap}. The $T_\mathrm{d}$~and $p_\mathrm{d}$~maps are shown after smoothing to the \HI4PI beam~($16.2'$) and downgrading to $N_\mathrm{side} = 256$.  The $\hat{\psi}$~values are in degrees and dust intensity in~$\mu\mathrm{K}_\mathrm{RJ}$.  We note the similarities in the large-scale features of~$\hat{\psi}$ and~$p_\mathrm{d}$.  \label{fig:psiTp panel}}
\end{center}
\end{figure*}

We note that the large-scale features of~$\hat{\psi}$ are similar to those of the dust polarization fraction~$p_\mathrm{d} \equiv \sqrt{Q_\mathrm{d}^2 + U_\mathrm{d}^2}/T_\mathrm{d}$. 
For visual comparison, we provide in \figref{fig:psiTp panel} a panel of maps displaying misalignment angle~$\hat{\psi}$, dust intensity~$T_\mathrm{d}$ and dust polarization fraction~$p_\mathrm{d}$.  
We omit an estimate of the correlation strengths, because the bright, large-scale modes induce covariances that are difficult to model.  The low-column (small-$T_\mathrm{d}$) sky regions in the northeast and southwest are also regions of increased variance in~$\hat{\psi}$,\rev{ as evidenced by the large fluctuations between neighboring patches}, but a much clearer visual correspondence appears between~$\hat{\psi}$ and~$p_\mathrm{d}$. The regions of larger~$p_\mathrm{d}$ show smaller variance in~$\hat{\psi}$, and those regions also tend to $\hat{\psi} > 0$. For both~$T_\mathrm{d}$ and~$p_\mathrm{d}$, the correspondence with regions of lower variance may be related to the signal-to-noise ratio of the misalignment measurement. In regions with higher polarized intensity, there is less variance in~$\hat{\psi}$.  

There may also be a connection to~$S_\mathrm{d}$, the angle dispersion of the dust polarization, and to~$S_\mathrm{\HI}$, that of the \HI~polarization template. The former anticorrelates with~$p_\mathrm{d}$, i.e., regions of greater polarization-angle coherence have larger polarization fractions~\citep{Planck2018XII}.  The variation in polarization-angle coherence may be related to the magnetic-field orientation relative to the line of sight~\citep[e.g.,][]{Hensley2019}.  The \HI-based dispersion~$S_\mathrm{\HI}$ and polarization fraction~$p_\mathrm{\HI}$ also anticorrelate, and the alignment of the dust polarization angle to the \HI~template anticorrelates with~$S_\mathrm{\HI}$~\citep{ClarkHensley2019}.  We would, therefore, expect that regions of larger polarization fraction correlate with regions of coherent magnetic misalignment, which is indeed what we observe.

\subsection{Large sky areas \label{sec:misalignmentEstimatorLargeSky}}

A major motivation for this study is to understand the origin of the parity-violating $TB$~correlation measured by \emph{Planck} on large fractions of the high-latitude sky~\citep{Clark2021}.  In addition to measuring the variation in misalignment angle across relatively small patches of sky~(\figref{fig:psimap}), we can apply our estimator~(\eqref{eq:psihat from B,A}) to large sky areas and compare to expectations based on measured cross-spectra~(\eqref{eq:tanpsi estimates}). On large sky areas, the variation in~$\hat{\psi}$ is suppressed, and we can safely make a small-angle approximation. Then we expect
\eq{ \hat{\psi}_\ell \approx  \frac{D_\ell^{E_\mathrm{\HI} B_\mathrm{d}}}{2 D_\ell^{E_\mathrm{\HI} E_\mathrm{d}}} = -  \frac{D_\ell^{B_\mathrm{\HI} E_\mathrm{d}}}{2 D_\ell^{B_\mathrm{\HI} B_\mathrm{d}}} = \frac{D_\ell^{T_\mathrm{\HI} B_\mathrm{d}}}{2 D_\ell^{T_\mathrm{\HI} E_\mathrm{d}}} = \frac{D_\ell^{T_\mathrm{d} B_\mathrm{d}}}{2 D_\ell^{T_\mathrm{d} E_\mathrm{d}}} .  \label{eq:psiEstimatesApprox} }
Noise in the denominators may bias these expressions, but we are here looking only for a broad consistency and for the approximate level of aggregate misalignment on large sky areas.  Since $\hat{\psi}_\ell$~is estimated by reference to the \HI~template, we expect greatest consistency with the dust-\HI\ cross-spectra, e.g., $E_\mathrm{\HI} B_\mathrm{d}$ as opposed to the \emph{Planck}-only~$T_\mathrm{d} B_\mathrm{d}$ and~$E_\mathrm{d} B_\mathrm{d}$.

\begin{figure}
\includegraphics[width=\columnwidth]{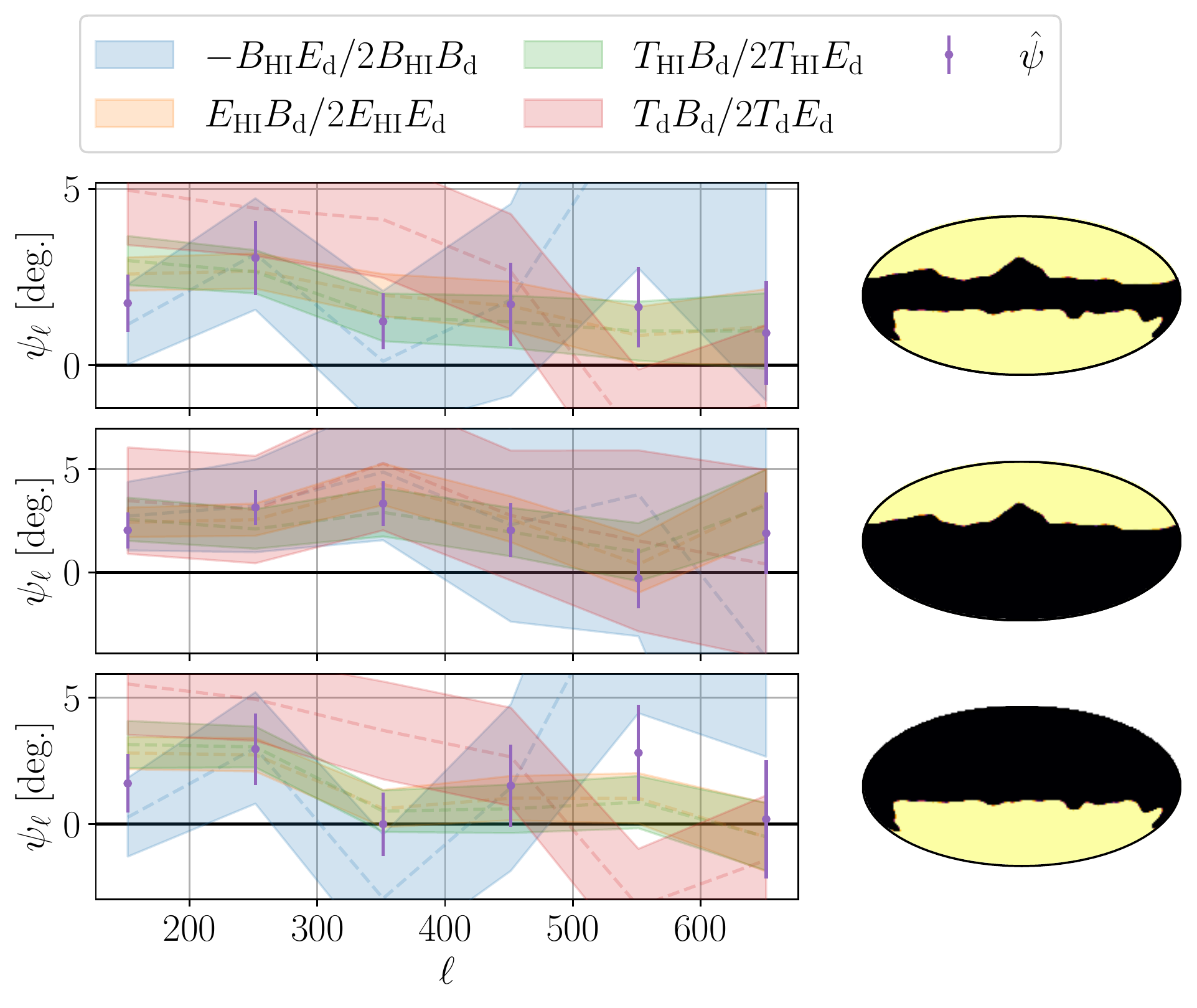}
\caption{Comparison of misalignment-angle estimates from $\hat{\psi}$~(\eqref{eq:psihat from B,A}) and from parity-violating cross-spectra~(\eqref{eq:psiEstimatesApprox}) for a 70\%~Galaxy mask~(\secref{sec:galaxyMasks}) and for each hemisphere independently~(bottom two rows).  The $\hat{\psi}$~estimates are formed after filtering the \emph{Planck} and \HI~maps to each multipole bin~($\Delta\ell = 100$). The parity-violating spectra are shown with Gaussian uncertainties.  The green and orange bands are nearly identical and difficult to separate visually. The error bars on the $\hat{\psi}$~points are from the standard deviations of our \rev{mock skies}~(\secref{sec:simulations}).  We find an approximately scale-independent \emph{positive}~$\hat{\psi}$ that persists at the same level in both hemispheres and is broadly consistent with the spectrum-based estimators, especially those incorporating the \HI~template. \label{fig:psiEstimates} 
}
\end{figure}

In \figref{fig:psiEstimates}, we compare the misalignment estimates from \eqref{eq:psiEstimatesApprox} for a $70\%$~Galaxy mask~(\secref{sec:galaxyMasks}).
We find a misalignment angle of~$\sim 2^\circ$ that is coherent in the range $101 < \ell < 702$. As expected,  $\hat{\psi}$~tends to be more consistent with the dust-\HI~cross-spectra, especially~$E_\mathrm{\HI} B_\mathrm{d}$ and~$T_\mathrm{\HI} B_\mathrm{d}$, which are more sensitive than~$B_\mathrm{\HI} E_\mathrm{d}$~(\secref{sec:implicationsCrossSpectra}). The \emph{Planck}-only $T_\mathrm{d} B_\mathrm{d}$~is more discrepant (though not dramatically so) but reproduces the coherently positive behavior for $\ell < 500$.  

The $\hat{\psi}$~estimates are broadly consistent between the northern and southern Galactic hemispheres.  In particular, the magnitude of the misalignment and its scale~(multipole) independence are consistent. The \emph{Planck}-only $T_\mathrm{d} B_\mathrm{d}/ 2 T_\mathrm{d} E_\mathrm{d}$~(red in \figref{fig:psiEstimates}) also shows a similar consistency between the hemispheres.  The positive~$\hat{\psi}$,  the approximate scale independence of~$\hat{\psi}$ and the consistency of~$\hat{\psi}$ between hemispheres are robust to the choice of Galaxy mask; we checked this for $f_\mathrm{sky} \in \{40\%,60\%,70\%,80\%,90\%\}$~(\secref{sec:galaxyMasks}) and present some related results below in \secref{sec:varyingSkyFrac}.  The consistency between hemispheres begs for an explanation, which should be a target for future investigations.  

The uncertainties on~$\hat{\psi}$ in \figref{fig:psiEstimates} are derived from the scatter of our \rev{mock skies}~(\secref{sec:simulations}) and are used only for visualization purposes, namely, to give a rough indication of the expected variance. We do \emph{not} rely on these uncertainties for statistical inference.

Because the large-scale (low-$\ell$) modes are difficult to reproduce in our \rev{mock-sky} framekwork (\secref{sec:simulations}), we have restricted \figref{fig:psiEstimates} to $101 < \ell < 702$. We can calculate~$\hat{\psi}$ for $\ell < 101$, but we cannot form a reliable \rev{uncertainty based on mock skies}.  Nevertheless, it is interesting to report the values for~$\hat{\psi}_{\ell < 101}$. We find~$1.4^\circ$ when using both hemispheres, $2.9^\circ$ in the northern hemisphere and $0.7^\circ$ in the southern hemisphere.  The statistical weight cannot be evaluated in the present analysis, but it is noteworthy that the large scales show the same tendency for misalignment to be \emph{positive}.

\subsection{Aggregate global misalignment? \label{sec:globalMisalignment}}

An intriguing possibility is that there is an aggregate global misalignment of $\sim 2^\circ$.  An aggregate misalignment would appear as an isotropic, multipole-independent rotation of the dust polarization relative to the filamentary structures, i.e., $\psi_\ell = \mathrm{const.}$  The implied magnetic-field structure relative to the dust intensity field would be qualitatively similar to that depicted by the green pseudovectors in \figref{fig:mapmisalign}.  A global polarization rotation can also be produced by a miscalibration of the absolute polarization angle or in the CMB by the phenomenon of \emph{cosmic birefringence}~(\secref{sec:IntroCB}). 

We consider miscalibration to be unlikely, since \emph{Planck} estimates a systematic uncertainty of~$0.28^\circ$~\citep{PlanckIntXLIX2016}, nearly an order of magnitude smaller than our measured misalignment.  In the following sections, we measure~$\hat{\psi}$ in small sky regions and search for correlated variations with other interesting quantities. The \emph{relative} variation from region to region is insensitive to an overall miscalibration.

A global misalignment signal in the dust, which acts as a foreground for CMB measurements, would need to be accounted for in searches for cosmic birefringence, especially with methods that rely on the symmetry properties of the dust polarization, e.g.,~\citet{Minami2019}. 

As a consistency check, we modified our \HI~template~(\secref{sec:HessianMethod}) by imposing a global polarization rotation of~$2^\circ$.  This rotation mixes $E$~and $B$~modes. Because the \HI~template is dominated by $E$~modes~($E_\mathrm{\HI} E_\mathrm{\HI}/B_\mathrm{\HI} B_\mathrm{\HI} \sim 5$ for $\ell > 100$), the effect is fractionally stronger in the $B$~modes.  We can estimate the expected impact of this modification by considering that $2^\circ \approx 0.03~\mathrm{rad}$, so this should produce a percent-level change in the correlations.  We correlate with the \emph{Planck} dust maps and find that the $B$-mode correlation for $100 < \ell < 700$ \emph{increases} by $0.1$-$0.5\%$ in addition to the original correlation of $10$-$25\%$, which is indeed a \emph{fractional} increase of~$\mathcal{O}(1\%)$. We performed the same exercise with the opposite rotation, i.e., by~$-2^\circ$, and we find an approximately symmetric \emph{decrease} in the \HI-\emph{Planck} correlation. These results are consistent with the estimates of \figref{fig:psiEstimates} and increase our confidence in a true on-sky aggregate misalignment of approximately~$2^\circ$.

The $T_\mathrm{d} B_\mathrm{d}$-based estimate of~$\psi_\ell$ (red in \figref{fig:psiEstimates}) is coherent only over the range $100 \lesssim \ell \lesssim 500$. Where it is nonzero, it also tends to be larger than~$\hat{\psi}$. The discrepancy may be an indication that $T_\mathrm{d} B_\mathrm{d}$~and $T_\mathrm{d} E_\mathrm{d}$~are affected by additional polarization sources that are missed by the filamentary misalignment model, and it is conceivable that the simultaneous positivity of~$T_\mathrm{d} B_\mathrm{d}$ and~$\hat{\psi}$ is merely coincidental.  In \secref{sec:parityViolatingCrossSpectra}, we seek further evidence of a relationship by analyzing small sky patches.

\subsection{Varying sky fraction \label{sec:varyingSkyFrac}}

We track the dependence of these misalignment estimates on the sky fraction~$f_\mathrm{sky}$.  In \figref{fig:psiEstimates}, we considered only $f_\mathrm{sky} = 70\%$, whereas we now allow $f_\mathrm{sky} \in \{20\%,40\%,60\%,70\%,80\%,90\%,97\%\}$~(\figref{fig:maskMap}). In \figref{fig:psiEstimates}, we considered multiple estimators for~$\psi_\ell$. For simplicity, we now downselect to only two. One is $\psi^{\mathrm{d} \times \mathrm{d}}_\ell \equiv \operatorname{atan2} \parens{ T_\mathrm{d} B_\mathrm{d} , T_\mathrm{d} E_\mathrm{d} }/2$~(red in \figref{fig:psiEstimates}), which uses only the dust field~\citep[cf.~Fig.~9 of][]{Clark2021}.  We contrast the dust-only estimator with one that includes some \HI\ filamentary information: $\psi^{\mathrm{\HI} \times \mathrm{d}}_\ell \equiv \operatorname{atan2} \parens{ T_\mathrm{\HI} B_\mathrm{d} , T_\mathrm{\HI} E_\mathrm{d} }/2$~(green in \figref{fig:psiEstimates}). Whereas $\psi^{\mathrm{d} \times \mathrm{d}}_\ell$~includes information from the entire dust field, $\psi^{\mathrm{\HI} \times \mathrm{d}}_\ell$~collapses the misalignment estimate onto the filamentary modes. When the two are in agreement, the filamentary magnetic misalignment is representative of the parity violation in the full dust field. When they deviate from each other, the \HI~template may be incomplete or inaccurate, or the full dust field may contain parity-violating contributions which are non-filamentary.

\begin{figure*}
\gridline{ \fig{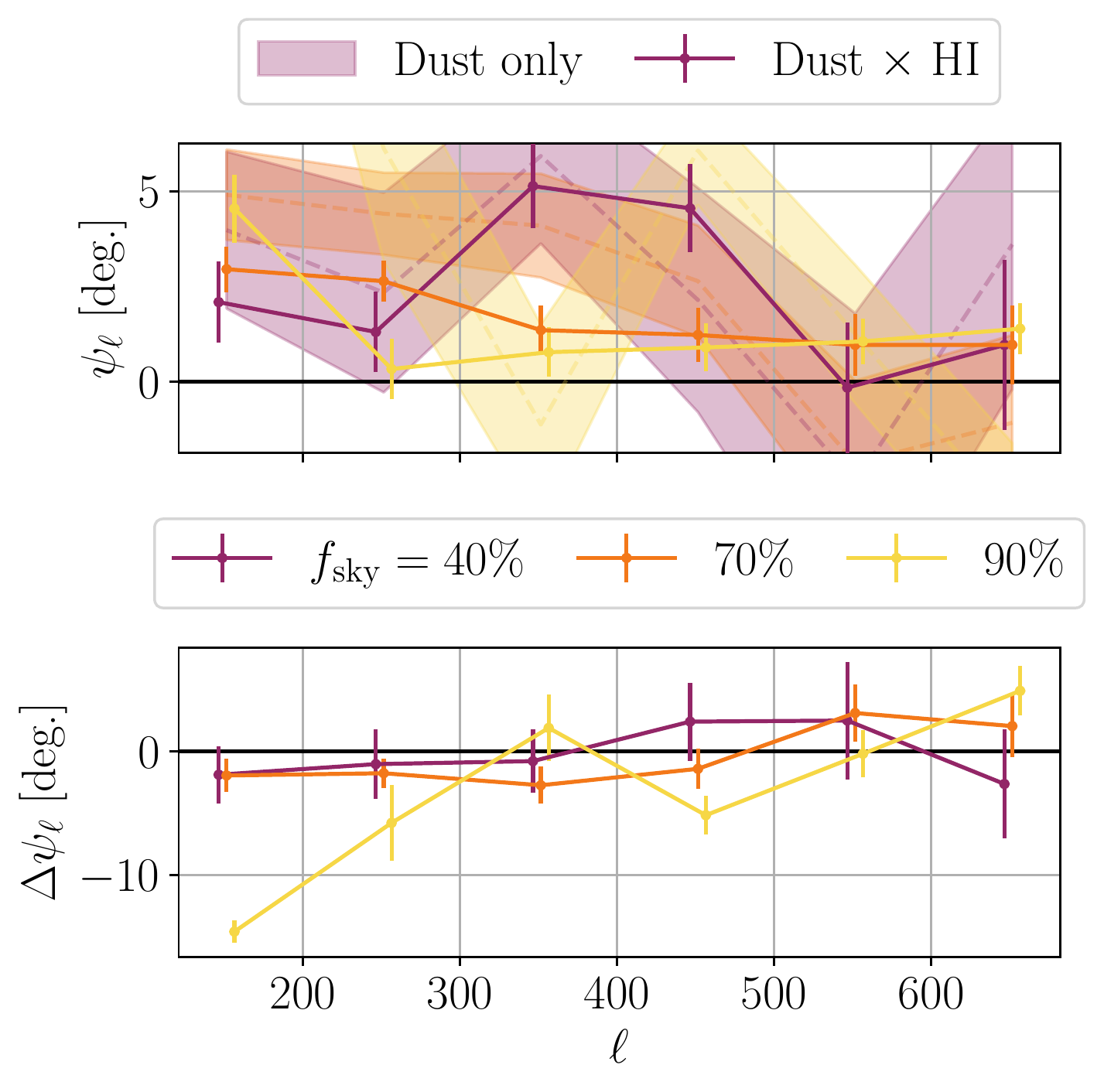}{0.45\textwidth}{Misalignment angle~$\psi_\ell$ as a function of multipole~$\ell$ in bins of width~$100$ for three example sky fractions~$f_\mathrm{sky}$. We note that the ``Dust only'' estimates are more variable over~$\ell$ and that the agreement with the ``Dust $\times$ \HI'' estimates tends to be better with smaller~$f_\mathrm{sky}$, i.e., at higher latitudes. The 70\%~curves in the upper panel also appeared in \figref{fig:psiEstimates}.} \fig{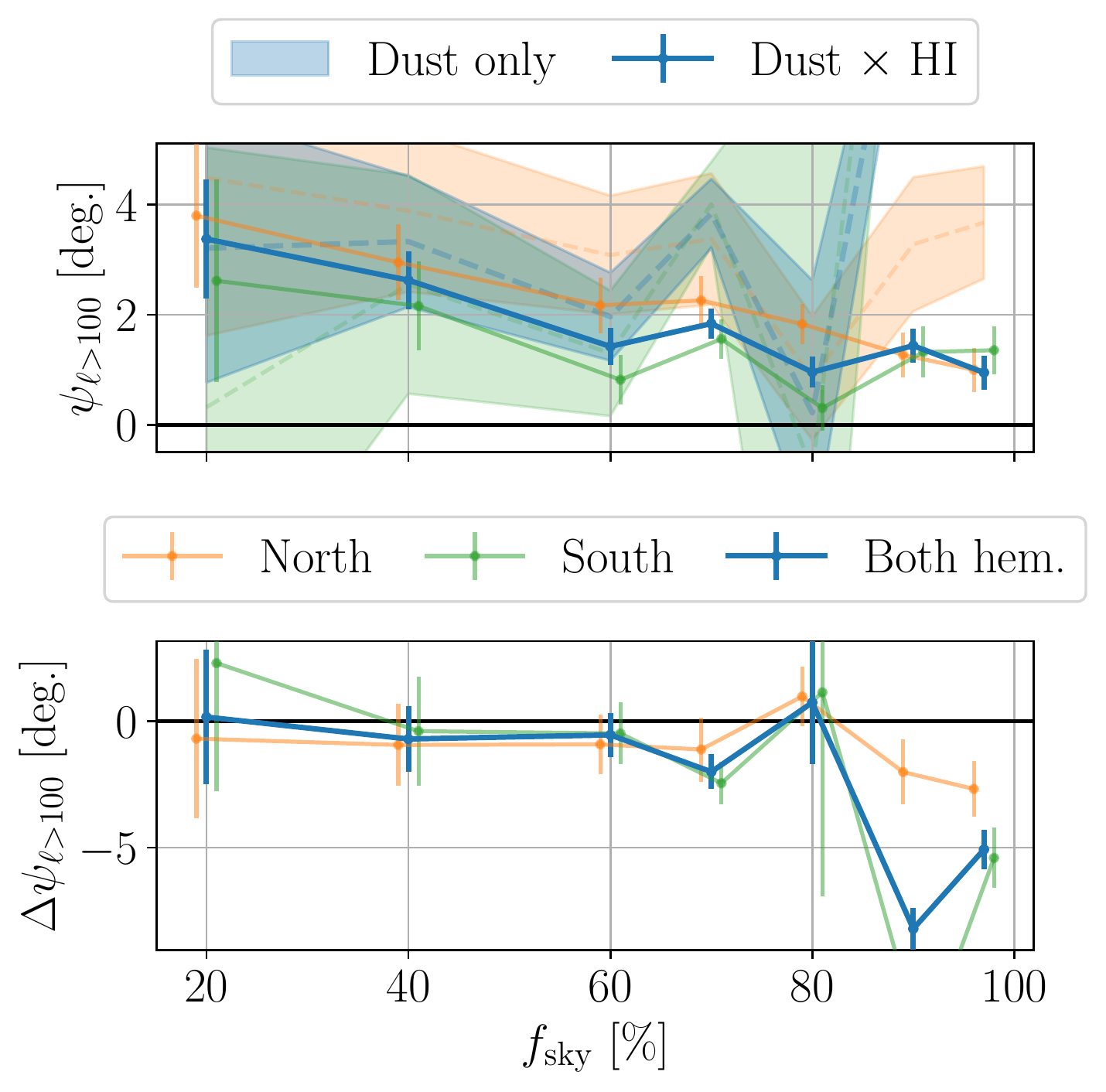}{0.45\textwidth}{Highpass-filtered misalignment angle~$\psi_{\ell > 100}$ as a function of sky fraction~$f_\mathrm{sky}$ with each hemisphere also considered individually. We note that the ``Dust $\times$ \HI'' estimates are all fairly consistent with a steady decline with increasing~$f_\mathrm{sky}$, while the ``Dust only'' estimates fluctuate strongly for $f_\mathrm{sky} \gtrsim 70\%$.} }
\caption{ Misalignment estimates $\psi^{x \times \mathrm{d}}_\ell = \operatorname{atan2} \parens{ T_x B_\mathrm{d} , T_x E_\mathrm{d} }/2$~(\eqref{eq:TBpsi relation}) with $x = \mathrm{d}$~(``Dust only'') and with $x = \mathrm{\HI}$~(``Dust $\times$ \HI'') for varying sky fraction~$f_\mathrm{sky}$.  The top panels show the two types of estimates; the bottom panels show the differences. \label{fig:psiVsFsky}}
\end{figure*}

In \figref{fig:psiVsFsky}, we show~$\psi^{\mathrm{d} \times \mathrm{d}}_\ell$ and~$\psi^{\mathrm{\HI} \times \mathrm{d}}_\ell$ for a variety of sky fractions~$f_\mathrm{sky}$. 
We show the estimates for individual multipole bins~(cf.~\figref{fig:psiEstimates}), and we also highpass filter to form the broadband~$\psi_{\ell > 100}$,  which may potentially average away signal but is less noisy. We find that $\psi^{\mathrm{\HI} \times \mathrm{d}}_\ell$~is consistently positive over all~$\ell$ and remains in the range of $0$-$5^\circ$, while $\psi^{\mathrm{d} \times \mathrm{d}}_\ell$~is much more variable, especially at large~$f_\mathrm{sky}$. We note that the two estimates display closer agreement at small~$f_\mathrm{sky}$, i.e., when restricting to high Galactic latitudes. At the same time, we find that $\psi^{\mathrm{\HI} \times \mathrm{d}}_{\ell>100}$~steadily decreases from~$\sim 3^\circ$ to~$\sim 1^\circ$ as $f_\mathrm{sky}$~increases from~$20\%$ to~$97\%$~(right plots of \figref{fig:psiVsFsky}),  a phenomenon which is observed in both hemispheres independently.  The decline may be related to the fact that the \HI~becomes a less robust tracer of dust at low Galactic latitudes where the column densities are relatively large~\citep[e.g.,][]{Lenz2017}, so it may be that the \HI~template is simply less representative of the dust field for large~$f_\mathrm{sky}$.

Interestingly, all of the variations considered in \figref{fig:psiVsFsky} produce $\psi^{\mathrm{\HI} \times \mathrm{d}}_\ell$ in the range of~$0$-$5^\circ$. This behavior persists for all of the considered multipoles and sky fractions and in both hemispheres independently.  Furthermore, we performed this analysis with the RHT-based \HI~template~\citep[][\secref{sec:comparisonWithRHT}]{ClarkHensley2019} instead of the Hessian, and we find consistent results.  These observations lend more weight to the speculations of \secref{sec:globalMisalignment} about a possible \emph{global} misalignment angle of~$\sim 2^\circ$.

\section{What is ``magnetic misalignment''? \label{sec:misalignment}}

While random deviations constitute a form of ``misalignment'' relative to the \HI-defined filaments, it is unsurprising that such deviations are detected. The \HI-based polarization templates \citep[presented here and in, e.g.,][]{ClarkHensley2019} correlate strongly with the \emph{Planck} dust maps, but they are not identical, even within the limits of the \emph{Planck} noise. If the term ``magnetic misalignment'' is to refer to \emph{any} kind of deviation from the \HI~template, then a detection of misalignment teaches us only that the \HI~template is an incomplete description of the dust polarization field.

As a result of these considerations, we focus much of the rest of our analysis on a search for magnetic misalignment that displays certain types of \emph{coherence}, which is less likely to be mimicked by random deviations from an \HI~template. We search for coherence both in harmonic and map space. Harmonic coherence indicates that $\psi_\ell$~is approximately constant with~$\ell$ and manifests as a uniform rotation of the dust polarization pseudovectors relative to the \HI~predictions~(\figref{fig:mapmisalign}).  We also refer to harmonic coherence as \emph{scale independence}.

\begin{figure}
\includegraphics[width=\columnwidth]{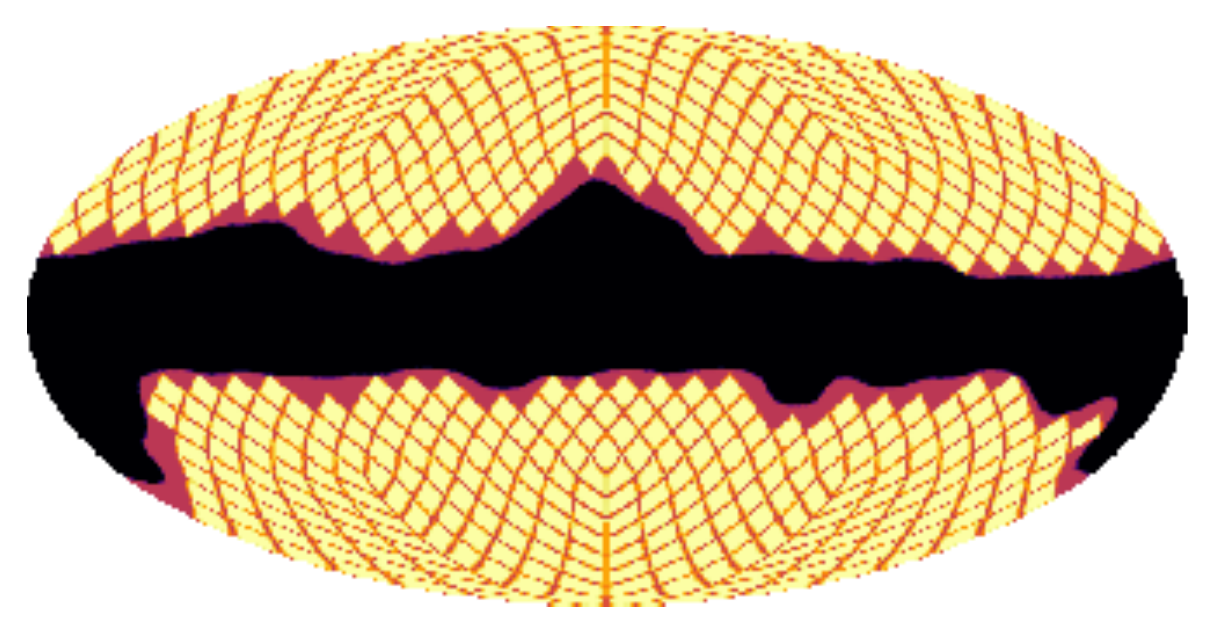}
\caption{Patches included in the analysis when the masks are defined by $N_\mathrm{side} = 8$ and an overall 70\%~Galaxy mask. The red area is allowed by the Galaxy mask, and the yellow patches are those that lie within the red area.  \label{fig:patches}}
\end{figure}

We restrict the analysis to the high-latitude sky by masking the Galactic plane to varying levels. Our fiducial choice is the \emph{Planck} 70\%~Galaxy mask. For example, when dividing the sky into patches defined by pixels with $N_\mathrm{side} = 8$, we consider only those shown in \figref{fig:patches}.
In subsequent sections, we will consider other, similarly-parameterized Galaxy masks and patch sizes.

\rev{In \secref{sec:correlationMetrics}, we introduce a number of cross-power and correlation metrics, which are used throughout the following sections.}

\subsection{Harmonic coherence \label{sec:psiHarmonicCoherence}}

The relative constancy of the \emph{Planck} $TB/TE$ at high latitudes~\citep[see, e.g.,  Fig.~9 of][]{Clark2021} is perhaps a hint that magnetic misalignment, if it is to address the mystery of the positive dust~$TB$, ought to display a harmonic coherence in the multipole range $100 \lesssim \ell \lesssim 500$. Indeed, we find that a direct calculation of the misalignment angle~$\hat{\psi}_\ell$ yields an apparent coherence over an even larger multipole range, tentatively across all~$\ell < 702$ (\figref{fig:psiEstimates} and \secref{sec:misalignmentEstimatorLargeSky}).  Is the apparent harmonic coherence an emergent phenomenon that appears only when aggregating large sky areas? In this section, we divide the sky into smaller patches and test whether harmonic coherence is a generic feature of magnetic misalignment.

To look for coherence in harmonic space, we bandpass filter the maps into two disjoint multipole ranges.  For all of our results, $101 < \ell < 702$, so $\ell_\mathrm{min} = 101$ and $\ell_\mathrm{max} = 702$ can be taken as lower and upper limits, respectively, on the multipole ranges.  We form a set of maps with $\ell < \ell_c - \Delta\ell/2$ and a set with $\ell > \ell_c + \Delta\ell/2$, where $\ell_c$~is a transition multipole and $\Delta\ell$~is a multipole buffer between the two ranges. We allow $\Delta\ell \in \{0,100,200\}$, and we sweep~$\ell_c$ across the range~$[101,702]$. 

Let $\hat{\psi}_\mathrm{lo}(\nhat)$~be the misalignment angle estimated in the patch centered on sky coordinate~$\nhat$ after filtering to $\ell < \ell_c - \Delta\ell/2$, and let $\hat{\psi}_\mathrm{hi}(\nhat)$ be similarly defined after filtering to $\ell > \ell_c + \Delta\ell/2$.  We will refer to~$\hat{\psi}_\mathrm{lo}(\nhat)$  and~$\hat{\psi}_\mathrm{hi}(\nhat)$, respectively, as the ``lowpass-filtered'' and ``highpass-filtered'' misalignment estimates. Recall, however, the multipole limits~$\ell_\mathrm{min}$ and~$\ell_\mathrm{max}$ mentioned above, so these estimates are, in fact,  products of \emph{bandpass} filtering.

Our correlation calculations must consider the circularity of the misalignment angle.  We expect $\hat{\psi}$~to cluster around~$0^\circ \mod 180^\circ$.  Even for the smallest masks of \figref{fig:psimap}, which are defined by $N_\mathrm{side} = 32$,  the majority of $\hat{\psi}$~values lie within $[-45^\circ,45^\circ] \mod 180^\circ$.  In our correlations, therefore, we ignore the circularity of~$\hat{\psi}$ and instead force the values to their physical equivalents in the range~$[-90^\circ,90^\circ]$.  In a small minority of cases, we will miss correlations between angles that lie at opposite extremes of this range. In testing for a correlation, our choice is \emph{conservative}.  Since we will be using \emph{Spearman} correlations, which operate on \emph{rank} variables, a convenient ordering strategy is to form~$\tan\hat{\psi}$.

\begin{figure}
\includegraphics[width=\columnwidth]{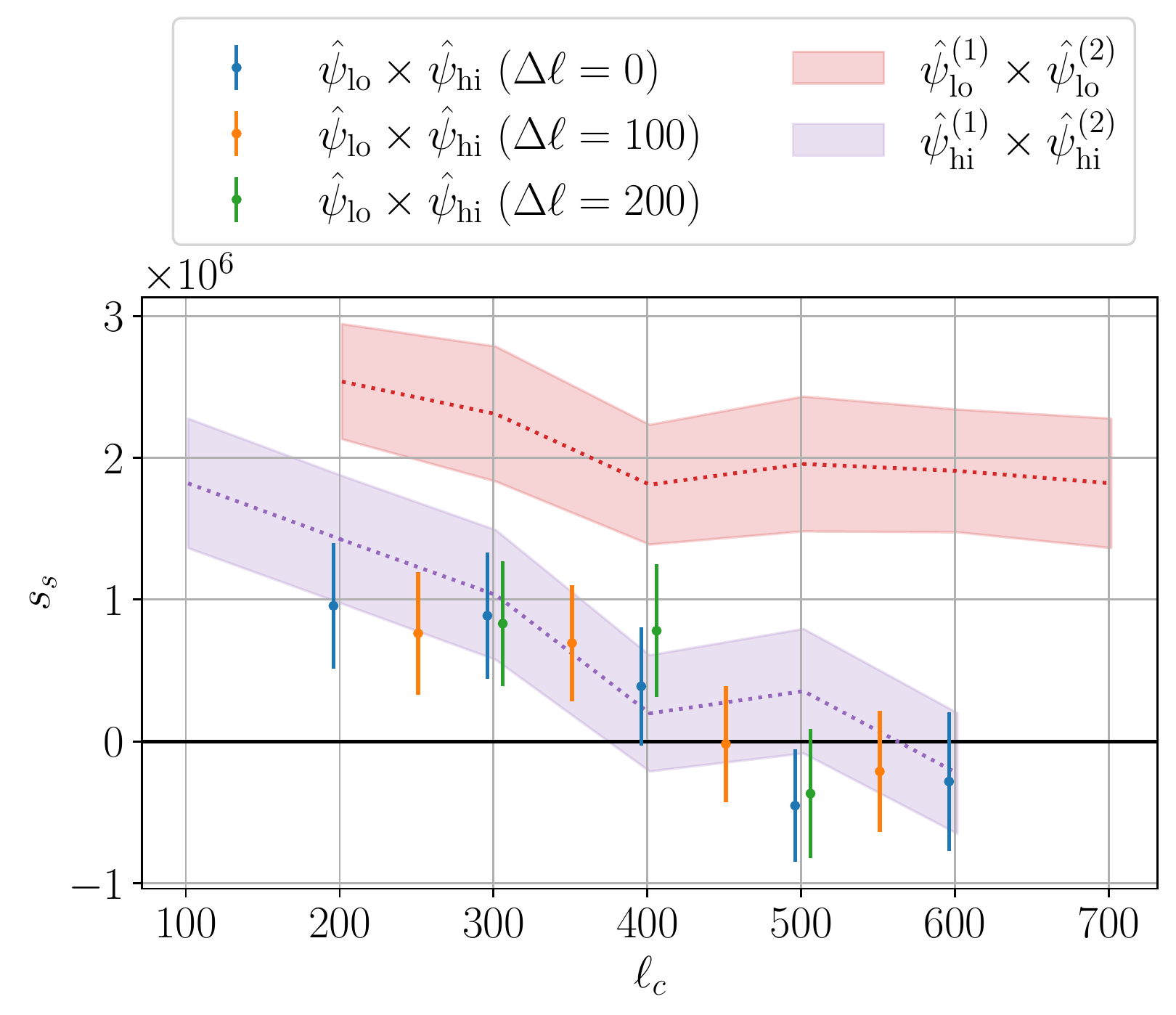}
\caption{Cross-power~$s_s$ between misalignment angles estimated from lowpass- and highpass-filtered maps (\eqref{eq:sspsilopsihi}): $\hat{\psi}_\mathrm{lo} \times \hat{\psi}_\mathrm{hi}$ for the patches of \figref{fig:patches}.  We consider three values for~$\Delta\ell$, and the data points have been offset from~$\ell_c$ for visual purposes. We also show the cross-power between half-mission splits in the shaded bands for lowpass-filtered~($\hat{\psi}^{(1)}_\mathrm{lo} \times \hat{\psi}^{(2)}_\mathrm{lo}$) and highpass-filtered estimates~($\hat{\psi}^{(1)}_\mathrm{hi} \times \hat{\psi}^{(2)}_\mathrm{hi}$). The $\hat{\psi}_\mathrm{hi}$~estimates become noise dominated for $\ell_c \gtrsim 400$.  This is a test for harmonic coherence (scale independence) of magnetic misalignment, and we find a positive signal for $\ell_c \lesssim 450$. \label{fig:psi_ellsplit_crosspower}}
\end{figure}

We form the Spearman cross-power (Spearman version of \eqref{eq:sp})
\eq{ \hat{\psi}_\mathrm{lo} \times \hat{\psi}_\mathrm{hi} \equiv s_s\parens{\tan \hat{\psi}_\mathrm{lo},  \tan \hat{\psi}_\mathrm{hi}} ,  \label{eq:sspsilopsihi} }
where the sum is taken over patches labeled by~$\nhat$.  Note that these Spearman cross-powers are \emph{not} correlation coefficients, so the numerical values range outside of~$[-1,1]$.  Correlation coefficients are less numerically stable in the presence of noise, so we prefer cross-powers for the purposes of establishing a relationship.  In \figref{fig:psi_ellsplit_crosspower}, we show $\hat{\psi}_\mathrm{lo} \times \hat{\psi}_\mathrm{hi}$ for several choices of~$\ell_c$ and~$\Delta\ell$ for the patches of \figref{fig:patches} (masks defined by $N_\mathrm{side}=8$ and $f_\mathrm{sky} = 70\%$).
We find a positive cross-power for $\ell_c \lesssim 450$ for all choices of~$\Delta\ell$. The noise in these measurements is mainly in the highpass-filtered misalignment estimates~$\hat{\psi}_\mathrm{hi}$.

We do \emph{not} expect our \rev{mock skies}~(\secref{sec:simulations}) to show a coherence over~$\ell$, because the Gaussian modes are resampled independently of each other and also independently of the \HI~template. One concern might be that the masking creates mode correlations, and this was part of the motivation for introducing the multipole buffer~$\Delta\ell$. As $\Delta\ell$~increases,  the two multipole passbands are further separated, and spurious correlations between the two are less likely. 

As a null test, we calculate $\hat{\psi}_\mathrm{lo} \times \hat{\psi}_\mathrm{hi}$ for an ensemble of \rev{mock skies}~(\secref{sec:simulations}), and we find the mean values to be consistent with zero.  Recall that these \rev{mock skies} are simplified in the sense that they are designed to reproduce only the 2-point statistics of the dust field, both in correlation with itself and with the \HI~template. Nonetheless, they are helpful in checking that our estimators produce sensible results.  The \HI~template appears in these \rev{mock skies} with the observed amplitude, and the only magnetic misalignment that has been input is due to random scatter.  We see no positive bias in the \rev{mock-sky} cross-powers.  The positive signal seen in the real map (\figref{fig:psi_ellsplit_crosspower}) must be due to a feature that it is absent in the \rev{mock skies}.

Due to noise in the $\hat{\psi}$~estimates, it is nontrivial to determine the \emph{fraction} of the misalignment signal that is harmonically coherent. In computing a correlation coefficient, noise tends to dilute the true signal. Using half-mission splits as in \eqref{eq:pearsonDenomDebiased} may lead to numerical pathologies when the denominators are small.  From the decay of $\hat{\psi}^{(1)}_\mathrm{hi} \times \hat{\psi}^{(2)}_\mathrm{hi}$ in \figref{fig:psi_ellsplit_crosspower}, we see that the highpass-filtered estimate~$\hat{\psi}_\mathrm{hi}$ is especially noisy.  The Spearman cross-powers~$\hat{\psi}^{(1)}_\mathrm{lo} \times \hat{\psi}^{(2)}_\mathrm{lo}$~(red in \figref{fig:psi_ellsplit_crosspower}) and~$\hat{\psi}^{(1)}_\mathrm{hi} \times \hat{\psi}^{(2)}_\mathrm{hi}$~(purple) are limited only by noise and set rough upper limits on the cross-power~$\hat{\psi}_\mathrm{lo} \times \hat{\psi}_\mathrm{hi}$.  Even if the misalignment angles were perfectly coherent across multipoles,  noise would suppress the cross-power.  That $\hat{\psi}_\mathrm{lo} \times  \hat{\psi}_\mathrm{hi}$~is of the same order as~$\hat{\psi}^{(1)}_\mathrm{lo} \times \hat{\psi}^{(2)}_\mathrm{lo}$ and~$\hat{\psi}^{(1)}_\mathrm{hi} \times \hat{\psi}^{(2)}_\mathrm{hi}$ is an indication that, within the limits of the noise, the harmonically coherent component is contributing a non-negligible fraction of the misalignment signal. 

We estimate the statistical significance of the apparently positive signal in \figref{fig:psi_ellsplit_crosspower}.  For each choice of~$\ell_c$ and~$\Delta\ell$, we construct permutation tests~(\secref{sec:statInf}), where covariances are preserved by using the same permutations for all choices. We combine the results with weights based on the half-mission cross-powers~(bands in \figref{fig:psi_ellsplit_crosspower}) and produce a single overall estimate of the statistical significance.
For the case of \figref{fig:psi_ellsplit_crosspower}, we estimate the statistical significance to be~$2.3\sigma$, where most of the sensitivity comes from the cross-powers with smaller~$\ell_c$ and smaller~$\Delta\ell$.  This is because $\hat{\psi}_\mathrm{hi}$~quickly becomes noise dominated as $\ell_c$~increases, and increasing~$\Delta\ell$ pushes the filter cutoff even higher. The data points in \figref{fig:psi_ellsplit_crosspower} are computed from the same maps but with different filtering parameters, so we expect them to be highly correlated. As such, combining the data points increases the overall significance only modestly. To a rough approximation, the overall significance can be estimated from the lowest-$\ell_c$ data point.

The results of \figref{fig:psi_ellsplit_crosspower} are based on the patches shown in \figref{fig:patches}, which are defined by $N_\mathrm{side}=8$ and $f_\mathrm{sky} = 70\%$. We can compute similar quantities for other values of~$N_\mathrm{side}$ and~$f_\mathrm{sky}$, and the results are compiled in \tabref{tab:psi_ellsplit_sig}, where we see that the significances are generally between~$2\sigma$ and~$4\sigma$ for $N_\mathrm{side} \in \{2,4,8,16,32\}$ and $f_\mathrm{sky} \in [60\%,90\%]$.
\begin{table}
\begin{center}
\begin{tabular}{c|c|c|c|c|c|c}
 & $40\%$ & $60\%$ & $70\%$ & $80\%$ & $90\%$ & $100\%$  \\
\hline
2 ($29.3^\circ$) & $1.1$ & $2.6$ & $3.0$ & $3.8$ & $3.6$ & $0.2$ \\
4 ($14.7^\circ$) & $3.7$ & $3.3$ & $3.5$ & $3.2$ & $2.8$ & $0.9$ \\
8 ($7.3^\circ$) & $1.9$ & $2.4$ & $2.3$ & $2.5$ & $3.9$ & $3.1$ \\
16 ($3.7^\circ$) & $1.2$ & $2.8$ & $2.7$ & $3.3$ & $2.2$ & $2.2$ \\
32 ($1.8^\circ$) & $3.8$ & $3.9$ & $4.2$ & $2.5$ & $1.5$ & $1.1$ \\
\end{tabular}
\caption{Statistical significance (in units of~$\sigma$) of measurements of harmonic coherence of~$\hat{\psi}$, e.g., those presented in \figref{fig:psi_ellsplit_crosspower},  for different values of~$N_\mathrm{side}$ (rows with side lengths provided parenthetically) and~$f_\mathrm{sky}$ (columns).  All of the results are positive with little dependence on~$N_\mathrm{side}$ and~$f_\mathrm{sky}$. \label{tab:psi_ellsplit_sig}}
\end{center}
\end{table}
With smaller~$f_\mathrm{sky}$, the significances tend to be smaller, but this may be simply a consequence of a decreased signal-to-noise ratio. On the full sky, the significances also tend to decrease, and this may be due to the inclusion of longer, denser sightlines at low Galactic latitudes.  
We do not attempt to combine the results from different choices of~$N_\mathrm{side}$ and~$f_\mathrm{sky}$, because the covariances are difficult to capture.

We consider the results of \figref{fig:psi_ellsplit_crosspower} and \tabref{tab:psi_ellsplit_sig} to represent \emph{tentative evidence for the harmonic-space coherence of misalignment angles} measured in small regions of sky away from the Galactic plane.

\subsection{Spatial coherence \label{sec:psiSpatialCoherence}}

We additionally search for \emph{spatial} coherence of misalignment angles by considering neighboring pairs of sky masks.  Although we bandpass filtered to $101 < \ell < 702$ in order to include only modes with wavelengths smaller than each patch, there is still a residual correlation between neighboring patches, which we detect with the \rev{mock skies} of \secref{sec:simulations}. To avoid the coherence due to common modes between neighboring patches, we again construct highpass- and lowpass-filtered maps as in \secref{sec:psiHarmonicCoherence}. We correlate the \emph{lowpass}-filtered estimate from each patch with the \emph{highpass}-filtered estimate from each of its neighbors, and we simultaneously correlate with the opposite application of filters. The misalignment estimates that enter the correlation calculations are separated in \emph{both} harmonic and map space. 

We utilize the Spearman version of the 4-variable cross-power (\eqref{eq:Sp})
\begin{equation}
\begin{split}
\hat{\psi}_\mathrm{lo}(\nhat) \times \hat{\psi}_\mathrm{hi}(\nhat')   & \equiv  S_s \left (  \tan \hat{\psi}_\mathrm{lo} (\nhat) ,  \tan \hat{\psi}_\mathrm{hi} (\nhat') ,  \right . \\
& \quad \quad \quad  \left . \tan \hat{\psi}_\mathrm{hi} (\nhat) ,  \tan \hat{\psi}_\mathrm{lo} (\nhat') \right ) ,  \label{eq:SsNeighbors} 
\end{split}
\end{equation}
where $\nhat$~and $\nhat'$~are the central sky coordinates of \emph{neighboring} patches.  As in \secref{sec:psiHarmonicCoherence}, we consider several choices for~$N_\mathrm{side}$ and~$f_\mathrm{sky}$, where \figref{fig:patches} shows one example.  The sum is taken over all pairs of neighboring patches. Each patch appears multiple times in this sum, but each pair appears only once.  Equation~\ref{eq:SsNeighbors} measures a \emph{simultaneous} correlation between~$\hat{\psi}_\mathrm{lo}(\nhat)$ and~$\hat{\psi}_\mathrm{hi}(\nhat')$ and between~$\hat{\psi}_\mathrm{hi}(\nhat)$ and~$\hat{\psi}_\mathrm{lo}(\nhat')$. The entire multipole range is being used in both patches but in two splits.  \emph{Without} multipole separation,  we cannot pass a null test based on our \rev{mock skies}~(\secref{sec:simulations}).  \emph{With} multipole separation, however, the \rev{mock skies} show no significant cross-power.

\begin{figure}
\includegraphics[width=\columnwidth]{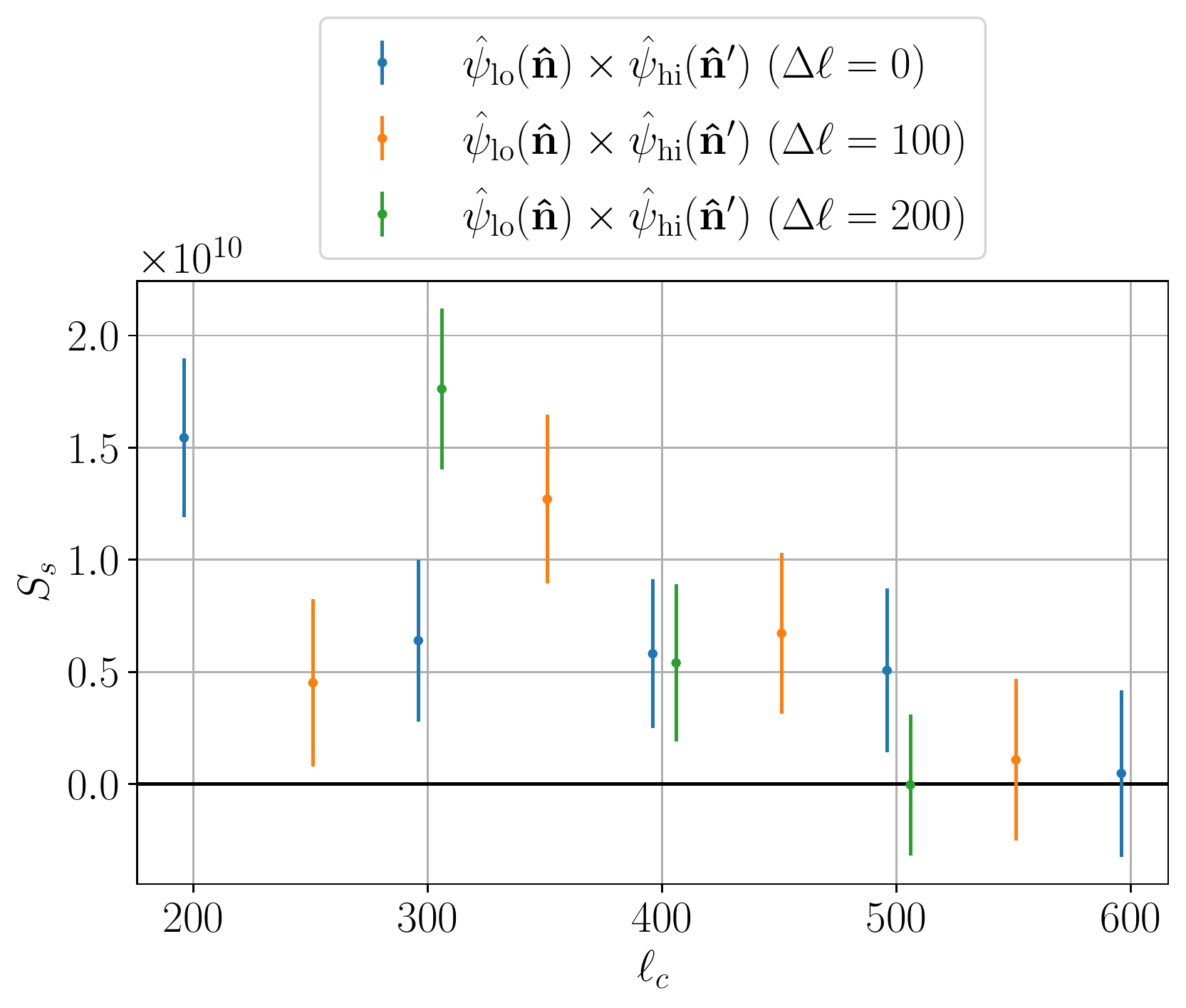}
\caption{Cross-power~$S_s$~(\eqref{eq:SsNeighbors}) between misalignment estimates from neighboring patches defined by $N_\mathrm{side}=16$ and $f_\mathrm{sky} = 70\%$.  This cross-power splits the misalignment estimates in \emph{both} harmonic and map space and is a test for spatial coherence.  The plotting conventions are the same as in \figref{fig:psi_ellsplit_crosspower}.  We find a positive signal for $\ell_c \lesssim 500$. \label{fig:neighbors_crosspower}}
\end{figure}

The results for $\hat{\psi}_\mathrm{lo}(\nhat) \times \hat{\psi}_\mathrm{hi}(\nhat')$~(\eqref{eq:SsNeighbors}) are shown in \figref{fig:neighbors_crosspower} for patches defined by $N_\mathrm{side}=16$. 
This patch area is 4~times smaller than that used for the measurement of harmonic coherence~(\figref{fig:psi_ellsplit_crosspower}). Measuring neighbor correlations at $N_\mathrm{side}=16$ probes the spatial coherence within patches defined by $N_\mathrm{side}=8$, so we are approximately measuring the spatial coherence \emph{within} the patches of \figref{fig:psi_ellsplit_crosspower}.  For the particular example of \figref{fig:neighbors_crosspower}, we find positive spatial coherence for $\ell_c \lesssim 500$.

We estimate the statistical significance of the positive signal shown in \figref{fig:neighbors_crosspower} by following a prescription similar to that of \secref{sec:psiHarmonicCoherence}.  Combining all of the measurements in a manner that accounts for covariances, we estimate the statistical significance to be~$3.6\sigma$, where most of the sensitivity, as for harmonic coherence, comes from the low-$\ell_c$, low-$\Delta\ell$ cross-powers. 

We can compute similar quantities with other values of~$N_\mathrm{side}$ and~$f_\mathrm{sky}$, and the results are compiled in \tabref{tab:neighbors_sig}, where we see that the spatial coherence tends to be stronger as the resolution is made finer.
\begin{table}
\begin{center}
\begin{tabular}{c|c|c|c|c|c|c}
 & $40\%$ & $60\%$ & $70\%$ & $80\%$ & $90\%$ & $100\%$ \\
\hline
2 ($29.3^\circ$)  & $-0.1$ & $-2.0$ & $-2.4$ & $-1.0$ & $-0.3$ & $-2.2$  \\
4 ($14.7^\circ$)  & $2.9$ & $1.7$ & $1.8$ & $1.7$ & $2.0$ & $-0.4$ \\
8 ($7.3^\circ$)  & $1.3$ & $0.8$ & $1.0$ & $0.5$ & $0.0$ & $-0.7$ \\
16 ($3.7^\circ$)  & $3.8$ & $3.9$ & $3.6$ & $3.0$ & $2.7$ & $3.4$ \\
32 ($1.8^\circ$)  & $2.9$ & $2.9$ & $4.0$ & $5.1$ & $4.7$  & $4.1$ \\
\end{tabular}
\caption{Statistical significance (in units of~$\sigma$) of measurements of spatial coherence of~$\hat{\psi}$~(\eqref{eq:SsNeighbors}), e.g., those presented in \figref{fig:neighbors_crosspower}, for different values of~$N_\mathrm{side}$ (rows with side lengths provided parenthetically) and~$f_\mathrm{sky}$ (columns). The significances are mostly positive and tend to increase with~$N_\mathrm{side}$.  \label{tab:neighbors_sig}}
\end{center}
\end{table}
For $N_\mathrm{side} \in \{16,32\}$, the significances are mostly between~$2\sigma$ and~$5\sigma$. As the side length associated with $N_\mathrm{side} = 32$ is~$1.8^\circ$, these results may imply that magnetic misalignment displays a coherence length of~$\mathcal{O}(1^\circ)$.

We consider the results of \figref{fig:neighbors_crosspower} and \tabref{tab:neighbors_sig} to represent \emph{tentative evidence for spatial coherence of misalignment angles}.  One perspective, however, is to consider the spatial coherence to be a necessary implication of the harmonic coherence that was established in \secref{sec:psiHarmonicCoherence}. We explained at the beginning of this section that the multipole split in \eqref{eq:SsNeighbors} helps to evade correlations between neighboring patches, which appear even in our statistically aligned \rev{mock skies}~(\secref{sec:simulations}).  The claim is that $\hat{\psi}_\mathrm{lo}(\nhat)$~is correlated with~$\hat{\psi}_\mathrm{lo}(\nhat')$ even in the \rev{mock skies}.  So we chose to correlate~$\hat{\psi}_\mathrm{lo}(\nhat)$ with~$\hat{\psi}_\mathrm{hi}(\nhat')$, and the \rev{mock skies} show no correlation in this case. But the \rev{mock skies} are also lacking harmonic coherence. In the real maps,  harmonic coherence appears to correlate~$\hat{\psi}_\mathrm{lo}(\nhat)$ with~$\hat{\psi}_\mathrm{hi}(\nhat)$ and~$\hat{\psi}_\mathrm{lo}(\nhat')$ with~$\hat{\psi}_\mathrm{hi}(\nhat')$~(\secref{sec:psiHarmonicCoherence}), but then we should expect, on the basis of the residual neighbor-to-neighbor correlations in the \rev{mock skies}, a correlation between~$\hat{\psi}_\mathrm{lo}(\nhat)$ and~$\hat{\psi}_\mathrm{hi}(\nhat')$ and between~$\hat{\psi}_\mathrm{hi}(\nhat)$ and~$\hat{\psi}_\mathrm{lo}(\nhat')$.  So there may indeed be a spatial coherence, but the crucial ingredient might be the \emph{harmonic} coherence.

\section{Parity-violating cross-spectra \label{sec:parityViolatingCrossSpectra}}

We now investigate connections between the misalignment angle~$\hat{\psi}$ and the parity-violating cross-spectra~$T_\mathrm{d} B_\mathrm{d}$, $E_\mathrm{d} B_\mathrm{d}$, $T_\mathrm{\HI} B_\mathrm{d}$, $E_\mathrm{\HI} B_\mathrm{d}$ and~$B_\mathrm{\HI} E_\mathrm{d}$. 
In Secs.~\ref{sec:misalignmentAnsatz} and~\ref{sec:misalignmentEstimatorLargeSky}, we described the expected relationships.  In Figs.~\ref{fig:EBfsky3} and~\ref{fig:psiEstimates}, we tested the implications of these relationships on relatively large sky areas.  In those particular cases, we used a 70\%~Galaxy mask and also checked the robustness of the results by restricting to the northern and southern hemisphere separately.

We now consider finer masks to search for coordinated variation in misalignment angle and parity-violating cross-spectra. 

\subsection{Random vs. harmonically coherent misalignment \label{sec:randomVsHarmCoh}}

As described in \secref{sec:misalignment}, we are interested in distinguishing between \emph{random} and \emph{harmonically coherent} misalignment. In \emph{both} cases, we expect to find that misalignment angle is correlated with~$TB$ and~$EB$,  but we can impose additional constraints to isolate harmonically coherent correlations.

Random misalignment is exemplified by the \rev{mock skies} of \secref{sec:simulations}. The \rev{mock skies} show deviations from the \HI~template, but the deviations are incoherent across multipoles, and there is no aggregate misalignment on large sky areas~(cf.~\secref{sec:misalignmentEstimatorLargeSky}).  We find the \rev{mock skies} show significant correlations between~$\hat{\psi}_\ell$ and~$D_\ell^{T_\mathrm{d}B_\mathrm{d}}$ and between~$\hat{\psi}_\ell$ and~$D_\ell^{E_\mathrm{d}B_\mathrm{d}}$. If we instead correlate between disjoint multipole bins, e.g.,  $\hat{\psi}_{\ell < \ell_c}$ with~$D_{\ell > \ell_c}^{T_\mathrm{d}B_\mathrm{d}}$ for some cutoff multipole~$\ell_c$, we find that the correlations vanish. The real data, as we will show below in Secs.~\ref{sec:corrMisalignSameMultipole} and~\ref{sec:harmCoherenceParityViol}, display \emph{both} types of correlations.

The \rev{mock skies} display correlations between~$\hat{\psi}_\ell$ and~$D_\ell^{T_\mathrm{d}B_\mathrm{d}}$ and between~$\hat{\psi}_\ell$ and~$D_\ell^{E_\mathrm{d}B_\mathrm{d}}$ as a direct consequence of the known strong correlation between the \emph{Planck} dust maps and the \HI~templates~\citep[e.g.,][]{ClarkHensley2019}. The \HI-dust correlation is maintained in the \rev{mock skies}. The \HI~component contributes non-negligibly to the dust polarization, and the \emph{Planck} maps can be viewed as \emph{perturbed} versions of the \HI~templates.  From \figref{fig:sim components}, we see that the perturbations need not be especially small; in fact, the Gaussian-dust component dominates over the \HI~component, though only modestly. If the perturbations are random, the dust polarization angles are symmetrically distributed relative to the \HI~template, and $\hat{\psi} = 0$. If, however, there is a region of sky in which the dust polarization angles are distributed \emph{asymmetrically} relative to the \HI~template, then $\hat{\psi} \not = 0$; in this case, there will be a net chirality, which will in turn produce non-zero contributions to~$T_\mathrm{d} B_\mathrm{d}$ and~$E_\mathrm{d} B_\mathrm{d}$.  Even in the \rev{mock skies}, there are regions of sky that fluctuate to non-zero~$\hat{\psi}$, and these regions tend to contribute non-zero~$T_\mathrm{d} B_\mathrm{d}$ and~$E_\mathrm{d} B_\mathrm{d}$ with a corresponding sign.  Our estimators avoid noise biases, so the relevant fluctuations are likely due to \emph{on-sky} dust components that deviate from the \HI~template. This is the expected contribution of magnetic misalignment to the parity-violating dust polarization quantities, but we aim to investigate whether the observed~$T_\mathrm{d} B_\mathrm{d}$ and~$E_\mathrm{d} B_\mathrm{d}$ are consistent with \emph{random} fluctuations away from the filament orientations -- as exemplified by the \rev{mock skies} -- or show evidence for harmonic or spatial coherence, which might be expected from a physical misalignment between the magnetic field and dusty filaments.

It is important to note that, while $\hat{\psi} > 0$ implies a tendency toward $T_\mathrm{d} B_\mathrm{d} , E_\mathrm{d} B_\mathrm{d} > 0$, the converse is \emph{not} guaranteed. It is possible to have $T_\mathrm{d} B_\mathrm{d}, E_\mathrm{d} B_\mathrm{d} > 0$ but $\hat{\psi} = 0$. Our \rev{mock skies}~(\secref{sec:simulations}) illustrate this point. They are constructed to retain the $T_\mathrm{d} B_\mathrm{d}$~spectrum of the true dust maps, but this property is placed entirely in the Gaussian component, which is statistically independent of the \HI~component. As such, the \rev{mock skies} display no aggregate misalignment (beyond realization-dependent scatter). For example,  on a 70\%~Galaxy mask, the ensemble mean of~$\hat{\psi}$ is \emph{zero}, although the mean $T_S B_S$~spectrum is \emph{positive} for $100 \lesssim \ell \lesssim 500$ as for the true~$T_\mathrm{d} B_\mathrm{d}$~(\secref{sec:IntroTB}). 

To the extent that the observed~$T_\mathrm{d} B_\mathrm{d}$ is related to magnetic misalignment, an outstanding question is whether the real dust~$TB$ is a consequence of physical misalignment or random scatter. Thus, we search for harmonically coherent relationships between misalignment angle and parity-violating cross-spectra. The \rev{mock skies} will help us to make the distinction, since harmonic coherence is \emph{not} included in them.

\subsection{Misalignment controls~$TB$ \label{sec:psiSelect}}

We begin the investigation with sky areas that are only modestly smaller than those of Secs.~\ref{sec:misalignmentAnsatz} and~\ref{sec:misalignmentEstimatorLargeSky}. In this limit,  the aggregate misalignment angles are small,  and the expected relationship to parity-violating cross-spectra can be approximated as (cf.~\eqref{eq:psiEstimatesApprox})
\eq{ \hat{\psi} \approx \frac{D_\ell^{T_\mathrm{d} B_\mathrm{d}}}{2 D_\ell^{T_\mathrm{d} E_\mathrm{d}}} \approx \frac{D_\ell^{E_\mathrm{d} B_\mathrm{d}}}{2 \parens{ D_\ell^{E_\mathrm{d} E_\mathrm{d}} - D_\ell^{B_\mathrm{d} B_\mathrm{d}} } } .  \label{eq:psiApproxTBEB} }
We will focus more on the dust-only spectra~$T_\mathrm{d} B_\mathrm{d}$ and~$E_\mathrm{d}B_\mathrm{d}$ as opposed to the dust-\HI\ spectra~$T_\mathrm{\HI}B_\mathrm{d}$, $E_\mathrm{\HI}B_\mathrm{d}$ and~$B_\mathrm{\HI}E_\mathrm{d}$, but similar operations can be performed for either set.

\begin{figure*}
\includegraphics[width=\textwidth]{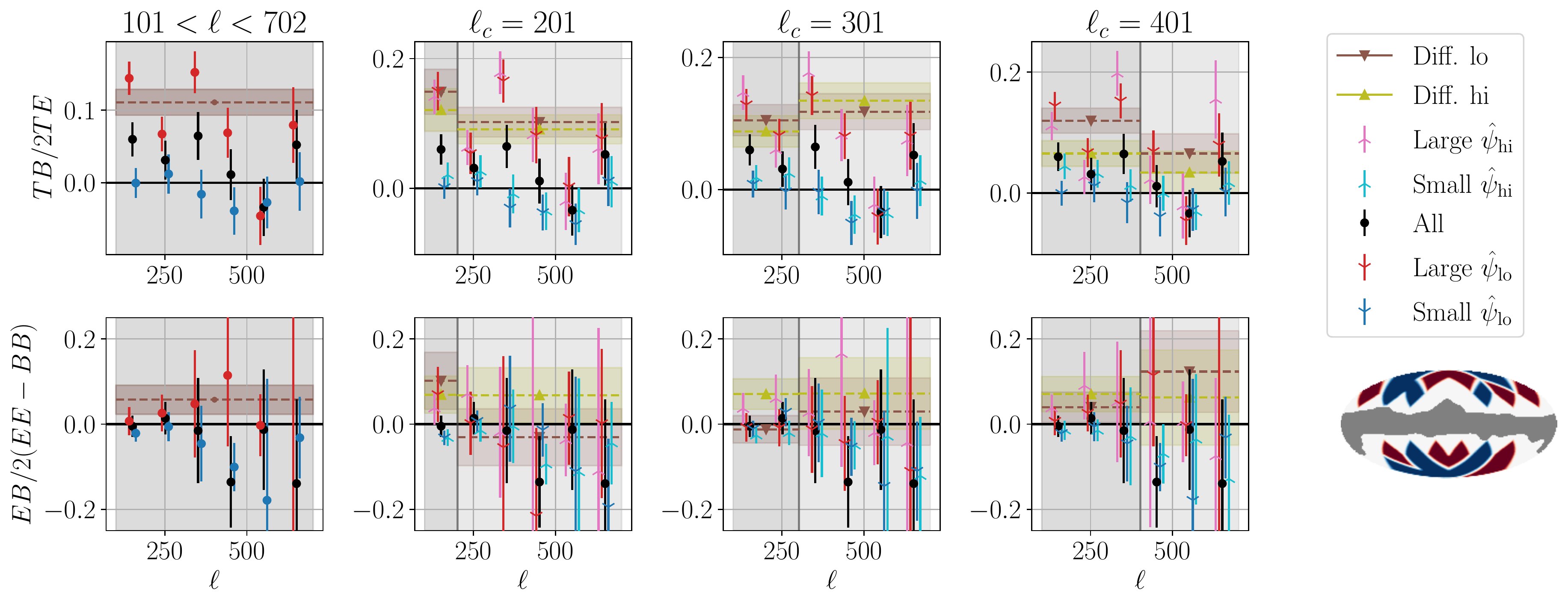}
\caption{Spectrum-based misalignment estimates (\eqref{eq:psiApproxTBEB}) of~$T_\mathrm{d} B_\mathrm{d}$~(top row) and~$E_\mathrm{d} B_\mathrm{d}$~(bottom row) for masks constructed by combining patches selected based on the value of~$\hat{\psi}$.  In general, we see an increase in the spectrum-based misalignment estimates for large-$\hat{\psi}$ masks and a decrease for small-$\hat{\psi}$. \rev{Black} points are for the total patch collection (combined red and blue in the lower-right map) and are the same across each row.  Red points are from large-$\hat{\psi}$ selections; blue from small-$\hat{\psi}$.  The leftmost column shows the results from filtering to $101 < \ell < 702$.  In the remaining columns,  the patch selections are based on~$\hat{\psi}_\mathrm{lo}$, computed after lowpass filtering to $101 < \ell < \ell_c$ (darker colors and downward markers), or~$\hat{\psi}_\mathrm{hi}$, computed after highpass filtering to $\ell_c < \ell < 702$ (lighter, upward), and the background shaded regions indicate the multipole splits.  The brown band shows the binned difference between the large- and small-$\hat{\psi}_\mathrm{lo}$ estimates; yellow shows the same for~$\hat{\psi}_\mathrm{hi}$.  (\emph{Lower right}) Patch selection based on~$\hat{\psi}$ after filtering to $101 < \ell < 702$.  Large-$\hat{\psi}$ patches are in red; small-$\hat{\psi}$ in blue.  The underlying white region is our fiducial 70\%~Galaxy mask.  \label{fig:psiSelect}}
\end{figure*}

We divide the sky into patches defined by $N_\mathrm{side} = 2$ within an overall 70\%~Galaxy mask.  In each patch, we measure~$\hat{\psi}$ as in \figref{fig:psimap} (with an additionally imposed Galaxy mask).  We form a combined mask from the patches with $\hat{\psi}$~larger than the median value, and we form an analogous mask for the patches with $\hat{\psi}$~smaller than the median; the masks are shown in the lower right of \figref{fig:psiSelect}, where the input $\hat{\psi}$~values are from maps that have been filtered to $101 < \ell < 702$ (as in the right column of \figref{fig:psimap}). 
We repeat for maps that have been lowpass filtered to $101 < \ell < \ell_c$~\rev{(which will be labeled by the subscript ``lo'')} and for maps that have been highpass filtered to $\ell_c < \ell < 702$~\rev{(subscript ``hi'')} for $\ell_c \in \{202,302,402\}$.

We calculate auto- and cross-spectra for the full combination of patches, for the large-$\hat{\psi}$ samples and for the small-$\hat{\psi}$ samples.  To avoid sharp mask features at shared vertices of the {\tt HEALPix}-defined patches, we use a conservative apodization scale of~$5^\circ$ for the results in this subsection. Note that, due to the {\tt HEALPix} pixelization and the increased apodization scale, the full combination of patches represents a \emph{smaller} overall sky area than produced by the fiducial 70\%~Galaxy mask of Secs.~\ref{sec:misalignmentAnsatz} and~\ref{sec:misalignmentEstimatorLargeSky}.  The reduction is somewhat severe and leaves a sky fraction of only~$37\%$. 

We convert the auto- and cross-spectra to misalignment estimates according to \eqref{eq:psiApproxTBEB}. The results are shown in the panel of spectra in \figref{fig:psiSelect}.

We find that the $T_\mathrm{d} B_\mathrm{d}$- and $E_\mathrm{d} B_\mathrm{d}$-based misalignment estimates increase and decrease in a manner consistent with the $\hat{\psi}$-based mask definition. The large-$\hat{\psi}$ masks tend to produce \emph{larger}~$T_\mathrm{d} B_\mathrm{d}/T_\mathrm{d} E_\mathrm{d}$ and~$E_\mathrm{d} B_\mathrm{d}/(E_\mathrm{d} E_\mathrm{d} - B_\mathrm{d} B_\mathrm{d})$, though the latter is much noisier. The small-$\hat{\psi}$ masks tend to produce \emph{smaller} spectrum-based estimates; interestingly, the resulting~$T_\mathrm{d} B_\mathrm{d}/T_\mathrm{d} E_\mathrm{d}$~(blue in the top row of \figref{fig:psiSelect}) is broadly consistent with zero rather than negative.  This may suggest that the positive~$T_\mathrm{d} B_\mathrm{d}$ measured at high Galactic latitudes is due to a few regions of sky with positive misalignment and that the rest of the sky respects parity.

When $\hat{\psi}$~is estimated over the same multipole range as the spectra,  as in the leftmost column of \figref{fig:psiSelect}, we cannot distinguish between the case of random fluctuations and that of harmonically coherent misalignment~(\secref{sec:randomVsHarmCoh}).  To isolate the harmonically coherent signal, we compare estimates from disjoint multipole ranges as in Secs.~\ref{sec:psiHarmonicCoherence} and~\ref{sec:psiSpatialCoherence}.  The three rightmost columns of the panel in \figref{fig:psiSelect} show the results when estimating~$\hat{\psi}$ from restricted multipole ranges.  Because the dust is brighter at low multipoles, the selections based on~$\hat{\psi}_\mathrm{lo}$ tend to be similar to those based on the unfiltered~$\hat{\psi}$.  For $T_\mathrm{d} B_\mathrm{d}$~and $E_\mathrm{d} B_\mathrm{d}$~to respond to the $\hat{\psi}$~selection in the disjoint multipole range is an indication of the harmonic coherence of magnetic misalignment, which was demonstrated in \secref{sec:psiHarmonicCoherence} but has now been explicitly connected to~$T_\mathrm{d} B_\mathrm{d}$.  With multipole splits, the data are too noisy to make a confident claim about~$E_\mathrm{d} B_\mathrm{d}$. 
The $T_\mathrm{d} B_\mathrm{d}$ results, which are less vulnerable to noise fluctuations, are similar for all choices of multipole filtering; furthermore, the small-$\hat{\psi}_\mathrm{lo}$ and small-$\hat{\psi}_\mathrm{hi}$ results tend to track each other, as do the large-$\hat{\psi}_\mathrm{lo}$ and large-$\hat{\psi}_\mathrm{hi}$ results. This may be yet another indication of the harmonic coherence of~$\hat{\psi}$, i.e., the $\hat{\psi}$-based patch selections are broadly similar in all multipole ranges.

We estimate the statistical significance of the multipole-split results by making \emph{random} patch selections to define the masks. We preserve covariances by using the same randomization for all~$\ell_c$.  In combining the results from all~$\ell_c$, we weight by a measure of signal-to-noise ratio~(cf.~\secref{sec:psiHarmonicCoherence}),  and we estimate the overall significance of the harmonic coherence to be~$2.2\sigma$ for~$T_\mathrm{d} B_\mathrm{d}$ and~$0.8\sigma$ for~$E_\mathrm{d} B_\mathrm{d}$.

\subsection{Correlations with misalignment angle \label{sec:corrMisalignSameMultipole} }

We search for correlations between misalignment angle~$\hat{\psi}$ and the parity-violating cross-spectra~$T_\mathrm{d} B_\mathrm{d}$, $E_\mathrm{d} B_\mathrm{d}$, $T_\mathrm{\HI} B_\mathrm{d}$, $E_\mathrm{\HI} B_\mathrm{d}$ and~$B_\mathrm{\HI} E_\mathrm{d}$.  We outlined our expectations in Eqs.~\ref{eq:EHBpsi relation}, \ref{eq:BHEpsi relation}, \ref{eq:TBpsi relation} and~\ref{eq:EBrelation}. For small angles, $\hat{\psi}_\ell$~is expected to track the parity-violating cross-spectra, but there are additional scaling factors. Rather than correlating the spectra with~$\hat{\psi}_\ell$ directly, we transform~$\hat{\psi}_\ell$ according to Eqs.~\ref{eq:EHBpsi relation}, \ref{eq:BHEpsi relation}, \ref{eq:TBpsi relation} and~\ref{eq:EBrelation}. 

\begin{figure*}
\includegraphics[width=\textwidth]{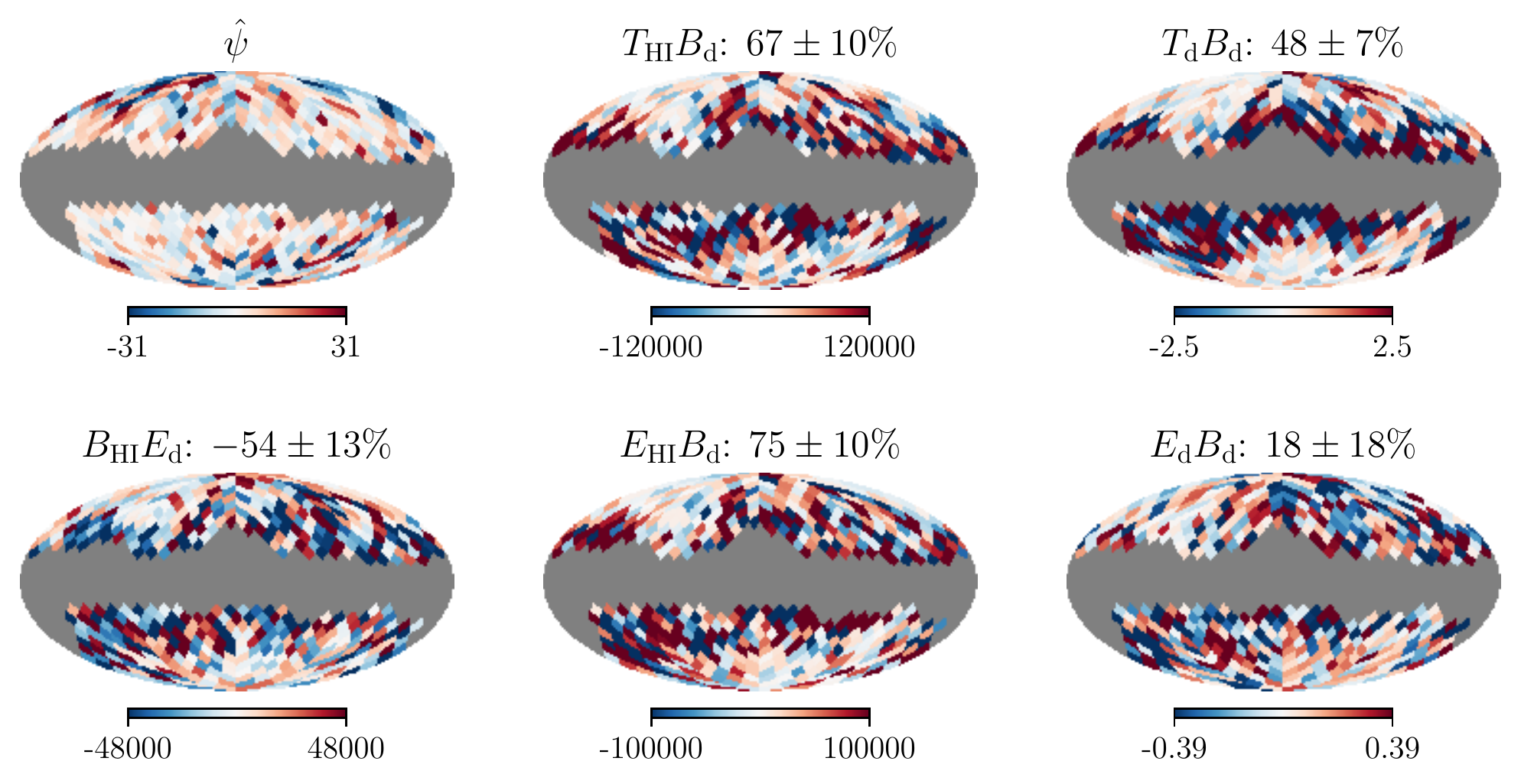}
\caption{Maps of misalignment angle~$\hat{\psi}$ and the five parity-violating cross-spectra binned to $101 < \ell < 702$ for patches defined by $N_\mathrm{side} = 8$ and $f_\mathrm{sky} = 70\%$.  For~$\hat{\psi}$, the units are degrees; for~$T_\mathrm{d} B_\mathrm{d}$ and~$E_\mathrm{d} B_\mathrm{d}$, ~$\mu\mathrm{K}_\mathrm{RJ}^2$; and, for~$T_\mathrm{\HI} B_\mathrm{d}$, $E_\mathrm{\HI} B_\mathrm{d}$ and~$B_\mathrm{\HI} E_\mathrm{d}$, ~$\mu\mathrm{K}_\mathrm{RJ}~\mathrm{K}~\mathrm{km}/\mathrm{s}$.  For the cross-spectra, the correlation with the transformed~$\hat{\psi}$ (Eqs.~\ref{eq:corr psiTB}, \ref{eq:corr psiEB}, \ref{eq:corr psiEHIB} and~\ref{eq:corr psiBHIE}) is given in the subtitle with a $1\sigma$~uncertainty (\secref{sec:statInf}).  We find all of the correlations predicted by the misalignment ansatz~(Eqs.~\ref{eq:EHBpsi relation}, \ref{eq:BHEpsi relation}, \ref{eq:TBpsi relation} and~\ref{eq:EBrelation}). \label{fig:map_panel}}
\end{figure*}

We compute Spearman half-mission correlation coefficients (cf.~\eqref{eq:pearsonHM}),  because both variables entering each of the following calculations are derived from the same \emph{Planck} dust modes and, therefore, subject to covariant noise fluctuations.  Due to the aforementioned transformations, the variables are different for each correlation calculation.  We compute~(cf.~Eqs.~\ref{eq:EHBpsi relation}, \ref{eq:BHEpsi relation}, \ref{eq:TBpsi relation} and~\ref{eq:EBrelation})
\eq{ \tilde{r}^{(\mathrm{HM})}_s\parens{ D_\ell^{E_\mathrm{\HI} B_\mathrm{d}} , D_\ell^{E_\mathrm{\HI} E_\mathrm{d}} \tan \parens{ 2 \hat{\psi}_\ell } } , \label{eq:corr psiEHIB} }
\eq{ \tilde{r}^{(\mathrm{HM})}_s\parens{ D_\ell^{B_\mathrm{\HI} E_\mathrm{d}} , D_\ell^{B_\mathrm{\HI} B_\mathrm{d}} \tan \parens{ 2 \hat{\psi}_\ell } } ,  \label{eq:corr psiBHIE} }
\eq{ \tilde{r}^{(\mathrm{HM})}_s\parens{ D_\ell^{T_x B_\mathrm{d}} , D_\ell^{T_x E_\mathrm{d}} \tan \parens{ 2 \hat{\psi}_\ell } } \label{eq:corr psiTB} }
for $x \in \{\mathrm{d},\mathrm{\HI}\}$ and
\eq{ \tilde{r}^{(\mathrm{HM})}_s\parens{ D_\ell^{E_\mathrm{d} B_\mathrm{d}} , \parens{ D_\ell^{E_\mathrm{d} E_\mathrm{d}} - D_\ell^{B_\mathrm{d} B_\mathrm{d}} } \tan \parens{ 4 \hat{\psi}_\ell } } ,  \label{eq:corr psiEB} }
where \eqref{eq:corr psiBHIE} is expected to be \emph{negative} and all others \emph{positive}.  We show these correlations and maps of the parity-violating quantities in \figref{fig:map_panel} for patches defined by~$N_\mathrm{side} = 8$ (as in \figref{fig:patches}).
The correlations have the expected sign in all cases.

Half-mission cross correlations~(\eqref{eq:pearsonHM}) avoid noise covariance but not \emph{sample variance} in the dust measurements.  The map features that produce positive~$\hat{\psi}$ also produce positive~$T_\mathrm{d} B_\mathrm{d}$, $E_\mathrm{d} B_\mathrm{d}$, $T_\mathrm{\HI} B_\mathrm{d}$ and~$E_\mathrm{\HI} B_\mathrm{d}$ and negative~$B_\mathrm{\HI} E_\mathrm{d}$~(\secref{sec:randomVsHarmCoh}). 
While the effective mode weighting in calculating~$\hat{\psi}_\ell$ is different than in the cross-spectra, we nevertheless find a correlation between the two in our \rev{mock skies} (\secref{sec:simulations}), for which the non-\HI~component is statistically independent of the \HI~component. In particular,  the \rev{mock skies} approximately reproduce the results of \figref{fig:map_panel}.

That our \rev{mock skies} show correlations between misalignment angle and parity-violating cross-spectra is an indication that the correlations of \figref{fig:map_panel} could be attributed to \emph{random} fluctuations away from the \HI~template~(\secref{sec:randomVsHarmCoh}). This is yet another motivation to restrict the search to signals that are \emph{coherent} in either harmonic or map space rather than correlating identical patches with identical multipole bins.

Independent of the distinction between random and harmonically coherent misalignment, the correlations of \figref{fig:map_panel} disfavor the presence of significant confounding contributions to parity violation in the polarization field.   \emph{A priori}, we might have expected non-filamentary contributions to dilute the relationship between \HI-based misalignment angle and parity-violating cross-spectra, especially when we consider that the \HI-correlated component is a minority contributor to the dust field~(\figref{fig:sim components}). Results like those of \figref{fig:map_panel} and the leftmost panel of \figref{fig:psiSelect} suggest that the \HI~template is sufficiently significant and representative to provide a reference for searches for parity violation.

\subsection{Harmonic coherence of parity violation \label{sec:harmCoherenceParityViol}}

Instead of directly correlating~$\hat{\psi}_\ell$ with, e.g.,~$D_\ell^{T_\mathrm{d} B_\mathrm{d}}$, we define disjoint multipole ranges and correlate the lowpass-filtered quantities with the highpass-filtered. This is similar to the multipole splits described in \secref{sec:psiHarmonicCoherence} and better extracts a signal that is coherent across multipoles. For the misalignment angle, we lowpass or highpass filter the map to form~$\hat{\psi}_\mathrm{lo}$ or~$\hat{\psi}_\mathrm{hi}$, respectively. For the spectra, we simply bin the lower or higher multipoles to form~$D_\mathrm{lo}^{XY}$ or~$D_\mathrm{hi}^{XY}$, respectively. As in \secref{sec:psiHarmonicCoherence}, the ``lowpass-filtered'' multipole range is $101 < \ell < \ell_c - \Delta\ell/2$, and the ``highpass-filtered'' is $\ell_c + \Delta\ell/2 < \ell < 702$, where $\ell_c$~is a transition multipole and $\Delta\ell$~is a multipole buffer between the two ranges.

We now modify Eqs.~\ref{eq:corr psiEHIB}, \ref{eq:corr psiBHIE} \ref{eq:corr psiTB} and~\ref{eq:corr psiEB} to correlate across multipole splits. We filter the spectra in the same way and the misalignment angle in the opposite way, e.g., we correlate~$D_\mathrm{lo}^{T_\mathrm{d} B_\mathrm{d}}$ with~$D_\mathrm{lo}^{T_\mathrm{d} E_\mathrm{d}} \tan \parens{ 2 \hat{\psi}_\mathrm{hi}}$.  We are seeking a relationship between~$D_\mathrm{lo}^{T_\mathrm{d} B_\mathrm{d}}$ and~$D_\mathrm{lo}^{T_\mathrm{d} E_\mathrm{d}}$, and our hypothesis is that the connection is provided by~$\hat{\psi}_\mathrm{hi}$, even though the latter is estimated in a disjoint multipole range. We look for a \emph{simultaneous} correlation when the multipole ranges are switched. 

\begin{figure*}
\includegraphics[width=\textwidth]{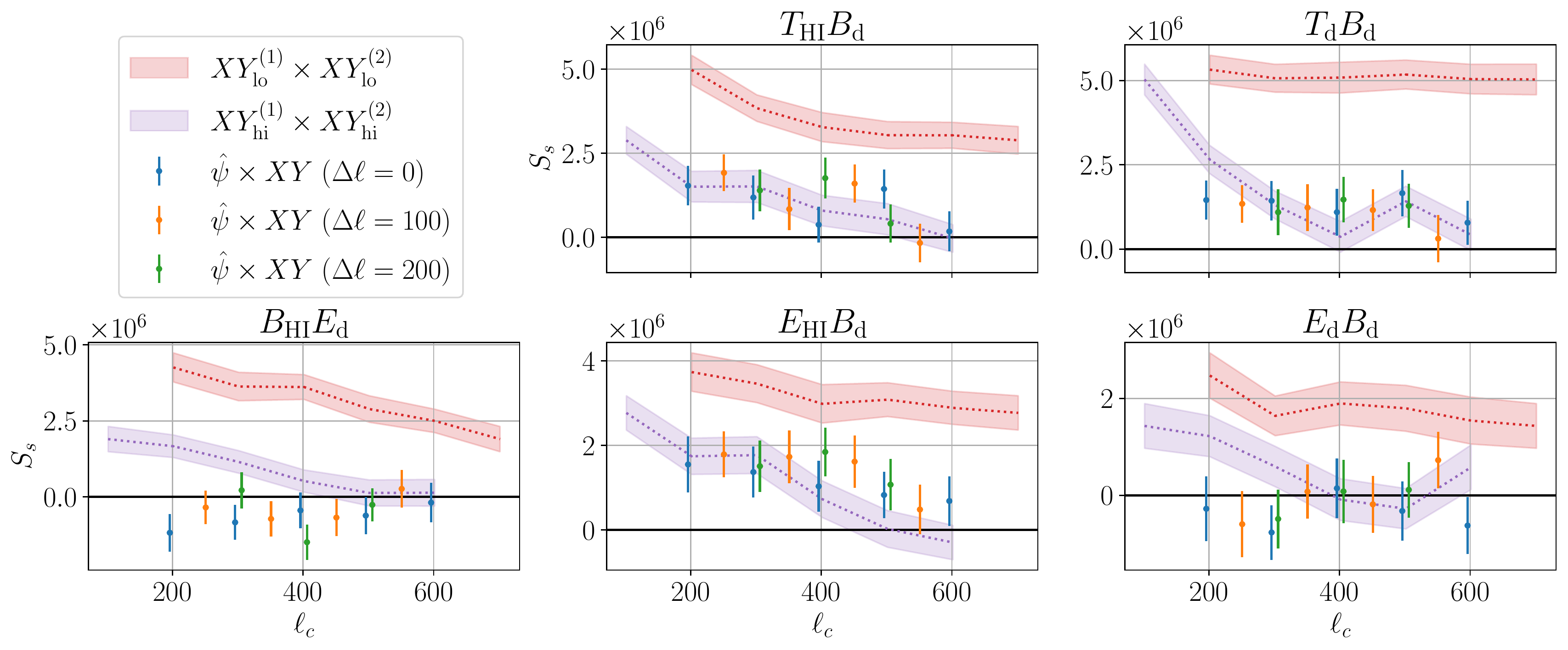}
\caption{Cross-power~$S_s$ between misalignment angle~$\hat{\psi}$ and parity-violating cross-spectra~$XY$~(e.g.,~\eqref{eq:SsTBpsi}) for patches defined by $N_\mathrm{side}=8$ and $f_\mathrm{sky} = 70\%$.  We consider three values for~$\Delta\ell$; the data points are offset from~$\ell_c$ for visual purposes.  The red and purple bands show, respectively the lowpass- and highpass-filtered half-mission cross-powers for the quantity in the subtitle~(Eqs.~\ref{eq:XYloCrossPower} and~\ref{eq:XYhiCrossPower}).  We find the expected correlations with~$E_\mathrm{\HI}B_\mathrm{d}$,  $B_\mathrm{\HI}E_\mathrm{d}$, $T_\mathrm{\HI}B_\mathrm{d}$ and~$T_\mathrm{d} B_\mathrm{d}$~(Eqs.~\ref{eq:EHBpsi relation}, \ref{eq:BHEpsi relation} and~\ref{eq:TBpsi relation}). \label{fig:rellpanel} }
\end{figure*}

For this purpose, we use the Spearman version of the 4-variable cross-power (\eqref{eq:Sp})
\begin{equation}
\begin{split}
\hat{\psi} \times T_\mathrm{d} B_\mathrm{d} & \equiv S_s \left ( D_\mathrm{lo}^{T_\mathrm{d} B_\mathrm{d}} , D_\mathrm{lo}^{T_\mathrm{d} E_\mathrm{d}} \tan \parens{ 2 \hat{\psi}_\mathrm{hi} } ,  \right . \\
& \quad \quad \quad  \left .  D_\mathrm{hi}^{T_\mathrm{d} B_\mathrm{d}} , D_\mathrm{hi}^{T_\mathrm{d} E_\mathrm{d}} \tan \parens{ 2 \hat{\psi}_\mathrm{lo} } \right ) \label{eq:SsTBpsi} 
\end{split}
\end{equation}
and similar combinations for the other four parity-violating cross-spectra.
The cross-powers are presented in \figref{fig:rellpanel} for patches defined by $N_\mathrm{side}=8$ (as in \figref{fig:patches}).
We also form these cross-powers for an ensemble of \rev{mock skies}. As noted earlier, our \rev{mock skies} \emph{cannot} be used for null-hypothesis testing, but they are useful for testing basic properties of our correlation metrics and misalignment estimators.  We find that the \rev{mock skies} produce null results within the realization-dependent scatter. 

To assess the noise level in the parity-violating cross-spectra, we consider the half-mission cross-powers~(Spearman version of \eqref{eq:sp})
\eq{XY_\mathrm{lo}^{(1)} \times XY_\mathrm{lo}^{(2)} \equiv s_s\parens{D^{X^{(1)} Y^{(1)}}_\mathrm{lo}, D^{X^{(2)} Y^{(2)}}_\mathrm{lo} } ,  \label{eq:XYloCrossPower} }
where $X^{(i)}$~and $Y^{(i)}$~are the $X$~and $Y$~fields, respectively, from the $i$th~half mission. We form a similar quantity for the highpass-filtered observables:
\eq{XY_\mathrm{hi}^{(1)} \times XY_\mathrm{hi}^{(2)} \equiv s_s\parens{D^{X^{(1)} Y^{(1)}}_\mathrm{hi}, D^{X^{(2)} Y^{(2)}}_\mathrm{hi} } .  \label{eq:XYhiCrossPower} }
Both $XY_\mathrm{lo}^{(1)} \times XY_\mathrm{lo}^{(2)}$ and $XY_\mathrm{hi}^{(1)} \times XY_\mathrm{hi}^{(2)}$ are limited only by noise. When noise is subdominant to the sky components, they will show strong positive signals; when noise is significant, they will decay to zero. We plot $XY_\mathrm{lo}^{(1)} \times XY_\mathrm{lo}^{(2)}$ and $XY_\mathrm{hi}^{(1)} \times XY_\mathrm{hi}^{(2)}$ as red and purple bands, respectively, in \figref{fig:rellpanel}, where we find that, in general, the high-$\ell$ quantities are substantially noisier than the low-$\ell$ quantities.

As discussed in \secref{sec:psiHarmonicCoherence}, half-mission cross-powers like $\hat{\psi}_\mathrm{hi}^{(1)} \times \hat{\psi}_\mathrm{hi} ^{(2)}$ and $XY_\mathrm{hi}^{(1)} \times XY_\mathrm{hi}^{(2)}$ set rough upper limits on the observable strength of the signals we are seeking.  In the case of \figref{fig:rellpanel}, we must consider the fidelity of both $\hat{\psi}$~(red/purple bands in \figref{fig:psi_ellsplit_crosspower}) and $XY$~(red/purple in \figref{fig:rellpanel}). When $\hat{\psi} \times XY$~is of the same order as the half-mission cross-powers, a non-negligible fraction of the variation in~$XY$ is associated with harmonically coherent misalignment. In \figref{fig:rellpanel}, this is the case for~$E_\mathrm{\HI}B_\mathrm{d}$,  $B_\mathrm{\HI}E_\mathrm{d}$, $T_\mathrm{\HI}B_\mathrm{d}$ and~$T_\mathrm{d} B_\mathrm{d}$.  For~$E_\mathrm{d} B_\mathrm{d}$, the half-mission cross-powers, especially~$\parens{E_\mathrm{d} B_\mathrm{d}}_\mathrm{hi}^{(1)} \times \parens{E_\mathrm{d} B_\mathrm{d}}_\mathrm{hi}^{(2)}$, are too noisy to make a reliable comparison.

We estimate the statistical significance of measurements like those of \figref{fig:rellpanel} by following a prescription similar to those of Secs.~\ref{sec:psiHarmonicCoherence} and~\ref{sec:psiSpatialCoherence}. The weights used in combining the measurements now account for noise in both~$\hat{\psi}$~(bands in \figref{fig:psi_ellsplit_crosspower}) and the cross-spectra~(bands in \figref{fig:rellpanel}).  For the example of \figref{fig:rellpanel}, we find $\hat{\psi} \times T_\mathrm{d} B_\mathrm{d}$ has a significance of~$2.9\sigma$, while $\hat{\psi} \times E_\mathrm{d} B_\mathrm{d}$ yields~$-0.8\sigma$, i.e., consistency with null. The cross-powers with~$T_\mathrm{\HI}B_\mathrm{d}$, $E_\mathrm{\HI}B_\mathrm{d}$ and~$B_\mathrm{\HI}E_\mathrm{d}$ yield, respectively,~$2.8\sigma$, $2.8\sigma$ and~$-1.75\sigma$, where the last value is expected to be negative~(\eqref{eq:BHEpsi relation}). Since the \emph{Planck}-\HI\ cross-spectra are approximate measures of the dust rotation relative to the \HI~template, the latter correlations can be considered further confirmation of the harmonic coherence of~$\hat{\psi}$~(\secref{sec:psiHarmonicCoherence}).

In \tabref{tab:psiXYsig}, we compile estimates of statistical significance for $\hat{\psi} \times T_\mathrm{d} B_\mathrm{d}$ and~$\hat{\psi} \times E_\mathrm{d} B_\mathrm{d}$ using different choices for~$N_\mathrm{side}$ and~$f_\mathrm{sky}$. 
\begin{table*}
\begin{center}
\begin{tabular}{c|c|c|c|c|c|c}
& $40\%$ & $60\%$ & $70\%$ & $80\%$ & $90\%$ & $100\%$ \\
\hline
2 ($29.3^\circ$) & $1.8$ ($1.4$) & $2.1$ ($1.2$) & $2.2$ ($1.8$) & $1.8$ ($1.9$) & $2.5$ ($1.3$) & $2.6$ ($1.5$) \\
4 ($14.7^\circ$) & $1.3$ ($-0.5$) & $2.5$ ($0.6$) & $2.8$ ($1.3$) & $3.0$ ($3.0$) & $2.3$ ($2.4$) & $1.7$ ($1.2$) \\
8 ($7.3^\circ$) & $1.9$ ($-0.9$) & $1.5$ ($-1.5$) & $2.9$ ($-0.8$) & $3.7$ ($-0.9$) & $3.9$ ($0.8$) & $4.4$ ($0.2$)
\end{tabular}
\end{center}
\caption{Statistical significance (in units of~$\sigma$) of measurements of harmonically coherent correlations between~$\hat{\psi}$ and~$T_\mathrm{d} B_\mathrm{d}$~($E_\mathrm{d} B_\mathrm{d}$), e.g., those shown in \figref{fig:rellpanel}, for different values of $N_\mathrm{side}$~(rows with side lengths provided parenthetically) and $f_\mathrm{sky}$~(columns).  The $\hat{\psi} \times T_\mathrm{d} B_\mathrm{d}$ significances are all positive and tend to increase with~$N_\mathrm{side}$ and~$f_\mathrm{sky}$. The $\hat{\psi} \times E_\mathrm{d} B_\mathrm{d}$ significances are mostly positive but are generally smaller.  \label{tab:psiXYsig} }
\end{table*}
The estimates are correlated with each other, and we have not attempted to estimate a global significance. What can be gleaned, however, is a tendency for \emph{positive} correlations with~$T_\mathrm{d} B_\mathrm{d}$ and mostly insignificant correlations with~$E_\mathrm{d} B_\mathrm{d}$. A few variations show a significance above~$2\sigma$ for~$E_\mathrm{d} B_\mathrm{d}$, and a majority are positive. But the overall picture is less compelling than in the case of~$T_\mathrm{d} B_\mathrm{d}$.  The $\parens{E_\mathrm{d} B_\mathrm{d}}_\mathrm{lo}^{(1)} \times \parens{E_\mathrm{d} B_\mathrm{d}}_\mathrm{lo}^{(2)}$ and $\parens{E_\mathrm{d} B_\mathrm{d}}_\mathrm{hi}^{(1)} \times \parens{E_\mathrm{d} B_\mathrm{d}}_\mathrm{hi}^{(2)}$ cross-powers~(Eqs.~\ref{eq:XYloCrossPower} and~\ref{eq:XYhiCrossPower} shown as red and purple bands in \figref{fig:rellpanel}) are 2-3~times smaller than the corresponding $T_\mathrm{d} B_\mathrm{d}$~quantities at low~$\ell_c$ and have fractionally larger uncertainties. Furthermore, the $\parens{E_\mathrm{d} B_\mathrm{d}}_\mathrm{hi}^{(1)} \times \parens{E_\mathrm{d} B_\mathrm{d}}_\mathrm{hi}^{(2)}$ signal becomes consistent with zero for $\ell_c \gtrsim 300$, i.e., $\parens{E_\mathrm{d} B_\mathrm{d}}_\mathrm{hi}$~becomes noise dominated. Given these considerations, it is consistent with our expectations that the $\sim 3\sigma$~results for $\hat{\psi} \times T_\mathrm{d} B_\mathrm{d}$ weaken to mostly null results for $\hat{\psi} \times E_\mathrm{d} B_\mathrm{d}$.

The real data, within the limits of the noise, are broadly consistent with our expectations, namely, positive correlations between~$\hat{\psi}$ and~$T_\mathrm{d} B_\mathrm{d}$,  $E_\mathrm{d} B_\mathrm{d}$, $T_\mathrm{\HI} B_\mathrm{d}$ and~$E_\mathrm{\HI} B_\mathrm{d}$ and negative between~$\hat{\psi}$ and~$B_\mathrm{\HI} E_\mathrm{d}$, though these signals disappear for some choices of~$N_\mathrm{side}$, $\ell_c$ and~$\Delta\ell$. In particular, the signal tends to decay as $\ell_c$~increases, which we expect due to increased noise in the highpass-filtered quantities.  For $N_\mathrm{side}=4$, the expected negative correlation with~$B_\mathrm{\HI} E_\mathrm{d}$ appears only for $\ell_c \gtrsim 350$ and is fairly weak.  For $N_\mathrm{side}=8$, the expected correlation with~$E_\mathrm{d} B_\mathrm{d}$ disappears. 

These results build confidence in our picture of harmonically coherent magnetic misalignment. Alternatively, these correlations can be considered \emph{necessary} implications of the harmonic coherence of~$\hat{\psi}$~(\secref{sec:psiHarmonicCoherence}) coupled with the expected relationship between~$\hat{\psi}$ and the parity-violating cross-spectra (Secs.~\ref{sec:randomVsHarmCoh} and~\ref{sec:corrMisalignSameMultipole}).  With this perspective, the cross-powers in Fig.~\ref{fig:rellpanel} are merely tests for consistency.

We draw special attention to the correlations between~$\hat{\psi}$ and~$T_\mathrm{d} B_\mathrm{d}$, as the positive~$T_\mathrm{d} B_\mathrm{d}$ measured by \emph{Planck} has been recently discussed in the literature~\citep{Huffenberger2020,Weiland2020,Clark2021,Huang2022}.
Although correlations between~$\hat{\psi}$ and~$E_\mathrm{d} B_\mathrm{d}$ yielded only weak results, we note the positivity of~$\parens{E_\mathrm{d} B_\mathrm{d}}_\mathrm{lo}^{(1)} \times \parens{E_\mathrm{d} B_\mathrm{d}}_\mathrm{lo}^{(2)}$~(\eqref{eq:XYloCrossPower} shown as the red band in \figref{fig:rellpanel}), which indicates on-sky variation in~$E_\mathrm{d} B_\mathrm{d}$ that rises above the noise level. Variation in~$E_\mathrm{d} B_\mathrm{d}$ may be attributable to sample-variance fluctuations of underlying parity-even statistical processes, but the particular dust realization that we observe is a foreground that must be mitigated for, e.g., measurements of the CMB.  Spatial variation in~$E_\mathrm{d} B_\mathrm{d}$ may be of relevance for measurements of cosmic birefringence~\citep{Minami2019,Minami2020,DiegoPalazuelos2022,Eskilt2022}. If future measurements can more confidently establish a relationship between~$\hat{\psi}$ and~$E_\mathrm{d} B_\mathrm{d}$,  foreground removal could be performed more robustly by, e.g., relating~$E_\mathrm{d} B_\mathrm{d}$ to~$T_\mathrm{d} B_\mathrm{d}$ and other observables~(e.g.,~Eqs.~\ref{eq:EHBpsi relation}, \ref{eq:BHEpsi relation}, \ref{eq:TBpsi relation} and~\ref{eq:EBrelation}).

\section{Conclusion and outlook \label{sec:conclusion}}

We have extended the work of \citet{Clark2021} in establishing a connection between dust $TB$~correlations and the magnetic misalignment of interstellar dust filaments.  We have introduced a new version of a Hessian-based \HI\ polarization template, which correlates more strongly with dust $B$~modes than the RHT-based template used previously~(\secref{sec:HessianMethod}).  We introduced several spectrum-based misalignment estimators formed from the auto- and cross-spectra of \emph{Planck} dust maps and \HI~polarization templates (\secref{sec:implicationsCrossSpectra}), and we also introduced a map-based estimator for misalignment angle~(\secref{sec:misalignmentEstimator}).  We have presented maps of the misalignment angle~(\secref{sec:misalignmentMaps}), which show a tendency to positive values and a visual correlation with dust polarization fraction.  We have provided evidence for the scale independence (harmonic coherence) of the misalignment angle for multipoles $\ell \lesssim 700$ (Secs.~\ref{sec:misalignmentEstimatorLargeSky} and \ref{sec:psiHarmonicCoherence}) and for spatial coherence on angular scales of~$\sim 1^\circ$~(\secref{sec:psiSpatialCoherence}).  On large sky areas at high Galactic latitudes, we find a scale-independent misalignment angle of~$\sim 2^\circ$, which is robust to a variety of masking choices~(\secref{sec:varyingSkyFrac}).  We have described a set of \rev{mock skies}~(\secref{sec:simulations}) containing \HI-based filamentary structure as well as Gaussian-random components, and we have used these \rev{mock skies} to refine our notion of magnetic misalignment. 
In particular, we have explored the question of whether the measured misalignment between \HI~filaments and the magnetic-field orientation is consistent with random fluctuations in the polarization field~(\secref{sec:randomVsHarmCoh}). This question motivated a search for scale independence~(harmonic coherence) as a salient physical property of magnetic misalignment.  We find evidence for a scale-independent correlation between misalignment angle and dust~$TB$~(\secref{sec:harmCoherenceParityViol}). With the noisier~$EB$, we find a correlation for some but not all of our masking choices.  We also find that the observed positive dust~$TB$ may be due to a few regions with strong positive misalignment while the rest of the sky largely respects parity~(\secref{sec:psiSelect}).

In general, the picture that is beginning to emerge contains the following features:
\begin{itemize}
\item \rev{On large scales at high Galactic latitudes, t}here is a global tendency toward an aggregate misalignment of~$\sim 2^\circ$~(Secs.~\ref{sec:misalignmentEstimatorLargeSky}, \ref{sec:globalMisalignment} and~\ref{sec:varyingSkyFrac}).  

\item Magnetic misalignment is a reliable predictor of parity violation in the dust polarization~(Secs.~\ref{sec:psiSelect} and~\ref{sec:corrMisalignSameMultipole}). 

\item Magnetic misalignment is partially scale independent~(harmonically coherent, Secs.~\ref{sec:psiHarmonicCoherence} and~\ref{sec:harmCoherenceParityViol}).
\end{itemize}

We now provide suggestions for potential improvements to our analysis.
\begin{enumerate}
\item The ansatz~(\secref{sec:misalignmentAnsatz}) could be modified to allow for only a fraction of the dust to participate in misalignment. In this work, it is assumed that all of the dust is misaligned, but this may dilute the sensitivity of our estimators.  In \citet{Clark2021}, this type of concern was addressed in estimating the misalignment-induced~$EB$ in Eq.~12.

\item The \HI~template~(\secref{sec:HessianMethod}) could be improved to correlate more strongly with the \emph{Planck} dust maps. In this work, we introduced a new Hessian-based template, which correlates more strongly with the dust $B$~modes than the RHT-based template used in \citet{Clark2021}, but the correlation is still less than~$20\%$ for $\ell \gtrsim 200$.  While the \HI-based filamentary model may be fundamentally limited due to diffuse non-filamentary dust or other dust morphologies, we consider it more likely that a dedicated exploration will yield stronger correlations with the measured dust polarization~\citep{HalalPrep}.  Magnetic misalignment is a perturbation to the filamentary model, so an increased correlation would improve the sensitivity of all of the \HI-related estimators presented in this work.

\item More realistic \rev{mock skies or} simulations~(\secref{sec:simulations}) will aid in the physical interpretation of our estimators. For example, the magnetohydrodynamic~(MHD) simulations of \citet{Kim2017}, which can be converted to dust polarization maps~\citep{Kim2019} that were  considered in \citet{Clark2021}, model the Solar neighborhood and are publicly available at a resolution of $N_\mathrm{side} = 128$.  Similar simulations with higher resolution and synthetic \HI~observations could be analyzed with our estimators.  Misalignment could also be investigated in synthetic dust polarization observations directly by searching for, e.g., scale independence in~$TB$ and~$EB$.  In Sec.~4 of \citet{Clark2021}, this type of analysis was performed on a limited multipole range ($60 < \ell < 120$). Higher-resolution simulations will enable an extension to higher multipoles and further investigation of the link to underlying physics.

\item The pixels weights~$w(\nhat)$~(\secref{sec:misalignmentEstimator}) that enter the calculation of~$\hat{\psi}$ are likely suboptimal.  We checked that our choice reduces variance relative to a uniform weighting, but we have not explored the full space of possibilities.  A better choice may be a Wiener filter that prevents a few bright pixels from dominating. Similarly, the correlation metrics used in Secs.~\ref{sec:misalignment} and~\ref{sec:parityViolatingCrossSpectra} could be defined with weights to suppress noisy regions of sky. 

\item The large-scale (low-$\ell$) misalignment should be considered more rigorously, because dust polarization is dominated by these modes. We have mostly limited our investigation to $\ell \gtrsim 100$ to avoid large-scale covariances. But there appears to be a strong positive misalignment on large scales~(\figref{fig:psimap}), and we speculate that this may be related to the magnetic-field structure in the vicinity of the Local Bubble~\citep[e.g.,][]{Lallement2003,Alves2018,Leike2020,Pelgrims2020,Vergely2022}.

\item Other sources of parity violation should be considered, since magnetic misalignment alone may be insufficient to account for, e.g., the observed~$TB$.  We mention in \secref{sec:intrinsic parity} that the distribution of dust filaments may itself display a chiral asymmetry even in the limit of perfect magnetic alignment.  Both the Hessian-based and RHT-based \HI~templates, which assume perfect \emph{alignment}, predict a rise in~$EB$ for $\ell \lesssim 100$~(\figref{fig:TBEBtemplate}), though the expected signal is below the \emph{Planck} noise levels.  We defer the investigation of this \emph{morphological} parity violation to future work.

\item Other magnetic-field tracers such as starlight polarization and Faraday rotation could be incorporated to better understand the three-dimensional manifestation of magnetic misalignment.  With stellar distance measurements from, e.g., \emph{Gaia}~\citep{Gaia2016}, starlight polarization measurements can enable a tomographic reconstruction of Galactic magnetic fields, though this technique is sensitive only to the plane-of-sky component~\citep{Panopoulou2019}. Faraday rotation measures probe the line-of-sight magnetic-field component~\citep{Hutschenreuter2020} and can be combined with model expectations or plane-of-sky observations to constrain the three-dimensional magnetic-field structure~\citep[e.g.,][]{Tahani2019,Tahani2022}.
\end{enumerate}

\rev{Our misalignment analysis can be applied to a variety of ISM environments. As a method of studying the relative orientations of magnetic fields~(not necessarily with dust polarization) and density structures~(not necessarily with~\HI), our approach is complementary to those of, e.g., \citet{PlanckXXXII2016}, \citet{Soler2017} and \citet{Fissel2019}, which consider both the diffuse ISM and molecular clouds.}

The study of parity violation in Galactic dust polarization is of central importance both for cosmology and for ISM physics. Our investigation has been limited by noise in the \emph{Planck} polarization maps, and we, therefore,  recommend follow-up surveys at millimeter and submillimeter wavelengths covering large sky fractions with resolution similar to or finer than the \HI4PI beam width~($16.2'$). More sensitive measurements will become available from upcoming projects including the space-based LiteBIRD~\citep{Hazumi2020} and the ground-based Simons Observatory~\citep{Hensley2022}, CCAT-prime~\citep{CCATPrime2021} and CMB Stage~4~\citep{CMBS42019}.

\section*{Acknowledgments}
We thank Dominic Beck, Federico Bianchini, Chao-Lin Kuo, Enrique Lopez-Rodriguez and Kimmy Wu for advice and helpful conversations.  
This work was partly supported by the National Science Foundation under Grant No. 2106607.
Some of the computing was performed on the Sherlock cluster; we thank Stanford University and the Stanford Research Computing Center.
This work makes use of observations obtained with \emph{Planck} (\url{http://www.esa.int/Planck}), an ESA science mission with instruments and contributions directly funded by ESA Member States, NASA, and Canada. \HI4PI is based on observations with the 100-m telescope
of the MPIfR at Effelsberg and the Parkes Radio Telescope,  part of the Australia Telescope National Facility which is funded by the Australian Government and managed by CSIRO.

\software{{\tt healpy}~\citep{Zonca2019},  {\tt matplotlib}~\citep{Hunter2007},  {\tt NaMaster}~\citep{Alonso2019},  {\tt numpy}~\citep{Harris2020},  {\tt scipy}~\citep{Virtanen2020}
          }

\appendix

\section{Hessian method: supplemental material \label{sec:HessianSupp}}

Here we provide additional information related to the Hessian method~(\secref{sec:HessianMethod}) that supports some of the analysis choices made in this work.  In the comparisons below, we will occasionally use the subscript~``$H$'' to refer to the Hessian method and~``$\mathrm{RHT}$'' for the Rolling Hough Transform. With the subscript~``$\mathrm{\HI}$'', we implicitly refer to the Hessian method as in the main text.

\subsection{Comparison with the Rolling Hough Transform \label{sec:comparisonWithRHT}}

The Rolling Hough Transform (RHT) is another filament-finding algorithm~\citep{Clark2014} from which we can produce polarization templates that correlate strongly with the dust polarization measured by \emph{Planck}~\citep{Clark2015,ClarkHensley2019}.  

\begin{figure}
\includegraphics[width=\columnwidth]{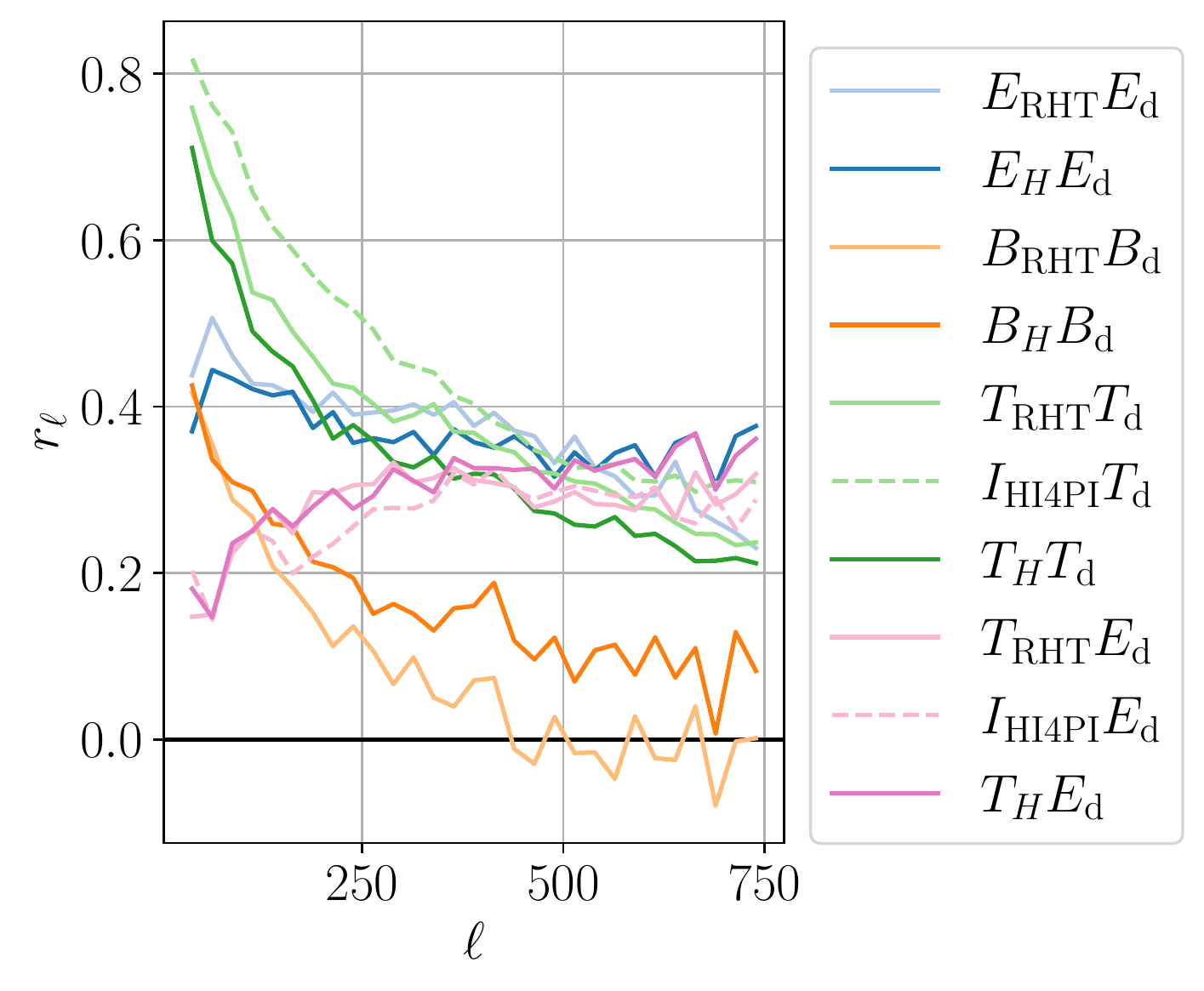}
\caption{Correlation~$r_\ell$ between \emph{Planck} dust maps~($\mathrm{d}$) and the RHT, the Hessian method~($H$) and the raw \HI4PI intensity map~($I_\mathrm{\HI4PI}$) with multipole bin width $\Delta\ell = 25$ on a 70\%~Galaxy mask.  \label{fig:RHTvsHess}}
\end{figure}

\rev{We find that the Hessian correlates more strongly with \emph{Planck} dust $B$~modes than the RHT for $\ell \gtrsim 100$, and we show a comparison in \figref{fig:RHTvsHess}.  In $E$~modes, the two perform similarly. In these comparisons, the RHT is constructed from velocities spanning~$-90$ to~$90~\mathrm{km}/\mathrm{s}$~\citep[as in][]{ClarkHensley2019}, while the Hessian template is constructed from the restricted range of~$-15$ to~$4~\mathrm{km}/\mathrm{s}$,~(\secref{sec:HessianPrescription}).}

For the $T$~template, we consider two choices for the~RHT. One might use $I_{\text{\HI4PI}}$, i.e.,  the \HI4PI intensity map~\citep{HI4PI2016} without any processing. Since we are especially interested in the \emph{filamentary component} of the dust intensity, it may be preferable to highpass filter the \HI4PI intensity as in the first step in the RHT algorithm of \citet{ClarkHensley2019}. The filter is implemented as an unsharp mask with $\mathrm{FWHM} = 30'$.  Denote the highpass-filtered intensity by~$T_\mathrm{RHT}$.  \rev{In \figref{fig:RHTvsHess}, we find that $I_\mathrm{\HI4PI}$~correlates more strongly than~$T_\mathrm{RHT}$ with \emph{Planck} $T$~modes.  This is not necessarily the relevant metric,  however, since we are specifically targeting \emph{filaments}. We also present in \figref{fig:RHTvsHess} the correlation with \emph{Planck} $E$~modes, since the $TE$~correlation is a signature of filamentary polarization. We find that $r_\ell^{T_\mathrm{RHT} E_\mathrm{d}}$~is generally larger than~$r_\ell^{I_{\text{\HI4PI}} E_\mathrm{d}}$.  For this reason, we prefer~$T_\mathrm{RHT}$ as a template for \emph{filamentary} dust intensity.}

The Hessian intensity~$T_H$ (Eqs.~\ref{eq:Hessian Tv} and \ref{eq:XsumDef}) is defined mainly by the Hessian eigenvalues.  \rev{In \figref{fig:RHTvsHess}, we find that $r_\ell^{T_H T_\mathrm{d}}$~is smaller than~$r_\ell^{T_\mathrm{RHT} T_\mathrm{d}}$ but that $r_\ell^{T_H E_\mathrm{d}}$~is similar to and, at high~$\ell$, slightly larger than~$r_\ell^{T_\mathrm{RHT} E_\mathrm{d}}$.}

On account of the greater correlation in $B$~modes, we have selected the Hessian-based template as our baseline.  The $T$~templates considered above all correlate strongly with both~$T_\mathrm{d}$ and~$E_\mathrm{d}$ and at roughly the same level. The $E$~templates for both algorithms correlate with~$E_\mathrm{d}$ at roughly the same level. 

We defer to future work a more detailed investigation of these and related filament-finding alogirthms for the construction of polarization templates~\citep{HalalPrep}. Each can be modified and tuned by making different choices for, e.g., velocity binning, weighting, spatial filtering,~etc.

\subsection{Transfer function \label{sec:TransferFunction}}

The \HI-based polarization templates have different mode structures than the \emph{Planck} dust maps. For example, the Hessian method upweights small-scale features; the $E_H$~and $B_H$~power spectra increase with~$\ell$. The RHT also upweights small-scale features but especially emphasizes the multipole range $300 \lesssim \ell \lesssim 500$ for~$E_\mathrm{RHT}$ and $150 \lesssim \ell \lesssim 350$ for~$B_\mathrm{RHT}$~\citep[with RHT parameters set to those of][]{ClarkHensley2019}.

Correlations, e.g., those presented in \citet{ClarkHensley2019}, are insensitive to differences in mode structure, because they are evaluated in individual multipole bins.  The upweighting of one multipole bin relative to another is normalized out of the calculation.  The difference in mode structure can be viewed as a multipole-dependent \emph{transfer function}.

For the purposes of converting our \HI-based polarization templates into quantities that are directly comparable to the \emph{Planck} dust maps, we assume a transfer function that depends only on multipole~$\ell$:
\eq{ k_\ell^{(X)} \equiv \frac{ D_\ell^{X_{\text{\HI}} X_\mathrm{d}} }{ D_\ell^{X_{\text{\HI}} X_{\text{\HI}}} } . \label{eq:kDef} }
This transfer function converts an \HI-based quantity into dust-intensity units with a mode structure that is directly comparable to the observed dust field.    We find that $k_\ell{(X)}$~rises strongly at low multipoles, which is an indication that the \HI~templates tend to \emph{under}predict the amplitude of large-scale dust polarization relative to small-scale. In spite of this underprediction, the \emph{correlations}, which normalize out the $\ell$~dependence, are actually stronger at low~$\ell$.

There is no guarantee that the transfer function~$k_\ell^{(X)}$ provides a representative estimate of the amplitude of \HI-based modes in the real dust maps.  The amplitudes may depend on both~$\ell$ and~$m$, the spherical-harmonic eigenvalues. We use~$k_\ell^{(X)}$ as a rough conversion factor to make direct comparisons between the \HI~templates and the real dust maps.

Because the Hessian-based template correlates non-vanishingly with \emph{Planck} up to at least $\ell = 750$, the transfer function remains usable across the entire multipole range considered in our analysis.

\subsection{Parity in the templates \label{sec:intrinsic parity}}

The \HI~templates are produced under the assumption of perfect magnetic \emph{alignment}. Even so,  chirality in the filament \emph{morphology} could produce parity-violating signatures such as non-zero $T_\mathrm{\HI} B_\mathrm{\HI}$~and $E_\mathrm{\HI} B_\mathrm{\HI}$. 

\begin{figure}
\centering
\includegraphics[width=0.5\textwidth]{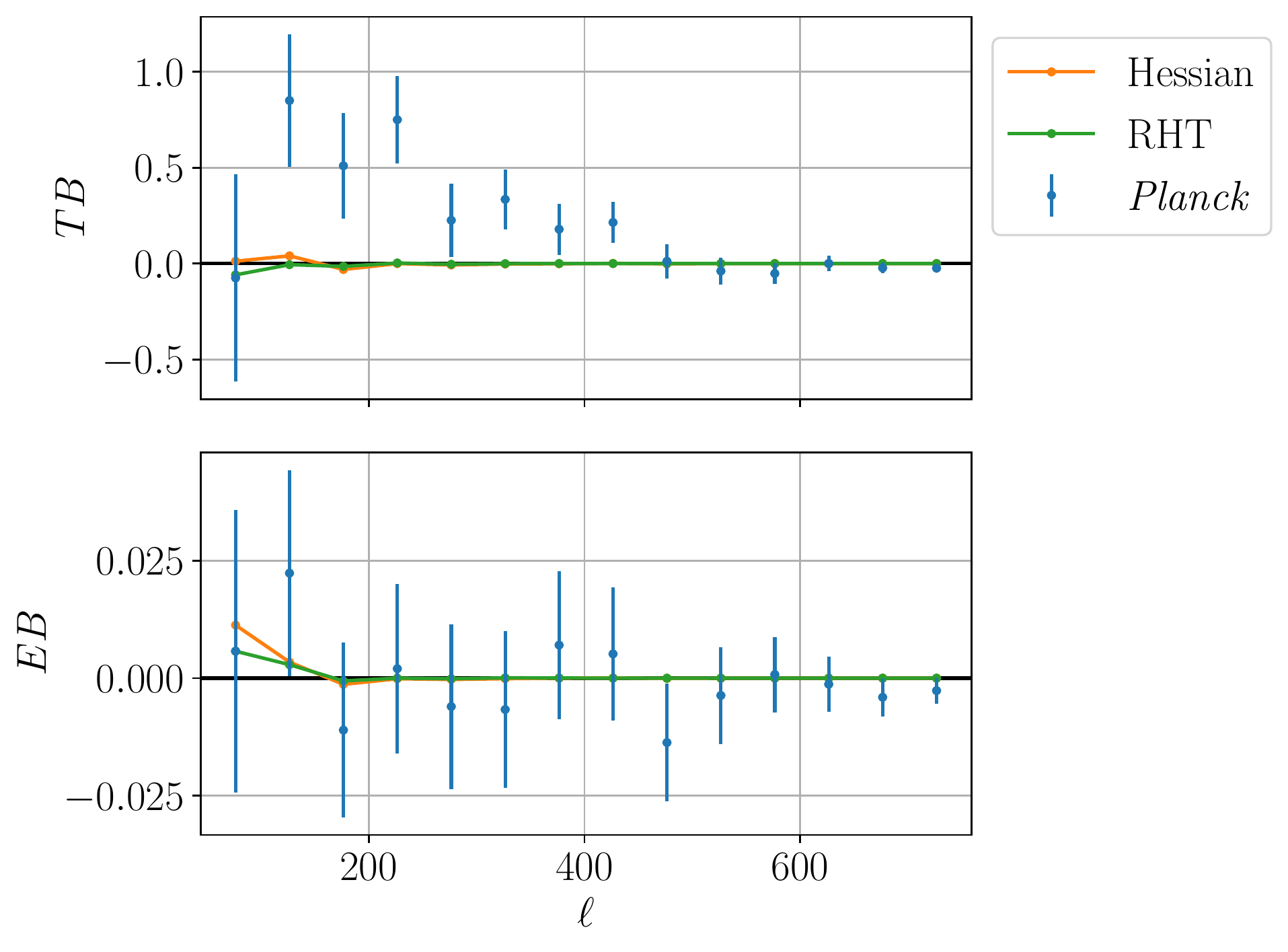}
\caption{Parity-violating $TB$~(top) and $EB$~(bottom) spectra measured by \emph{Planck}~(blue) and predicted by the Hessian \HI\ template~(orange) and by the RHT~(green). The multipole bin width is~$\Delta\ell = 50$.  The units are~$\mu\mathrm{K}_\mathrm{RJ}^2$. The error bars are derived from Gaussian variances. The \HI-based predictions include the transfer function of \secref{sec:TransferFunction}. For $\ell > 100$,  the multipole range used in most of our analysis, the \HI-based predictions are negligible in comparison with the \emph{Planck} measurements. \label{fig:TBEBtemplate} }
\end{figure}

We computed these parity-violating cross-spectra for both the Hessian and the RHT templates.\footnote{Our RHT implementation has been updated since \citet{ClarkHensley2019}. We call this new version the ``spherical RHT'', because it employs a convolution on the sphere. This both speeds up the computation and removes a spurious $E_\mathrm{RHT} B_\mathrm{RHT}$~correlation that is present at the $5\%$~level in the \HI~templates of~\citet{ClarkHensley2019}. We will report on the spherical RHT in greater detail in future work~\citep{HalalPrep}.}  To determine if the results are significant, we compare to the $T_\mathrm{d} B_\mathrm{d}$~and $E_\mathrm{d} B_\mathrm{d}$~spectra from the \emph{Planck} dust maps.  To make this comparison, we applied the transfer function~$k_\ell^{(X)}$ introduced in \secref{sec:TransferFunction}, which converts the \HI~templates into dust-intensity units.  
The results are shown in \figref{fig:TBEBtemplate}.
The \emph{Planck}~$T_\mathrm{d} B_\mathrm{d}$ displays a positive signal, and $T_\mathrm{\HI} B_\mathrm{\HI}$~is negligible in comparison. The \emph{Planck}~$E_\mathrm{d} B_\mathrm{d}$ appears to be consistent with noise, and we find that $E_\mathrm{\HI} B_\mathrm{\HI}$~is negligible in comparison with the fluctuations.  The above comparisons are restricted to $\ell > 100$, which is the target multipole range of the analysis presented in this work. 

Based on these observations, the \emph{intrinsic} (or \emph{morphological}) \HI-based $T_\mathrm{\HI} B_\mathrm{\HI}$ and~$E_\mathrm{\HI} B_\mathrm{\HI}$ are assumed in our analysis to vanish.

Intriguingly,  however, the \HI-based~$E_\mathrm{\HI} B_\mathrm{\HI}$ shows a rise for $\ell < 150$.  With finer multipole binning~$\Delta\ell$, we find that this rise persists down to $\ell = 17$ with $\Delta\ell=10$, the lowest bin center with the finest binning that we checked.  There is a corresponding rise in~$T_\mathrm{\HI}B_\mathrm{\HI}$ that persists down to~$\ell = 27$.  In all cases, the \HI-based predictions are subdominant to the expected noise in the \emph{Planck} measurements but only by a factor of~$\sim 10$.  We defer to future work an investigation of these low-$\ell$ \HI-based predictions, which could represent a source of parity violation independent of magnetic misalignment.

\section{Cross-power and correlation metrics \label{sec:correlationMetrics}}

We make use of a variety of correlation metrics. We wish for these metrics to be numerically stable,  unbiased by noise or other covariances and, in some cases,  sensitive to two different effects simultaneously.

Let~$\overline{X}$ be the sample mean for a set of $n$~measurements~$\{X_i\}$.  The mean-subtracted observable is
\eq{ x_i \equiv X_i - \overline{X} . }
The index~$i$ will be labeling the central sky coordinate of small patches.  Unless the sky mask has been chosen to retain relatively isotropic dust power,  the dust intensity can vary dramatically across the observing region. It is, therefore, likely that the set of observables~$\{x_i\}$ is dominated by the brightest patches.  Because many of the quantities of interest range over several orders of magnitude, we prefer metrics related to the Spearman rank correlation coefficient, for which the observables are converted to \emph{rank variables}. This avoids overweighting bright sightlines.  By collapsing the observables onto ranks, the absolute magnitudes are less important, and both large and small values of~$X$ contribute equally.

We consider correlation metrics for two data vectors~$X$ and~$Y$.
When the data vectors are noisy, it is more useful to consider a \emph{cross-power}, which is essentially the numerator of a Pearson correlation coefficient. We define the Pearson cross-power of~$X$ and~$Y$ as
\eq{ s_p(X,Y) \equiv \sum_i x_i y_i .  \label{eq:sp} }
As mentioned above,  we will often prefer quantities related to the Spearman rank correlation coefficient.  For the Spearman cross-power, which we denote~$s_s(X,Y)$, we simply convert~$x_i$ and~$y_i$ to rank variables in \eqref{eq:sp}.

Both Pearson and Spearman correlation coefficients are biased low by noise. We can use data splits to avoid positive-definite quantities like~$x_i^2$.  For data split~$j \in \{1,2\}$, define the mean-subtracted observable as $x_i^{(j)} \equiv X_i^{(j)} - \overline{X}$, 
i.e., the mean that is subtracted is from the \emph{full} data.  We modify the denominator of the Pearson correlation coefficient to include data-split cross-powers (\eqref{eq:sp}), which are unbiased by noise: 
\eq{ \tilde{r}_p(X,Y) \equiv \frac{ s_p(X,Y) }{ \sqrt{ \abs{ s_p(X^{(1)},X^{(2)}) } \abs{ s_p(Y^{(1)},Y^{(2)}) } } } ,  \label{eq:pearsonDenomDebiased}}
where the tilde serves to indicate that this is a modification to the conventional definition of a Pearson correlation coefficient. To avoid numerical pathologies in the case of noisy data, we have taken the absolute values of the cross-powers in the denominator.  When noise dominated, $\tilde{r}_p(X,Y)$~can return values larger in magnitude than unity.  In such cases, we will typically avoid correlation coefficients~(\eqref{eq:pearsonDenomDebiased}) in favor of cross-powers~(\eqref{eq:sp}). We can form a Spearman version of \eqref{eq:pearsonDenomDebiased}, which we denote~$\tilde{r}_s(X,Y)$, by using~$s_s$~(Spearman version of \eqref{eq:sp}) in place of~$s_p$.

While $\tilde{r}_p(X,Y)$~(\eqref{eq:pearsonDenomDebiased}) avoids a suppression due to noise biases, it may still be vulnerable to noise covariances between~$X$ and~$Y$, though this is only a concern if $X$~and $Y$ are drawn from related data sets.  For instance, both may be derived from \emph{Planck} dust maps, so noise covariances must be considered seriously. We define the half-mission cross-correlation coefficient
\eq{ \tilde{r}_p^{(\mathrm{HM})}(X,Y) \equiv \frac{ s_p\parens{X^{(1)},Y^{(2)}} + s_p\parens{X^{(2)},Y^{(1)}} }{2 \sqrt{ \abs{ s_p(X^{(1)},X^{(2)}) } \abs{ s_p(Y^{(1)},Y^{(2)}) } } } \label{eq:pearsonHM} , }
which measures a simultaneous correlation between opposite data splits of~$X$ and~$Y$. The convenient splits for \emph{Planck} are half-mission (HM) maps.  As above, we form the Spearman version simply by using~$s_s$ in place of~$s_p$.

In some cases, we will wish to measure simultaneous correlations between two \emph{pairs} of observables, e.g., a correlation between~$W$ and~$X$ and a correlation between~$Y$ and~$Z$.  We define the 4-variable cross-power
\eq{ S_p(W,X,Y,Z) \equiv \sum_i \parens{ w_i x_i + y_i z_i}  \label{eq:Sp} . }
As above, we can construct a Spearman version of this cross-power, which we denote~$S_s(W,X,Y,Z)$, by converting to rank variables.

\subsection{Statistical inference \label{sec:statInf} }

For statistical inference regarding our correlation metrics, we use \emph{permutation tests}.  For 2-variable metrics, we randomly  permute~$Y$ to obtain~$\pi(Y)$, where the function~$\pi$ defines a random permutation that here acts on the data vector~$Y$.  We then compute, e.g., $s_s(X,\pi(Y))$. With a large number of permutations, we can build a null-hypothesis distribution for~$s_s(X,Y)$. Similar permutation tests can be formed for the other 2-variable correlation metrics. In this work, each ensemble of permutations contains 200~realizations. Our results change only negligibly with larger permutation ensembles.

For the 4-variable correlation metrics, we also appeal to permutation tests, but we coordinate the permutations of~$X$ and~$Z$. We form many realizations of, e.g., $S_s(W,\pi(X),Y,\pi(Z))$. 

We will often estimate uncertainties on our correlation coefficients by converting to $z$~scores and taking half the difference between the value at~$1\sigma$ and the value at~$-1\sigma$.  For~$s_s(X,Y)$,  call this uncertainty estimate~$\sigma[s_s(X,Y)]$, and we maintain similar notational conventions for the other correlation metrics.

\bibliography{MagneticMisalignment}{}
\bibliographystyle{aasjournal}

\end{document}